%% file: main_arxiv_update_Aug_2022.tex
\pdfoutput=1
\documentclass[11pt,a4paper]{article}
\usepackage[top=75pt,bottom=75pt,left=60pt,right=60pt]{geometry}
\usepackage{authblk}
\usepackage[utf8]{inputenc}
\usepackage{caption}
\usepackage{subcaption}
\usepackage{units}
\usepackage{bigdelim}

\usepackage{xcolor}
\usepackage{xfrac}
\usepackage{mathtools}
\usepackage[acronym]{glossaries}
\usepackage{amssymb}
\usepackage{amsmath}
\usepackage{comment}
\usepackage{appendix}
\usepackage{hyperref}
\usepackage[square,sort]{natbib}
\usepackage{algpseudocode}

\newcommand{\RomanNumeralCaps}[1]
\linenumbers
\usepackage{array}
\newcolumntype{C}[1]{>{\PreserveBackslash\centering}p{#1}}
\newcolumntype{R}[1]{>{\PreserveBackslash\raggedleft}p{#1}}
\newcolumntype{L}[1]{>{\PreserveBackslash\raggedright}p{#1}}

\title{Learned Turbulence Modelling with Differentiable Fluid Solvers: Physics-based Loss-functions and Optimisation Horizons}
\author{Björn List,
        Li-Wei Chen,
        Nils Thuerey}
\affil{Departement of Informatics, Technical University of Munich}

\begin{document}
\newacronym{nn}{NN}{Neural Network}
\newacronym{gpu}{GPU}{Graphics Processing Unit}
\newacronym{tf}{TF}{TensorFlow}
\newacronym{cnn}{CNN}{Convolutional Neural Network}
\newacronym{dns}{DNS}{Direct Numerical Simulation}
\newacronym{cfl}{CFL}{Courant-Friedrichs-Lewy}
\newacronym{rans}{RANS}{Reynolds-Averaged-Navier-Stokes}
\newacronym{les}{LES}{Large Eddy Simulation}
\newacronym{sgs}{SGS}{subgrid-scale}

\hyphenation{sig-ni-fi-cance}
\hyphenation{ex-po-sing}

\maketitle

\begin{abstract}
In this paper, we train turbulence models based on convolutional neural networks. These learned turbulence models improve under-resolved low resolution solutions to the incompressible Navier-Stokes equations at simulation time. Our study involves the development of a differentiable numerical solver that supports the propagation of optimisation gradients through multiple solver steps.
The significance of this property is demonstrated by the superior stability and accuracy of those models that unroll more solver steps during training.
Furthermore, we introduce loss terms based on turbulence physics that further improve the model accuracy. This approach is applied to three two-dimensional turbulence flow scenarios, a homogeneous decaying turbulence case, a temporally evolving mixing layer, and a spatially evolving mixing layer. Our models achieve significant improvements of long-term \textit{a-posteriori} statistics when compared to no-model simulations, without requiring these statistics to be directly included in the learning targets. At inference time, our proposed  method also gains substantial performance improvements over similarly accurate, purely numerical methods.
\end{abstract}

\setcounter{page}{1}

\input{sections/introduction.tex}
\input{sections/learning.tex}
\input{sections/isotropic_turbulence.tex}
\input{sections/planar_mixing_layers.tex}
\input{sections/backpropagation.tex}
\input{sections/performance.tex}
\input{sections/conclusion_arxiv_v2.tex}

\bibliographystyle{plainnat}
\newpage
\addcontentsline{toc}{section}{Bibliography}
\bibliography{main_arxiv_update_Aug_2022.bbl}

\newpage

\input{sections/appendix.tex}

\end{document}

%% file: sections/introduction.tex
\section{Introduction}
\label{section:introduction}
Obtaining accurate numerical solutions to turbulent fluid flows remains a challenging task, and is subject to active research efforts 
in fluid dynamics \citep{Argyropoulos2015RecentFlows} 
and adjacent fields including climate research \citep{Aizinger2015Large-scaleSimulations} and the medical sciences \citep{Bozzi2021TheActivation}. \gls{dns}, which attempts to fully resolve the vast scale of turbulent motion, is prohibitively expensive in many flow scenarios and is thus often adverted by using turbulence models. For instance, \gls{rans} modelling has successfully been deployed to complex flow problems such as aircraft shape design and optimisation of turbo-machinery \citep{Argyropoulos2015RecentFlows}. However, the temporally averaged solutions from \gls{rans} simulations lack concrete information about instantaneous vortex movements in the flow. Thus, \gls{les} constitutes another common choice for turbulence modelling, providing a time-sensitive perspective to the turbulent flows \citep{Pope2004}. The computational expense of \gls{les} is nevertheless still substantial, and their applicability remains restricted \citep{choi2012grid,Zhiyin2015, slotnick2014cfd}.

The persistent challenges of traditional approaches motivate the use of machine learning, in particular deep learning, for turbulence modelling \citep{Duraisamy2019}. The reduced complexity of steady-state \gls{rans} made these setups a promising target for early efforts of machine learning based turbulence. As a result, substantial progress has been made towards data-driven prediction of \gls{rans} flow fields, vastly outperforming pure numerical solvers in the process \citep{Thuerey2020DeepFlows,Bhatnagar2019PredictionNetworks, ling2016reynolds}.

Contrasting data-driven \gls{rans} modelling, further studies were motivated by the additional challenges of predicting transient turbulence. Some of these target performance gains over numerical models by moving the temporal advancement to a reduced order embedding, where Koopman-based approaches have been an effective choice for constructing these latent spaces \citep{lusch2018deep, EIVAZI2021108816}. In the domain of deep learning based fluid mechanics, these studies are also among the first to explore the effects of recurrent application of neural networks on training. A related approach by \cite{li2020fourier} moved the learned temporal integrator to Fourier space, with successful applications to a range of problems including Navier-Stokes flow. An extensive comparison of different turbulence prediction architectures is provided by \cite{stachenfeld2021learned}, and includes applications to multiple flow scenarios.

While turbulence prediction aims to remove the numerical solver at inference time, other concepts on machine learning turbulence try to integrate a learned model in the solver. In the following, we will refer to approaches characterised by this integration of neural networks into numerical solvers as \textit{hybrid methods}. Some of these efforts target the data-driven development of \gls{les} models. An early work showcased the capability of neural networks to reproduce the turbulent viscosity coefficient \citep{Sarghini2003}. Furthermore, \cite{Maulik2019} proposed a supervised machine learning method to infer the \gls{sgs}
stress tensor from the flow field, and achieved promising results on the two-dimensional decaying turbulence test cases. Herein, the \textit{a-priori} evaluations served as a learning target and could be accurately reproduced, however \textit{a-posteriori} evaluations were not always in direct agreement. \cite{Beck2019DeepModels} trained a data-driven closure-model based on a \gls{cnn} and demonstrated good accuracy at predicting the closure on a three-dimensional homogeneous turbulence case, albeit stating that using their trained model in \gls{les} is not yet possible. Related prediction capabilities with trade-offs in terms of model stability of a similar supervised approach were reported by \cite{Cheng2019}. \cite{xie2019modeling} utilised a similar approach on compressible flows, later expanding their method to multi-scale filtering \citep{xie2020spatially}. \cite{park2021toward} studied possible formulations for the input to the neural network and evaluated their results on a turbulent channel flow.

Beyond the supervised learning methods covered so far, \cite{Novati2021AutomatingLearning} proposed a multi-agent reinforcement learning approach, where the \gls{les} viscosity coefficient was inferred by local agents distributed in the numerical domain. Their hybrid solver achieved good results when applied to a forward simulation. These previous studies on machine learning based turbulence models lead to two fundamental observations. Firstly, sufficiently large networks parameterise a wide range of highly non-linear functions. Their parameters, i.e. network weights, can be trained to identify and differentiate turbulent structures and draw modelling conclusions from these structures, which yields high accuracy towards \textit{a-priori} statistics. Secondly, the feedback from supervised training formulations cannot express the long term effects of these modelling decisions, and thus cannot provide information about the temporal
stability of a model. While reinforcement learning provides long temporal evolutions, its explorative nature 
makes this method computationally expensive. To exploit the benefits of data-driven training like supervised models, and simultaneously provide training feedback over long time horizons, a deeper integration of neural network models in numerical solvers is necessary.

Further research achieved this deep integration by training networks through differentiable solvers and adjoint optimisation for partial differential equations. Such works initially focused on learning-based control tasks \citep{Holl2020,Belbute-Peres2018}. By combining differentiable solvers with neural network models, optimisation gradients can propagate through solver steps and network evaluations \citep{thuerey2021physics}. This allows for targeting loss formulations that require a temporal evolution of the underlying partial differential equation.
These techniques were shown to overcome the stability issues of supervised methods, and thus provided a basis for hybrid methods in unsteady simulations. By integrating \gls{cnn}s into the numerical solver, \cite{Um2020Solver-in-the-loop:PDE-solvers} found models to improve with increased time horizons seen during training, which resulted in a stable learned correction function that was capable of efficiently improving numerical solutions to various partial differential equations. Similarly, \cite{Kochkov} found differentiable solver architectures to be beneficial for training turbulence models.
While this work estimates substantial performance gains over traditional techniques for first-order time integration schemes, we will evaluate a different solver that is second-order in time, putting more emphasis on an evaluation with appropriate metrics from fluid mechanics.

In another related approach, \cite{Sirignano2020} proposed a learned correction motivated by turbulence predictions in  \gls{les} of isotropic turbulence, and later expanded on this by studying similar models in planar jets \citep{MacArt2021}. Here, \textit{a-posteriori} statistics served as a training target, and the authors also compared the performance of models trained on temporally averaged and instantaneous data. However, the study did not investigate the temporal effects of hybrid solvers and their training methodologies in more detail.

In this paper, we seek to develop further understanding of turbulence modelling with hybrid approaches. In an effort to bridge the gap between previously mentioned papers we want to address a series of open questions. 
Firstly, no previous adjoint-based learning approach has been evaluated on a range of turbulent flow scenarios. While this has been done for other, purely predictive learning tasks \citep{li2020fourier,stachenfeld2021learned}, we will demonstrate the applicability of adjoint-based training of hybrid methods in multiple different scenarios.
Secondly, there is little information on the choice of loss functions for turbulence models in specific flow scenarios. Previous studies have focused on matching ground truth data. Their optimisation procedures did not emphasise specific fluid dynamical features that might be particularly important in the light of long term model accuracy and stability. 
Thirdly, previous works on adjoint optimisation have not studied in detail how the number of unrolled steps seen during training  affects the neural network models' \textit{a-posteriori} behaviour. While previous work on \textit{flow prediction} reported good results when using multiple prediction steps during training \citep{EIVAZI2021108816,lusch2018deep}, we want to explore how this approach behaves with learned turbulence models in \textit{hybrid solvers}.
In order to provide insights to these questions, we utilise a \gls{cnn} to train a corrective forcing term through a differentiable solver, which allows an end-to-end training that is flexible towards the number of unrolled steps, loss formulations and training targets. We then show that the same network architecture can achieve good accuracy with respect to \textit{a-posteriori} metrics of three different flow scenarios. In our method, we relax the timestep requirements usually found in unsteady turbulence modelling, such as \gls{les}, by downscaling our simulations such that a constant \gls{cfl} ratio is maintained. By implication, a learned model is trained to (i) take the classical sub-grid scale closure into account, (ii) approximate temporal effects, and (iii) correct for discretisation errors. 
It is worth noting that a network trained for these three targets combines their treatment into one output, with the result that these treatments cannot be separated at a network-output level. Instead, our \textit{a-posteriori} evaluations show that neural network models can learn to account for all three of these elements.

The turbulence models are trained and evaluated on three different, two-dimensional flow cases: the isotropic decaying turbulence, a temporally developing mixing layer as well as the spatially developing mixing layer. We show that in all cases, training a turbulence model through an increasing number of unrolled solver steps enhances the model accuracy and thus demonstrate the benefits of a differentiable solver. Unless stated otherwise, all of the evaluations in the coming sections were performed on out-of-sample data and show the improved generalising capabilities of models trained with the proposed unrollment strategy.

Our unrollment study extends to 60 simulation steps during training. The long solver unrollments involve recurrent network applications, which can lead to training insabilities caused by exploding and diminishing gradients. We introduce a custom gradient stopping technique that splits the gradient calculations into non-overlapping subranges, for which the gradients are evaluated individually. This techniques keeps the long term information from all unrolled steps, but stops the propagation of gradients through a large number of steps and thus avoids the training instabilities.

Furthermore, our results indicate that accurate models with respect to \textit{a-posteriori} turbulence statistics are achieved without directly using them as training targets. Nonetheless, a newly-designed loss formulation inspired by \textit{a-posteriori} evaluations and flow physics is shown to yield further improvements. Finally, we provide a performance analysis of our models that measures speed-ups of up to $14$ with respect to comparably accurate solutions from traditional solvers. 

The remainder of this paper is organised as follows. In section \ref{section:learning_turbulence_models}, we give an overview of our methodology and the solver-network interaction. A description and evaluation of experiments with the isotropic decaying turbulence case is found in section \ref{section:isotropic_turbulence}, which is followed by similar studies regarding the temporally developing mixing layer and the spatially developing mixing layer in sections \ref{subsection:temporal_mixing_layers} and \ref{subsection:spatial_mixing_layers} respectively. Section \ref{section:backpropagation} studies the effect our method of splitting back-propagated gradients into subranges. A comparison of computational costs at inference time can be found in section \ref{section:perfromance}, while section \ref{section:conclusion} contains concluding thoughts. 

%% file: sections/learning.tex
\section{Learning Turbulence Models}
\label{section:learning_turbulence_models}
In this paper, we study neural networks for turbulence modelling in in\-com\-pressible fluids. These flows are governed by the Navier-Stokes equations
\begin{equation} \label{equation:navier_stokes}
\begin{aligned}
    &\frac{\partial \mathbf{u}}{\partial t} + \mathbf{u}\cdot\nabla\mathbf{u} = -\nabla p + \frac{1}{\text{Re}}\nabla^2\mathbf{u} + \mathbf{f} ,\\
    &\nabla \cdot \mathbf{u} = 0,
\end{aligned}
\end{equation}
where $\mathbf{u}= [u \hspace{0.5em} v]^T$, $p$ and $Re$ are the velocity field, pressure field and Reynolds number respectively. The term $\mathbf{f} = [f_x \hspace{0.5em} f_y]^T$ represents an external force on the fluid. In the context of turbulent flows, an accurate solution to these equations entails either resolving and numerically simulating all turbulent scales, or modelling the turbulence physics through an approximative model.

Our aim is to develop a method that enhances fluid simulations by the means of a machine learning model. In particular, we aim to improve the handling of fine temporal and spatial turbulence scales that are potentially under-resolved, such that the influence of these scales on the larger resolved motions needs to be modelled. The function that approximates these effects is solely based on the low resolution data and is herein parameterised by a \gls{cnn}. The network is then trained to correct a low-resolution numerical solution during the simulation, such that the results coincides with a downsampled high-resolution dataset. Within this hybrid approach, the turbulence model directly interacts with the numerical solver at training and at inference time. To achieve this objective, we utilise differentiable solvers, i.e. solvers which provide derivatives with respect to their output state.
Such solvers can be seen as part of the differentiable programming methodology in deep learning, which is equivalent to employing the adjoint method from classical optimisation \citep{giles2003algorithm} in the context of neural networks.
The differentiability of the solver enables the propagation of optimisation gradients through multiple solver steps and neural network evaluations.

\subsection{Differentiable PISO solver} 
\label{subsection:PISO}
Our differentiable solver is based on the semi-implicit PISO-scheme introduced by \cite{Issa1986}, which has been used for a wide range of flow scenarios \citep{Barton1998ComparisonFlows,Kim1992ComparisonFlows}. Each second-order time integration step is split into an implicit predictor step solving the discretised momentum equation, followed by two corrector steps that ensure the incompressibility of the numerical solution to the velocity field. The Navier-Stokes equations are discretised using the Finite-Volume method, while all cell fluxes are computed to second-order accuracy.

The solver is implemented on the basis of \textit{TensorFlow} \citep{199317},  which facilitates parallel execution of linear algebra operations on the \gls{gpu}, as well as differentiability of said operations. Additional functions exceeding the scope of \gls{tf} are written as custom operations and implemented using CUDA. This approach  allows us to seamlessly integrate initially unsupported features such as sparse matrix operations in the \gls{tf}-graph. More details about the solver can be found in appendix \ref{appendix:piso_solver}, where the solver equations are listed in \ref{appendix:solver_equations}, implementation details in \ref{appendix:implementation}, and a verification is conducted in \ref{appendix:verification}. Figure \ref{fig:piso_one_figure} gives a brief overview of the solver procedure.
In the following, we will denote a PISO solver step $\mathcal{S}$ as
\begin{equation}\label{eq:piso_notation}
    (\mathbf{u}_{n+1}, p_{n+1}) = \mathcal{S}(\mathbf{u}_n, p_n,\mathbf{f}_n)\
\end{equation}
where $\mathbf{u}_{n}$, $p_{n}$ and $\mathbf{f}_n$ represent discretised velocity, pressure and forcing fields at time $t_n$.

\begin{figure}
\centering
\includegraphics[width=0.9\textwidth]{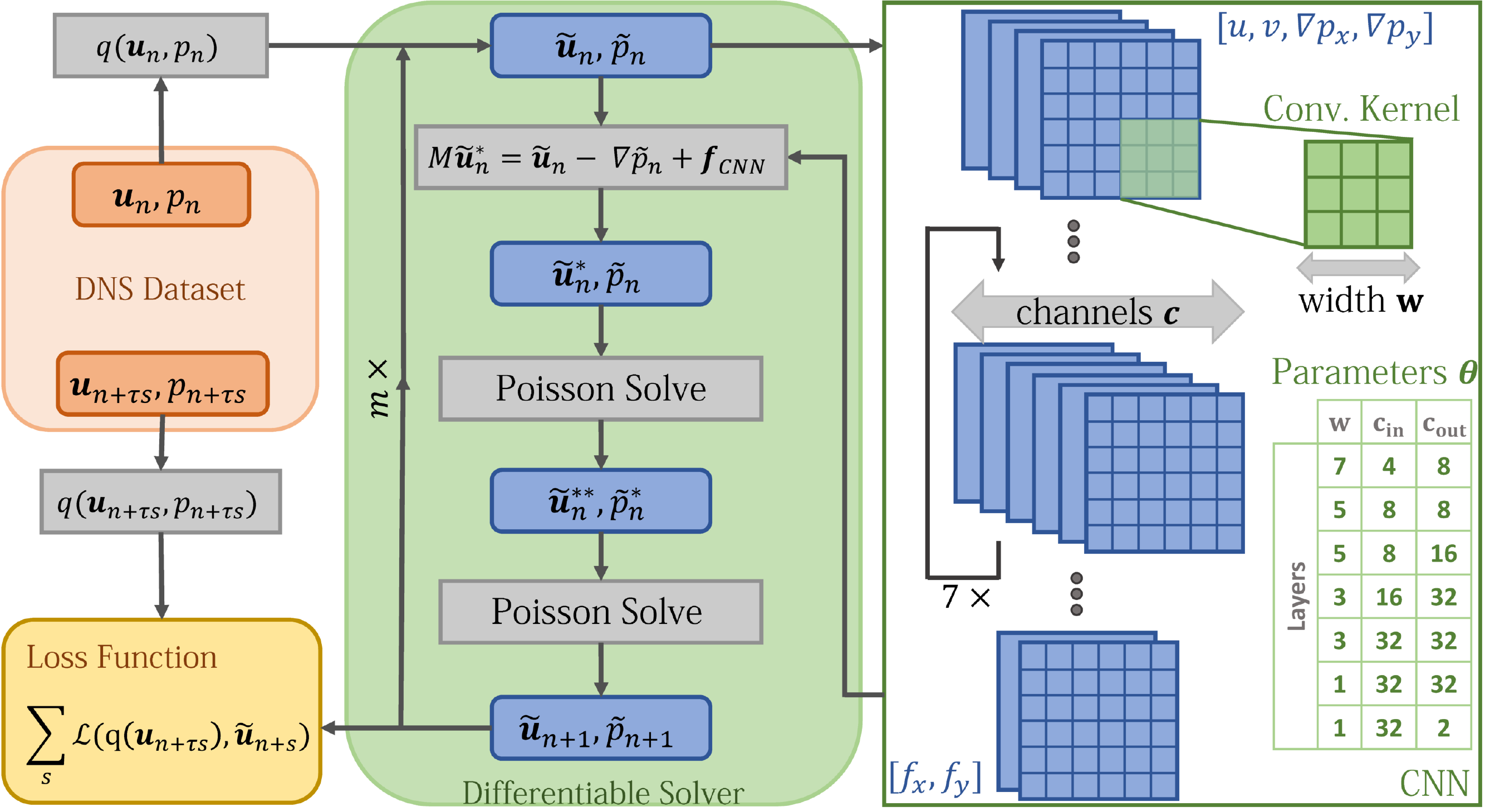}
\caption{Solver procedure of the PISO-scheme and its interaction with the convolutional neural network; data at time $t_n$ is taken from the \gls{dns} dataset and processed by the resampling $q$ before entering the differentiable solver; the solver unrollment performs $m$ steps, each of which is corrected by the \gls{cnn}, and is equivalent to $\tau$ high-resolution steps; the optimisation loss takes all resulting (intermediate) timesteps}
\label{fig:piso_one_figure}
\end{figure}

\subsection{Neural Network Architecture} 
\label{subsection:cnn}
Turbulence physics strongly depends on the local neighbourhood. Thus, the network has to infer the influence of unresolved scales for each discrete location based on the surrounding flow fields. This physical relation can be represented by discrete convolutions, where each output value is computed based solely on the surrounding computational cells as well as a convolutional weighting kernel.  This formulation introduces a restricted receptive field for the convolution and ensures the local dependence of its output \citep{Luo2016}.
Chaining multiple of these operations results in a deep \gls{cnn}, which was successfully used in many applications ranging from computer vision and image recognition \citep{Albawi2018} to fluid mechanics and turbulence research \citep{Guastoni2020,Beck2019DeepModels, Lapeyre2019}.
We use a fully convolutional network with 7 convolutional layers and leaky ReLU activations, containing $\sim 82\times10^3$ trainable parameters.
As illustrated in figure \ref{fig:piso_one_figure}, our \gls{cnn} takes the discretised velocity and pressure gradient fields as input.  This formulation contains full information of the field variable states, and enables the modelling of both temporal and spatial effects of turbulence, as well as correction of numerical inaccuracies. However, any principles of the modelled physics, like Galilean invariance in the case of \gls{sgs}-closure, must be learnt by the network itself. The choice of network inputs is by no means trivial, but shall not be further studied in this paper. Refer to \citep{choi2012grid,xie2019modeling,xie2020spatially,MacArt2021} for in-depth analyses. The output of our networks is conditioned on its weights $\theta$, and can be interpreted as a corrective force $\mathbf{f}_{\text{CNN}}(\tilde{\mathbf{u}}_{n}, \nabla \tilde{p}_{n}|\theta):\mathbb{R}^{\tilde N_x\times\tilde N_y \times 4}\xrightarrow{}\mathbb{R}^{\tilde N_x \times\tilde N_y\times 2}$
to the under-resolved simulation of the Navier-Stokes equations \eqref{equation:navier_stokes} with domain-size $\tilde N_x\times\tilde N_y$. This force directly enters the computational chain at PISO's implicit predictor step. As a consequence, the continuity equation is still satisfied at the end of a solver step, even if the simulation is manipulated by the network forcing.
For a detailed description of the network structure, including \gls{cnn} kernel sizes, intialisations and padding, refer to appendix \ref{appendix:cnn}.

\subsection{Unrolling Timesteps for Training}
Our method combines the numerical solver introduced in section \ref{subsection:PISO} with the modelling capabilities of \gls{cnn}s as outlined in \ref{subsection:cnn}. As also illustrated in figure \ref{fig:piso_one_figure}, the resulting data-driven training algorithm works based on a dataset $(\mathbf{u}(t_n), p(t_n))$ consisting of high-resolution  ($N_x\times N_y$) velocity fields $\mathbf{u}(t_n)\in \mathbb{R}^{N_x\times N_y \times 2}$ and corresponding pressure fields $p(t_n)\in \mathbb{R}^{N_x\times N_y}$ for the discrete time $t_n$. In order to use this \gls{dns} data for training underresolved simulations on different grid resolutions, we define a down-sampling procedure $q(\mathbf{u}, p):\mathbb{R}^{ N_x\times N_y \times 3}\xrightarrow{}\mathbb{R}^{\tilde N_x \times\tilde N_y\times 3}$
, that takes samples from the dataset and outputs the data $(\tilde{\mathbf{u}}_n,\tilde{p}_n)$ at a lower target-resolution ($\tilde N_x\times \tilde N_y$) via bilinear interpolation. This interpolation provides a simple method of acquiring data at the shifted cell locations of different discretisations. It can be seen as a repeated linear interpolation to take care of two spatial dimensions.
The resampling of \gls{dns} data is used to generate input and target frames of an optimisation step. For the sake of simplicity, we will denote a downsampled member of the dataset consisting of velocities and pressure as $\tilde{q}_n=q(\mathbf{u}(t_n), p(t_n))$. Similarly, we will write $\tilde{\mathbf{f}}_n= \mathbf{f}_\text{CNN}(\tilde{\mathbf{u}}_{n},\nabla \tilde{p}_{n}|\theta)$. Note that the network operates solely on low-resolution data and introduces a corrective forcing to the low-resolution simulation, with the goal of reproducing the behaviour of a \gls{dns}.
We formulate the training objective as
\begin{equation}\label{eq:general_train}
    \min_\theta(\mathcal{L}(\tilde{q}_{n+\tau}, \mathcal{S}_\tau(\tilde{q}_n,\tilde{\mathbf{f}}_n))),
\end{equation}
for a loss function $\mathcal{L}$ that satisfies $\mathcal{L}(x,y)\xrightarrow{}0$ for $x\approx y$. By this formulation, the network takes a downsampled \gls{dns} snapshot and should output a forcing which makes the flow fields after a low-resolution solver step closely resemble the next downsampled frame. The temporal increment $\tau$ between these subsequent frames is set to match the timesteps in the low-resolution solver $\mathcal{S}$, which in turn are tuned to maintain Courant numbers identical to the \gls{dns}.

\cite{Um2020Solver-in-the-loop:PDE-solvers} showed that similar training tasks benefit from unrolling multiple temporal integration steps in the optimisation loop.
The optimisation can then account for longer term effects of the network output on the temporal evolution of the solution, increasing accuracy and stability in the process. We utilise the same technique and find it to be critical for the long-term stability of  turbulence models. Our notation from equations \eqref{eq:piso_notation} and \eqref{eq:general_train} is extended to generalise the formulation towards multiple subsequent snapshots.
When training a model through $m$ unrolled steps, the optimisation objective becomes
\begin{equation}\label{eq:multistep_train}
    \min_\theta\big(\sum_{s=0}^m\mathcal{L}(\tilde{q}_{n+s\tau},\mathcal{S}_\tau^s(\tilde{q}_n,\tilde{\mathbf{f}}_n))\big),
\end{equation}
where $\mathcal{S}^s$ denotes the successive execution of $s$ solver steps including network updates, starting with the initial fields $q_i$. By this formulation the optimisation works towards matching not only the final, but also all intermediate frames. Refer to appendix \ref{appendix:solver_equations} for a detailed explanation of this approach including equations for optimisation and loss differentiation.

\subsection{Loss Functions}
As introduced in equation \eqref{eq:general_train}, the training of deep \gls{cnn}s is an optimisation of its parameters. The loss function $\mathcal{L}$ serves as the optimisation objective and thus has to assess the quality of the network output. Since our approach targets the reproduction of \gls{dns}-like behaviour on a coarse gird, the chosen loss function should consequently aim to minimise the distance between the state of a modelled coarse simulation and the \gls{dns}. In this context, a natural choice is the $\mathcal{L}_2$-loss on the $s$-th unrolled solver step
\begin{equation}\label{eq:l2_loss}
    \mathcal{L}_2 = \sqrt{(\tilde{\mathbf{u}}_{s}-q(\mathbf{u}_{s\tau}))\cdot(\tilde{\mathbf{u}}_{s}-q(\mathbf{u}_{s\tau}))}\hspace{.2em},
\end{equation}
since this formulation drives the optimisation towards resembling a desired outcome. Therefore, the $\mathcal{L}_2$-loss trains the network to directly reproduce the downsampled high-resolution fields, and the perfect reproduction from an ideal model gives $\mathcal{L}_2=0$. Since the differentiable solver allows us to unroll multiple simulation frames, we apply this loss formulation across a medium-term time horizon and thus also optimise towards multi-step effects.
By repeatedly taking frames from a large \gls{dns} dataset in a stochastic sampling process, a range of downsampled instances are fed to the training procedure.
While the \gls{dns} dataset captures all turbulence statistics, they are typically lost in an individual training iteration. This is due to the fact that training mini-batches do not generally include sufficient samples to represent converged statistics, and no specific method is used to satisfy this criterion.
This means that data in one training iteration solely carries instantaneous information. Only the repeated stochastic sampling from the dataset lets the network recover awareness of the underlying turbulence statistics. The repeated matching of instantaneous \gls{dns} behaviour thus encodes the turbulence statistics in the training procedure.
While the $\mathcal{L}_2$-loss described in equation \eqref{eq:l2_loss} has its global minimum when the \gls{dns} behaviour is perfectly reproduced,
in practice, we find that it can neglect the time evolution of certain fine scale, low amplitude features of the solutions. This property of the $\mathcal{L}_2$-loss is not unique to turbulence modelling and has previously been observed in machine learning in other scientific fields such as computer vision \citep{Yu2018Super-ResolvingAttributes}.
To alleviate these shortcomings, we include additional loss-formulations, which alter the loss-landscape to avoid these local minima.

We define a spectral energy loss $\mathcal{L}_{E}$, designed to improve the accuracy of the learned model on fine spatial scales. It is formulated as the log-spectral distance of the spectral kinetic energies at the $s$-th step \begin{equation}\label{eq:spectral_energy_loss}
    \mathcal{L}_E = \sqrt{\int_k \log\Big(\frac{\tilde{E}_{s}(k)}{E_{s\tau}^q(k)}\Big)^2 dk},
\end{equation}
where $\tilde{E}_s(k)$ denotes the spectral kinetic energy of the low-resolution velocity field at wavenumber $k$, and $E^q_{s\tau}$ represents the same quantity for the downsampled \gls{dns} data. In practice, this loss formulation seeks to equalise the kinetic energy in the velocity field for each discrete wavenumber. The $\log$-rescaling of the two respective spectra regularises the relative influence of different spatial scales. This energy loss elevates the relative importance of fine scale features.

Our final aim is to train a model that can be applied to a standalone forward simulation. The result of a neural network modelled low-resolution simulation step should therefore transfer all essential turbulence information, such that the same model can once again be applied in the subsequent step. The premises of modelling the unresolved behaviour are found in the conservation equation for the implicitly filtered low-resolution kinetic energy in tensor notation
\begin{equation}
\begin{aligned}
\frac{\partial \tilde{E_f}}{\partial t} + \tilde{u}_i\frac{\partial \tilde{E_f}}{\partial x_i}+\frac{\partial}{\partial x_j}  \tilde{u}_i \big[ \delta_{ij}\tilde{p}+ \tau_{ij}^r - \frac{2}{\text{Re}}\tilde{\mathit{S}}_{ij} \big] = -\epsilon_f-\mathcal{P}^r,
\end{aligned}
\end{equation}
where $\tilde{E}_f$ denotes the kinetic energy of the filtered velocity field, $\tau_{ij}^r$ represents the \gls{sgs} stress tensor, $\mathit{\tilde{S}}_{ij}= \frac{1}{2}\big(\frac{\partial\tilde{u}_i}{\partial x_j} + \frac{\partial\tilde{u}_j}{\partial x_i}\big)$ is the resolved rate of strain, whereas $\epsilon_f$ and $\mathcal{P}^r$ are sink and source terms for the filtered kinetic energy. These terms are defined as $\epsilon_f = \frac{2}{Re}\mathit{\tilde{S}}_{ij}\mathit{\tilde{S}}_{ij}$ and $\mathcal{P}^r=-\tau_{ij}^r\mathit{\tilde{S}}_{ij}$. The viscous sink $\epsilon_f$ represents the dissipation of kinetic energy due to molecular viscous stresses at grid-resolved scales. In hybrid methods, this viscous dissipation at grid level $\epsilon_f$ is fully captured by the numerical solver. On the contrary, the source term $\mathcal{P}^r$ representing the energy transfer from resolved scales to residual motions cannot be computed, because the \gls{sgs} stresses $\tau_{ij}^r$ are unknown.
One part of the modelling objective is to estimate these unresolved stresses and the interaction of filtered and \gls{sgs} motions. Since the energy transfer between these scales $\mathcal{P}^r$ depends on the filtered rate of strain $\mathit{\tilde S}_{ij}$, special emphasis is required to accurately reproduce the filtered rate of strain tensor. This motivates the following rate of strain loss at the $s$-th unrolled solver step
\begin{equation}\label{eg:strain_loss}
    \mathcal{L}_{\mathit{S}}= \sum_{i,j}|\tilde{\mathit{S}}_{ij, s} - \mathit{S}^q_{ij, s\tau}| ,
\end{equation}
where $\mathit{S}^q_{ij, s}$ denotes the rate of strain of the downsampled high-resolution velocity field. This loss term insures that the output of a hybrid solver step carries the information necessary to infer an accurate forcing in the subsequent step.

While our models primarily focus on influences of small scale motions on the large scale resolved quantities, we now draw attention to the largest scale, the mean flow.
To account for the mean flow at training time, an additional loss contribution is constructed to match the multi-step statistics and written as
\begin{equation}\label{eq:thickness_loss}
   \mathcal{L}_\text{MS}= ||<\mathbf{u}_s>_{s=0}^m - <\tilde{\mathbf{u}}_{s\tau}>_{s=0}^m||_1,
\end{equation}
where $<>_{s=0}^m$ denotes an averaging over the $m$ unrolled training steps with iterator $s$. This notation resembles Reynolds-averaging, albeit being focused on the shorter time-horizon unrolled during training. Matching the averaged quantities is essential to achieving long-term accuracy of the modelled simulations for statistically steady simulations, but lacks physical meaning in transient cases. Therefore, this loss contribution is solely applied to statistically steady simulations. In this case, the rolling average $<>_{s=0}^m$ approaches the steady mean flow for increasing values of $m$.

Our combined turbulence loss formulation as used in the network optimisations additively combines the aforementioned terms as
\begin{equation}\label{eq:total_loss}
    \mathcal{L}_\text{T} = \lambda_2\mathcal{L}_2 + \lambda_E\mathcal{L}_E + \lambda_\mathit{S}\mathcal{L}_\mathbf{\mathit{S}} + \lambda_\text{MS}\mathcal{L}_\text{MS},
\end{equation}
where $\lambda$ denotes the respective loss factor. Their exact values are mentioned in the flow scenario specific chapters.
Note that these loss terms, similar to the temporal unrolling, do not influence the architecture or computational performance of the trained neural network at inference time. They only exist at training time to guide the network to an improved state with respect to its trainable parameters. In the following main sections of this paper, we use three different turbulence scenarios to study effects of the number of unrolled steps and the whole turbulence loss $\mathcal{L}_\text{T}$. An ablation on the individual components of $\mathcal{L}_\text{T}$ is provided in appendix \ref{appendix:loss_ablation}. We start with employing the training strategy on isotropic decaying turbulence.

%% file: sections/isotropic_turbulence.tex
\section{Two-Dimensional Isotropic Decaying Turbulence}
\label{section:isotropic_turbulence}
Isotropic decaying turbulence in two dimensions provides an idealised flow scenario \citep{Lilly1971}, and is frequently used for evaluating model performance \citep{Maulik2019,Kochkov,San2014ATurbulence}. It is characterised by a large number of vortices that merge at the large spatial scales whilst the small scales decay in intensity over time. We use this flow configuration to explore and evaluate the relevant parameters, most notably the number of unrolled simulation steps as well as the effects of loss formulations.

In order to generate training data, we ran a simulation on a square domain with periodic boundaries in both spatial directions. The initial velocity and pressure fields were generated using the initialisation procedure by \cite{San2012}. The Reynolds numbers are calculated as $Re = (\hat{e} \hat{l})/\nu$, with the kinetic energy $\hat{e}=(<u_iu_i>)^{1/2}$ and the integral length scale $\hat{l} = \hat{u}/\hat{\omega}$ and $\hat{\omega}=(<\omega_i\omega_i>)^{1/2}$. The Reynolds number of this initialisation was $Re=126$. The simulation was run for a duration of $T=10^4\Delta t_\text{\gls{dns}}= 100\hat{t}$, where the integral timescale is calculated as $\hat{t}=1/\hat{\omega}$ at the initial state. During the simulation, the backscatter effect transfers turbulence energy to the larger scales, which increases the Reynolds number \citep{Chasnov1997, kraichnan1967inertial}. In our dataset, the final Reynolds number was $Re=296$. Note that despite this change in Reynolds number, the turbulence kinetic energy is still decreasing and the flow velocities will decay to zero.

Our aim is to estimate the effects of small scale turbulent features on a coarser grid based on fully resolved simulation data. Consequently, the dataset should consist of a fully resolved \gls{dns} and suffice the resolution requirements. In this case the square domain $(L_x,L_y)=(2\pi,2\pi)$ was discretised by $(N_x,N_y) = (1024,1024)$ grid cells and the simulation evolved with a timestep satisfying \gls{cfl}$=0.3$.

We trained a  series of models on downsampled data, where spatial and temporal resolution were decreased by a factor of $8$ resulting in an effective simulation size reduction of $8^3=512$. Our best performing model was trained through $30$ unrolled simulation steps. This is equivalent to $1.96\hat t$ for the initial simulation state. Due to the decaying nature of this test case, the integral timescale increases over the course of the simulation, while the number of unrolled timesteps is kept constant.
As a consequence, the longest unrollments of 30 steps cover a temporal horizon similar to the integral timescale. All our simulation setups will study unrollment horizons ranging up to the relevant integral timescale, and best results are achieved when the unrollment approaches the integral timescale. For training the present setup,
the loss factors from equation \eqref{eq:total_loss} were chosen as $(\lambda_2 , \hspace{0.5em} \lambda_E, \hspace{0.5em} \lambda_\mathit{S}, \hspace{0.5em} \lambda_\text{MS})= (10, \hspace{0.5em} 5\times10^{-2}, \hspace{0.5em} 1\times10^{-5}, \hspace{0.5em} 0)$.

To evaluate the influence of the choice of loss function and the number of unrolled steps, several alternative models were evaluated.
Additionally, we trained a model with a traditional supervised approach. In this setting, the differentiable solver is not used, and the training is performed purely on the basis of the training data set. In this case, the corrective forcing is added after a solver step is computed. The optimisation becomes
\begin{equation}\label{eq:supervised}
    \min_\theta (\mathcal{L}(q_{n+\tau},\mathbf{f}_\text{CNN}(\mathcal{S}_\tau(q_n))).
\end{equation}
The equations for the supervised training approach are detailed in appendix \ref{appendix:solver_equations}.
Furthermore, a \gls{les} with the standard Smagorinsky model was included in the comparison. A parameter study targeting the Smagorinsky coefficient revealed that a value of $C_s=0.008$ handles the physical behaviour of our setup best. See appendix \ref{appendix:smagorinsky} for details. An overview of all models and their parameters is given in table \ref{table:randomised_l2_comparison}.

\begin{table}
\centering
    \begin{tabular}{ l c c c r r r}
    Name                        & \ Loss \              &\ Steps\ & $\hat t$ &  \ \ MSE at $t_1$         & MSE at $t_2$ \\
    \hline \hline
    $\text{NoModel}$            & -    & -   &-  & $2.78\mathrm{e}{-3}$ & $0.057$ \\
    $\text{LES}$                & -       & -      &-  & $2.69\mathrm{e}{-3}$ & $0.051$  \\
    $\text{NN}_\text{sup,T}$    &$\mathcal{L}_\text{T}$ &1 & 0.07 & $1.52\mathrm{e}{-3}$ & $0.369$ \\
    $\text{NN}_{1,\text{T}}$    &$\mathcal{L}_\text{T}$ &1 & 0.07 & $1.65\mathrm{e}{-3}$ & $0.046$ \\
     $\text{NN}_{10}$           &$\mathcal{L}_2$        &10 & 0.66 & $4.23\mathrm{e}{-4}$ & $0.018$ \\
     $\text{NN}_{10,\text{T}}$  &$\mathcal{L}_\text{T}$ &10 & 0.66 & $4.25\mathrm{e}{-4}$ & $0.022$\\
     $\text{NN}_{30,\text{T}}$  &$\mathcal{L}_\text{T}$ &30 & 1.98 & $4.09\mathrm{e}{-4}$ & $0.021$\\
    \end{tabular}
    \caption{Training details for models trained on the isotropic turbulence case, MSE evaluated at $t_1 = 64\Delta t\approx5\hat{t}$ and $t_2 = 512\Delta t\approx40\hat{t}$  }
    \label{table:randomised_l2_comparison}
\end{table}
\begin{figure}
    \centering
    \begin{subfigure}[b]{\textwidth}
        \centering
        \includegraphics[width=\textwidth]{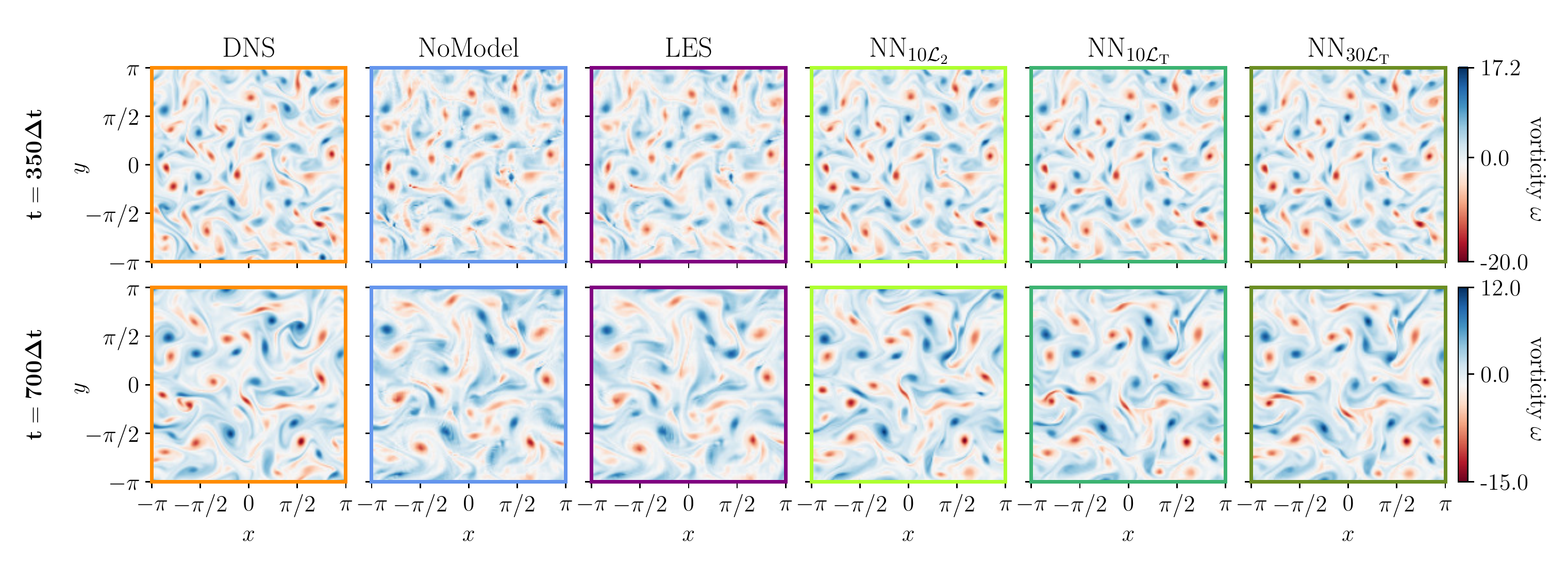}
    \end{subfigure}
    \begin{subfigure}[b]{\textwidth}
        \centering
        \includegraphics[width=\textwidth]{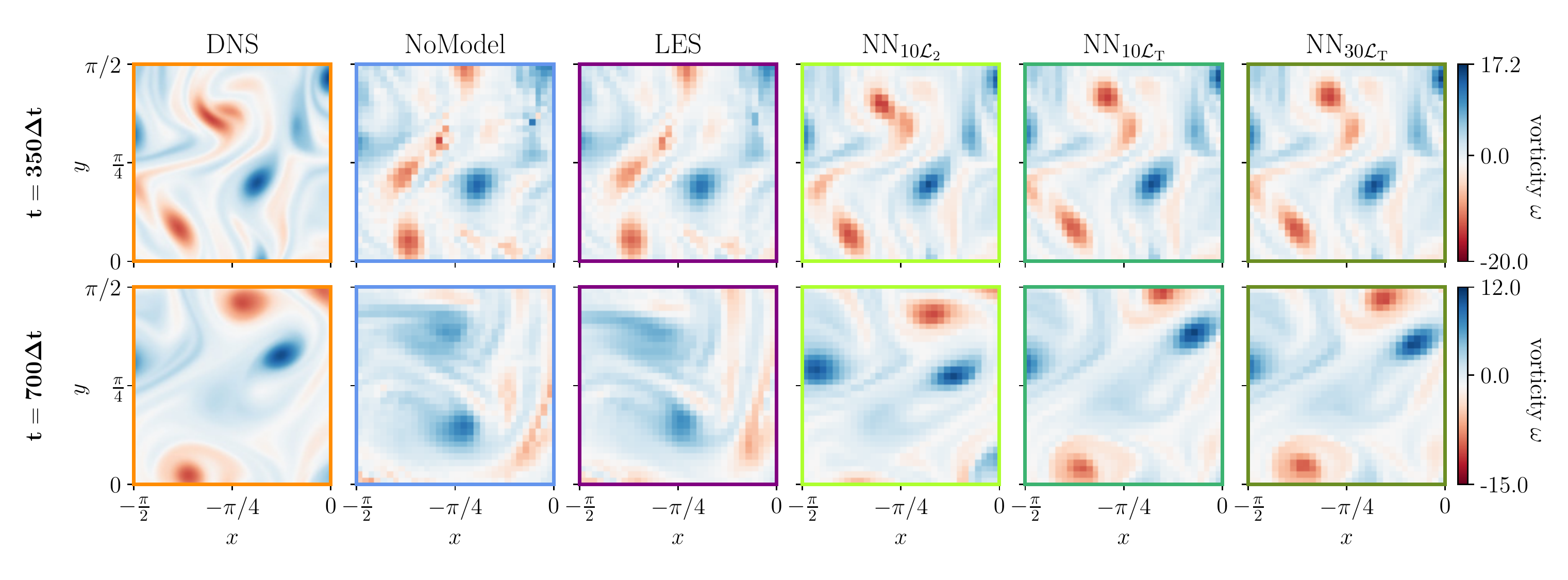}
    \end{subfigure}
    \caption{Vorticity visualisations of \gls{dns}, no-model,
    LES, and learned model simulations at $t = (350,700) \Delta t$ on the test dataset, zoomed-in version below}
    \label{fig:randomised_turbulence_vorticity}
\end{figure}

After training, a forward simulation was run for comparison with a test dataset. For the test-data, an entirely different, randomly generated initialisation was used, resulting in a velocity field different from the simulation used for training.
The test simulations were advanced for $1000\Delta t = 80\hat t$.

Note that the temporal advancement of the forward simulations greatly surpasses the unrolled training horizon, which leads to instabilities with the supervised and 1-step model, and ultimately to the divergence of their simulations. Consequently, we conclude that more unrolled steps are critical for the applicability of the learned models and do not include the 1-step model in further evaluations. While an unrollment of multiple steps also improves the accuracy of supervised models, these models nevertheless fall short of their differentiable counterparts, as seen in a deeper study in appendix \ref{appendix:supervised_models}.

\begin{figure}
    \centering
    \includegraphics[width=\textwidth/3*2]{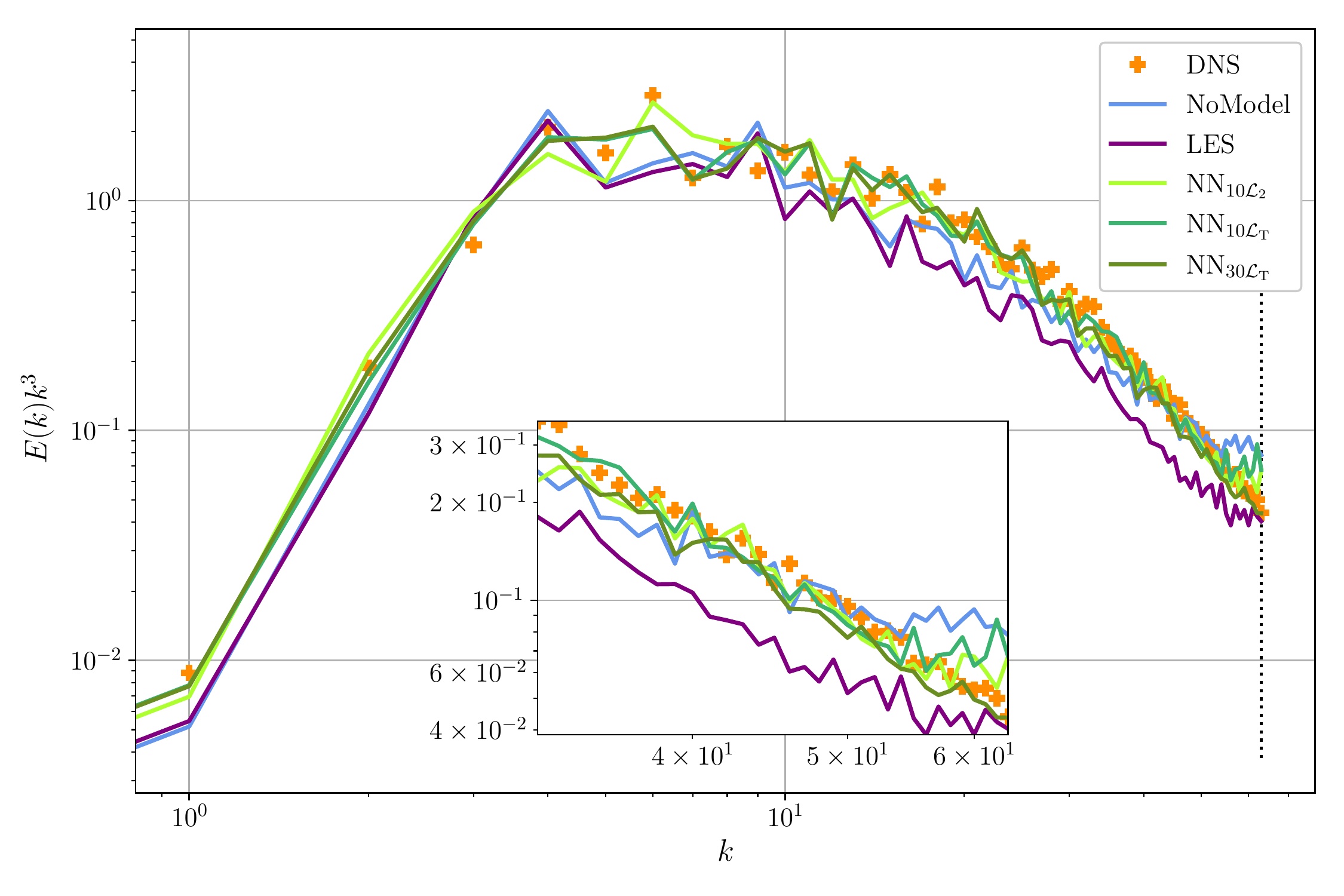}

    \caption{Resolved turbulence kinetic energy spectra of the downsampled \gls{dns}, no-model, LES, and learned model simulations; the learned 30-step model matches the energy distribution of downsampled \gls{dns} data; the vertical line represents the Nyquist-wavenumber of the low-resolution grid}
    \label{fig:randomised_turbulence_spectral_energy}
\end{figure}

\begin{figure}
\centering
\begin{subfigure}[b]{.5\textwidth}
    \centering
    \includegraphics[width=\textwidth]{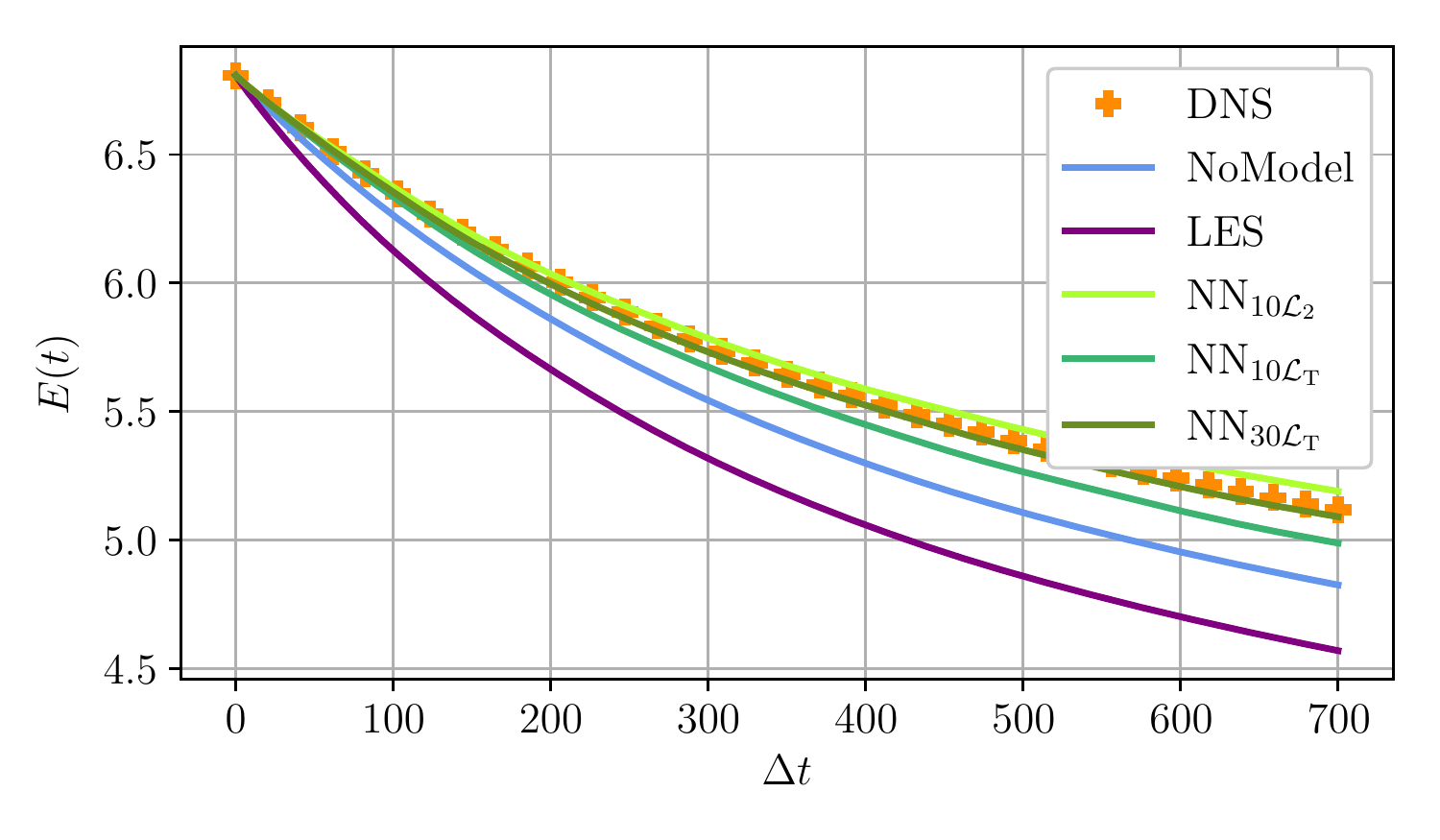}
    \caption{}
\end{subfigure}%
\begin{subfigure}[b]{.5\textwidth}
    \centering
    \includegraphics[width=\textwidth]{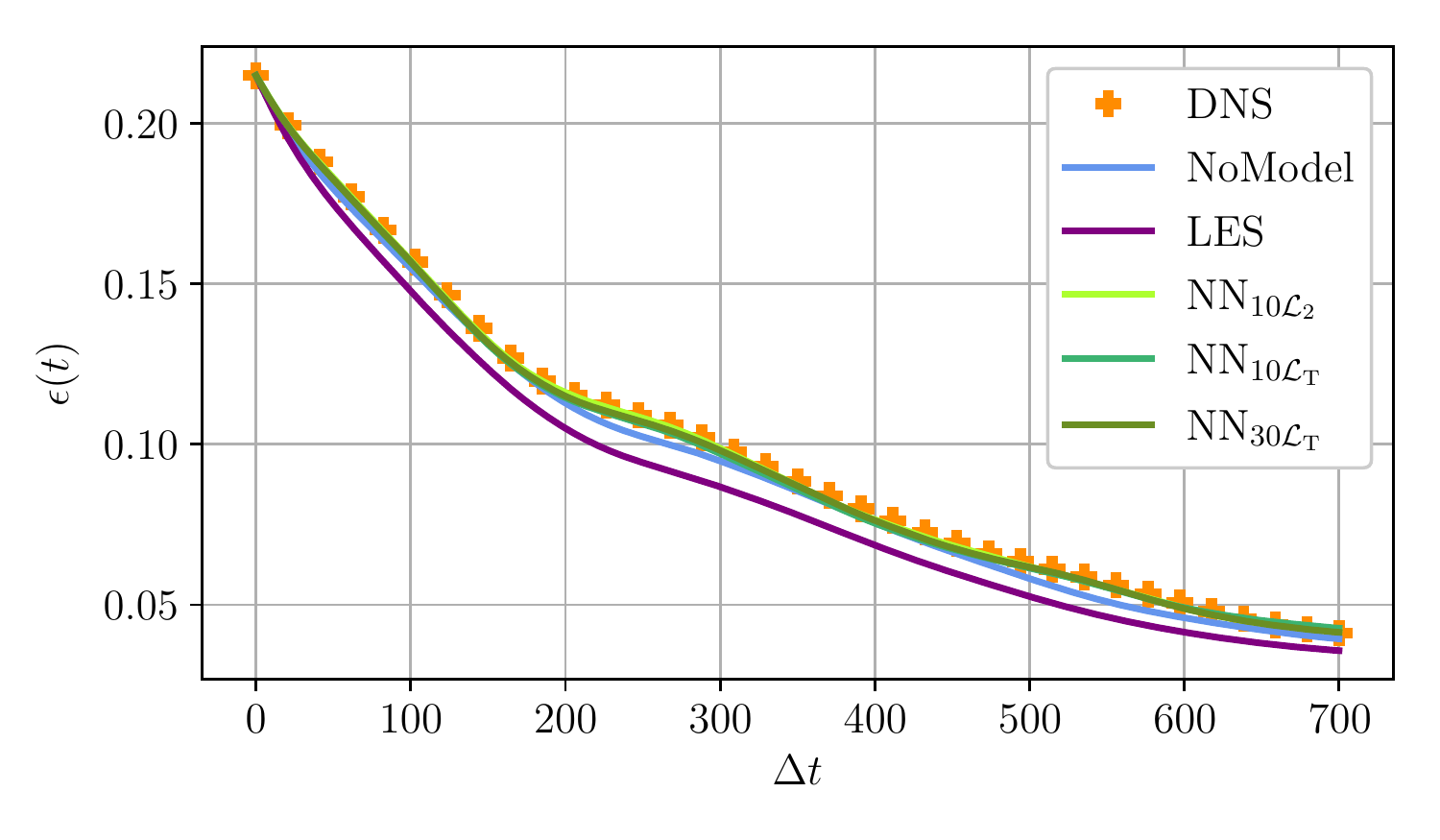}
    \caption{}
\end{subfigure}
    \caption{Comparison of \gls{dns}, no-model, LES, and learned model simulations with respect to resolved turbulence kinetic energy over time (a); and turbulence dissipation rate over time (b)}
    \label{fig:randomised_turbulence_tke_dissipation}
\end{figure}

We provide visual comparisons of vorticity snapshots in figure \ref{fig:randomised_turbulence_vorticity}, where our method's improvements become apparent. The network-modelled simulations produce a highly accurate evolution of vorticity centers, and comparable performance cannot be achieved without a model for the same resolution.  We also investigate the resolved turbulence kinetic energy spectra in figure \ref{fig:randomised_turbulence_spectral_energy}. Whilst the no-model simulation overshoots the \gls{dns} energy at its smallest resolved scales, the learned model simulations perform better and match the desired target spectrum. Figure \ref{fig:randomised_turbulence_tke_dissipation} shows temporal evolutions of the domain-wide resolved turbulence energy and the domain-wide resolved turbulence dissipation rate. The turbulence energy is evaluated according to $E(t)= \int u_i'(t)  u_i'(t) d\Omega$, where $u_i'$ is the turbulent fluctuation. We calculate the turbulence dissipation as $\epsilon(t)=\int <\mu \frac{\partial u'_i\partial u_i'}{\partial x_j \partial x_j}>d\Omega$. Simulations with our \gls{cnn} models strongly agree with the downsampled \gls{dns}.

All remaining learned models stay in close proximity to the desired high-resolution evolutions, whereas the LES-modelled and no-model simulations significantly deviate from the target. Overall, the neural network models trained with more unrolled steps outperformed others, while the turbulence loss formulation $\mathcal{L}_\text{T}$ also had a positive effect.

\begin{figure}
    \centering
    \includegraphics[width=.6\textwidth]{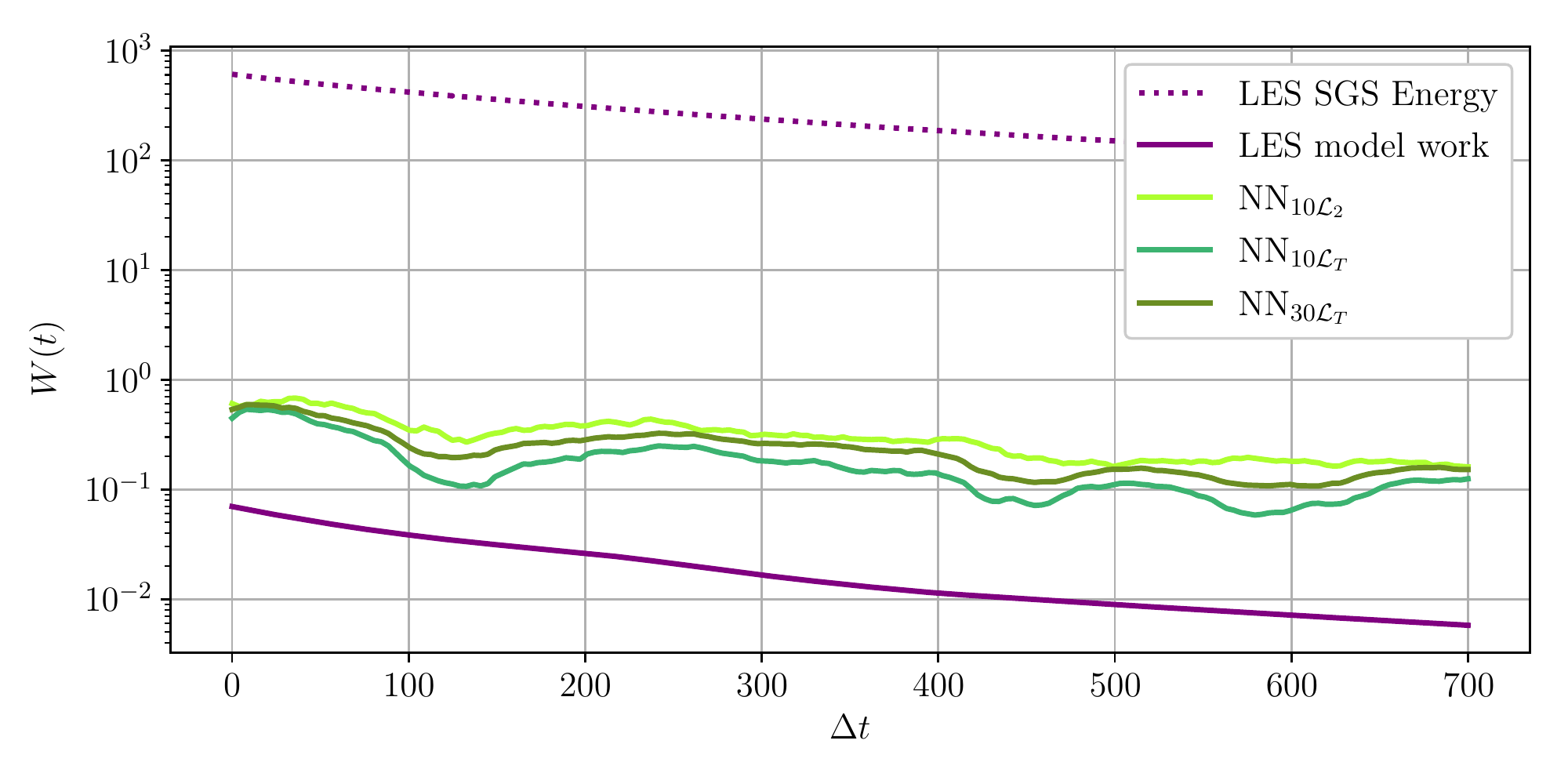}
    \caption{NN-model work on the flow field, work by the \gls{les} model and the estimated \gls{sgs} energies from \gls{les}}
    \label{fig:randomised_turbulence_nn_work}
\end{figure}

In particular, the backscatter effect is crucial for simulations of decaying turbulence \citep{kraichnan1967inertial, smith1996crossover}. The CNN adequately dampens the finest scales as seen in the high wavenumber section of the energy spectrum (figure \ref{fig:randomised_turbulence_spectral_energy}), it also successfully boosts larger scale motions. In contrast, the no-model simulation lacks dampening in the finest scales and cannot reproduce the backscatter effect on the larger ones. On the other hand, the dissipative nature of the Smagorinsky model used in the \gls{les} leads to undersized spectral energies across all scales. Especially the spectral energies of no-model and \gls{les} around wavenumber $k=10$ show large deviations form the ground truth, while the \gls{cnn} model accurately reproduces its behaviour. These large turbulent scales are the most relevant to the resolved turbulence energy and dissipation statistics, which is reflected in figure \ref{fig:randomised_turbulence_tke_dissipation}. Herein, the neural-network models maintain the best approximations, and high numbers of unrolled steps show the best performance at long simulation horizons. The higher total energy of the neural network modelled simulations can be attributed to the work done by the network forcing, which is visualised together with the SGS stress tensor work from the LES simulation as well as its SGS energy in figure \ref{fig:randomised_turbulence_nn_work}. This analysis reveals that the neural networks do more work on the system as the \gls{les} model does, which explains the higher and more accurate turbulence energy in figure \ref{fig:randomised_turbulence_tke_dissipation} and the spectral energy behavior at large scales in figure \ref{fig:randomised_turbulence_spectral_energy}.

%% file: sections/planar_mixing_layers.tex
\section{Temporally Developing Planar Mixing Layers}
\label{subsection:temporal_mixing_layers}
Next, we apply our method to the simulation of two-dimensional planar mixing layers. Due to their relevance to applications such as chemical mixing or combustion, mixing layers have been the focus of theoretical and numerical studies in the fluid-mechanics community. These studies have brought forth a large set and good understanding of \textit{a-posteriori} evaluations, like the Reynolds-averaged turbulent statistics or the vorticity and momentum thickness. Herein, we use these evaluations to assess the accuracy of our learned models with respect to metrics that are not directly part of the learning targets.

Temporally evolving planar mixing layers are the simplest numerical representation of a process driven by the Kelvin-Helmholtz instability in the shear layer. They are sufficiently defined by the Reynolds number, domain sizes, boundary conditions, and an initial condition. Aside from the shear layer represented by a $\tanh$-profile, the initial flow fields feature an oscillatory disturbance that triggers the instability leading to the roll up of the shear layer. This has been investigated by theoretical studies involving linear stability analysis \citep{Michalke1964} or numerical simulation \citep{Rogers1994}. Our setup is based on the work by \cite{Michalke1964}, who studied the stability of the shear layer and proposed initialisations that lead to shear layer roll up. As initial condition, we add randomised modes to the mean profile, resulting in the stream-function
\begin{equation}\label{eq:temp_mixing_layer_stream_function}
    \Psi(x,y) = y + \frac{1}{2}\ln(1+e^{-4y})+a((\alpha y)^2+1)e^{-(\alpha y)^2}\cos(\omega_\Psi x),
\end{equation}
where $a$ is the amplitude of the perturbation, $\alpha$ parameterises the decay of the perturbation in $y$-direction, and $\omega_\Psi$ represents the perturbation frequency. The initial flow field can then be calculated by
\begin{equation}
    u(x,y) = \frac{\partial \Psi}{\partial y}, \hspace{5em} v(x,y) = - \frac{\partial \Psi}{\partial x}.
\end{equation}
\begin{figure}
    \centering
    \includegraphics[width=\textwidth]{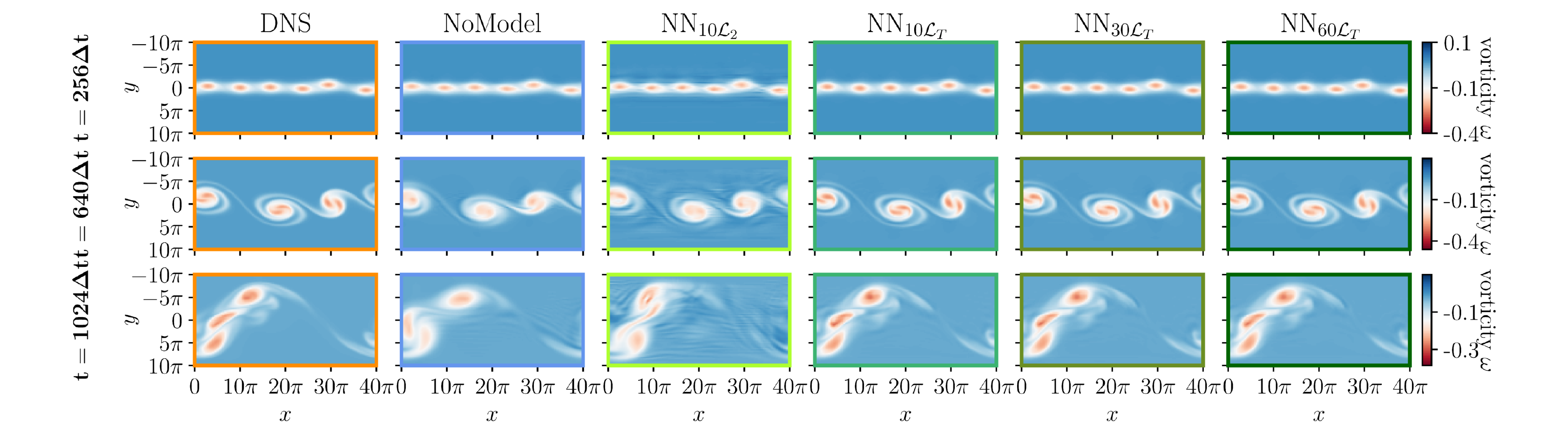}
    \caption{Vorticity visualisations of \gls{dns}, no-model, and learned model simulations at $t=(256,640,1024)\Delta t$ on the test dataset}
    \label{fig:temporal_vorticity}
\end{figure}
At the initial state this results in a velocity step $\Delta U= {U_2}-{U_1}=1$ and a vorticity thickness of $\delta_\omega=\sfrac{\Delta U}{\frac{\partial U }{\partial y}\big|_{\text{max}}}= 1$, where velocities marked as $U$ represent mean-stream quantities. Thus, $U_2$ and $U_1$ are the fast and slow mean velocities of the shear layer. The computational domain of size $(L_x,L_y)=(40\pi,20\pi)$ is discretised by $(N_x,N_y)=(1024,512)$ grid cells for the high-resolution dataset generation. The streamwise boundaries are periodic, while the spanwise boundaries in $y$-direction are set to a free-slip boundary where $\frac{\partial u}{\partial y}\big |_{\Omega_y}=0$, $v|_{\Omega_y}=0$ and $p|_{\Omega_y}=0$. The Reynolds number based on the unperturbed mean profile and the vorticity thickness is calculated to be $Re=\frac{\Delta U \delta_\omega}{\nu}=250$ for all randomised initialisations. The simulations are run for $T=420=12000\Delta t_\text{DNS}$. Our dataset consists of three simulations based on different initialisations. Their perturbation details are found in table \ref{table:temporal_perturbation_parameters}. Two of these simulations were used as training datasets, while all of our evaluation is performed on the remaining one as extrapolation test dataset.

Following the approach in section \ref{section:isotropic_turbulence}, the model training uses a $8\times$ downscaling in space and time. The loss composition was set to $(\lambda_2 , \hspace{0.5em} \lambda_E, \hspace{0.5em} \lambda_\mathit{S}, \hspace{0.5em} \lambda_\text{MS})= (100, \hspace{0.5em} 2, \hspace{0.5em} 5\times10^{-2}, \hspace{0.5em} 0)$. We used the same \gls{cnn} architecture as introduced earlier, though due to the difference in  boundary conditions a different padding procedure was chosen (see appendix \ref{appendix:cnn}).
To illustrate the impact of the turbulence loss $\mathcal{L}_\text{T}$ and an unrolling of $60$ numerical steps, we compare to several variants with reduced loss formulations and fewer unrolling steps.  The maximum number of $60$ unrolled steps corresponds to $16 t_{\delta_\theta}$ integral timescales computed on the momentum thickness as $t_{\delta_\theta} = \delta_\theta/\Delta U$. With the shear layer growing, the momentum thickness increases $7$-fold, which decreases the number of integral timescales
to $2$ for 60 steps of unrollment.
Table \ref{table:temporal_l2_comparison} shows details of the model parameterisations. To avoid instabilities in gradient calculation that could ultimately lead to unstable training, we split the back-propagation into subranges for the $60$-step model. This method stabilises an otherwise unstable training of the $60$-step model, and a split into $30$-step long back-propagation subranges performs best. Such a model is added to the present evaluations as $\text{NN}_{60,\mathcal{L}_\mathrm{T}}$. Detailed results regarding the back-propagation subranges are discussed in section \ref{section:backpropagation}.
\begin{table}
\parbox{.40\textwidth}{
\centering
    \begin{tabular}{l c r r }
    & \multicolumn{1}{c}{\ Setup \ } & \multicolumn{1}{c}{\ \ \ \ \ $a$} & \multicolumn{1}{c}{\ \ \ \ \ $\omega_{\Psi}$}  \\
    \hline
    \parbox[t]{7mm}{\multirow{2}{*}{train }} &$1$& $6.0$ & $0.7$\\

    & $2$& $3.3$ & $1.5$\\
    \hline
    \parbox[t]{7mm}{\multirow{1}{*}{test}} &$3$& $9.0$ & $0.3$\\
    \end{tabular}
    \caption{Perturbation details for initial conditions of temporal mixing layer training and test datasets}
    \label{table:temporal_perturbation_parameters}
}
\hfill
\parbox{.50\textwidth}{
\centering
    \begin{tabular}{l c c r r}
    Name\  & \  Loss \ & \ Steps \ & \ $t_{\delta_\theta}$ \ & MSE at $ t_e$ \\
    \hline
    $\text{NoModel}$ & - & - & -  & $1.25\mathrm{e}{-3}$\\
    $\text{NN}_{10}$            & $\mathcal{L}_2$        & 10 & $2.7$ &  $3.19\mathrm{e}{-4}$\\
    $\text{NN}_{10,\mathcal{L}_\text{T}}$   & $\mathcal{L}_\text{T}$ & 10 & $2.7$ & $3.31\mathrm{e}{-5}$\\
    $\text{NN}_{30,\mathcal{L}_\text{T}}$   & $\mathcal{L}_\text{T}$ & 30 & $8.1$ & $2.26\mathrm{e}{-5}$\\
    $\text{NN}_{60,\mathcal{L}_\text{T}}$   & $\mathcal{L}_\text{T}$ & 60 & $16.1$ & $1.93\mathrm{e}{-5}$\\
    \end{tabular}
    \caption{Model details for unrollment study; MSE w.r.t. \gls{dns} from test-data at $t_e = 512\Delta t$}
    \label{table:temporal_l2_comparison}
}
\end{table}

The trained models were compared to a downsampled \gls{dns} and a no-model simulation, all sharing the same starting frame from the test-dataset. This test-dataset changes the initial condition, where different perturbation frequencies and amplitudes result in a variation in vortex roll-up and vortex merging behaviour of the mixing layer. The resulting numerical solutions were compared at three different evolution times $t=[256  \hspace{.5em} 640  \hspace{.5em} 1024]\Delta t$. Figure \ref{fig:temporal_vorticity} shows the vorticity heatmap of the solutions. Qualitatively, the simulations corrected by the \gls{cnn} exhibit close visual proximity to the \gls{dns} by boosting peaks in vorticity where applicable, and additionally achieve a dampening of spurious oscillations.

\begin{figure}
\centering
\begin{subfigure}[b]{\textwidth}
    \centering
    \includegraphics[width=\textwidth]{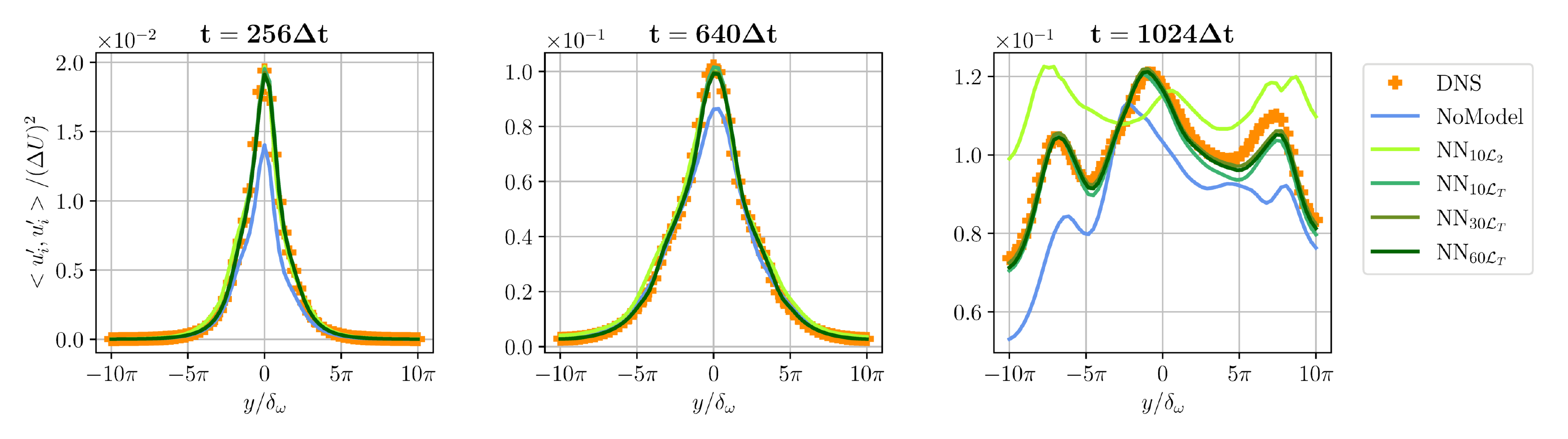}
    \caption{}
\end{subfigure}
\centering
\begin{subfigure}[b]{\textwidth}
    \centering
    \includegraphics[width=\textwidth]{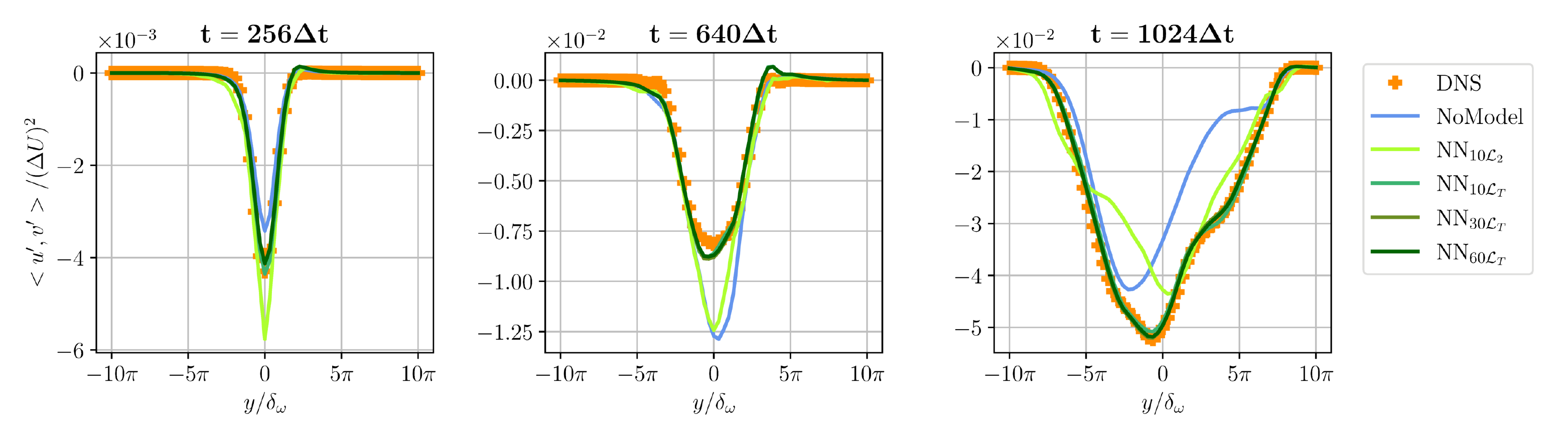}
    \caption{}
\end{subfigure}
\caption{Comparison of \gls{dns}, no-model, and learned model simulations with respect to resolved turbulence kinetic energy (a), and Reynolds stresses (b)}
\label{fig:temporal_rans_tke_reyStr}
\end{figure}

\begin{figure}
    \centering
    \includegraphics[width=\textwidth]{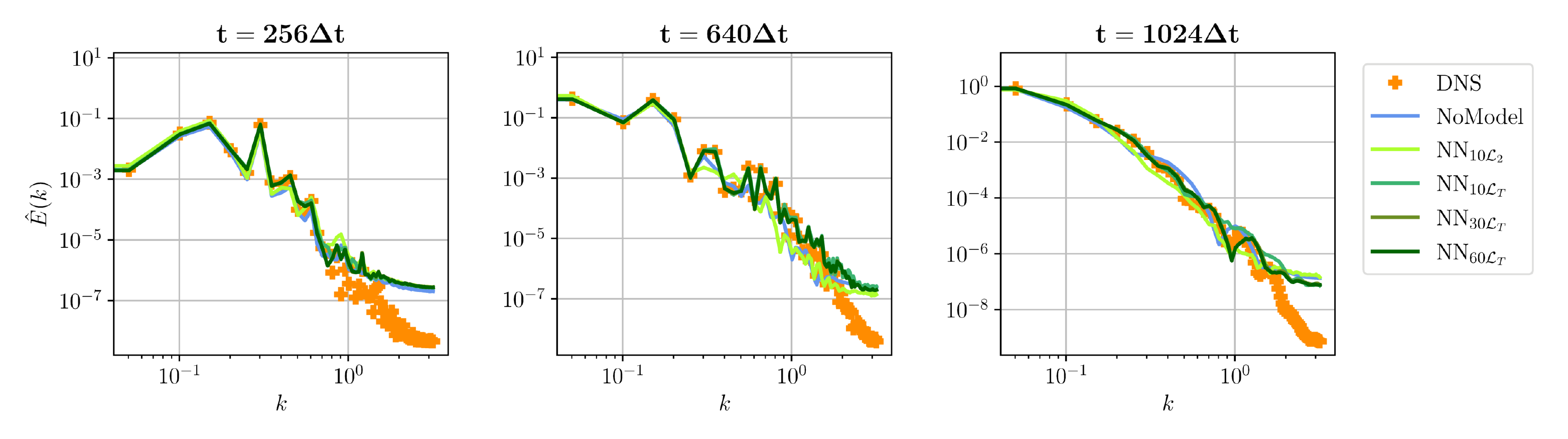}
    \caption{Centerline kinetic energy spectra for the downsampled \gls{dns}, no-model, and learned model simulations}
    \label{fig:temporal_centerline_spectrum}
\end{figure}

\begin{figure}
    \centering
    \includegraphics[width=\textwidth]{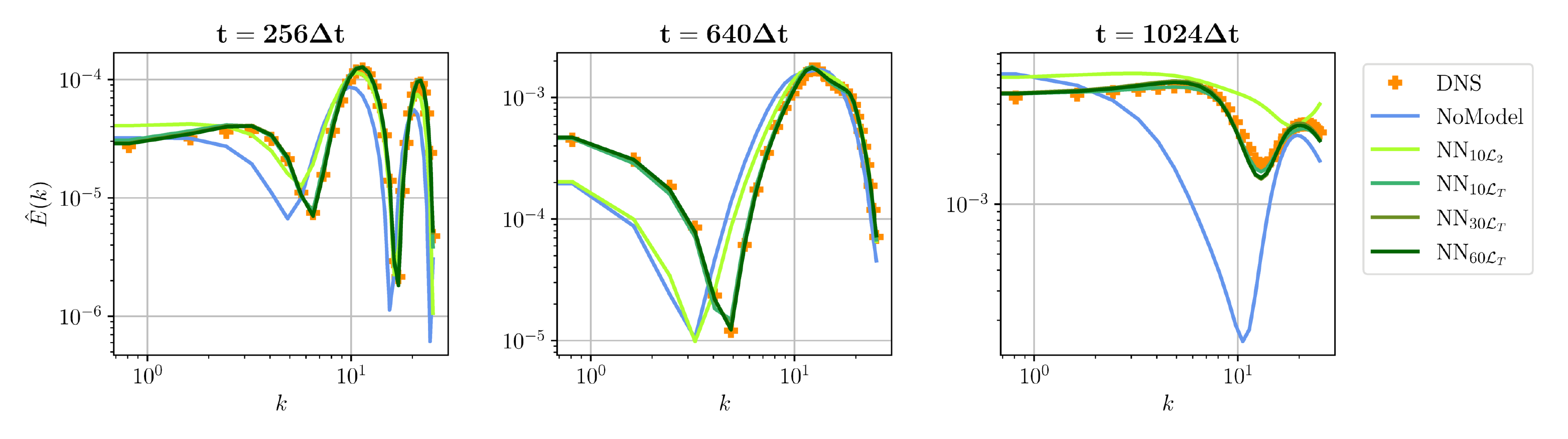}
    \caption{Cross-sectional kinetic energy spectra of the downsampled \gls{dns}, no-model, and learned model simulations}
    \label{fig:temporal_crosssection_spectrum}
\end{figure}

These observations are matched by corresponding statistical evaluations. The statistics are obtained by averaging the simulation snapshots along their streamwise axis and the resulting turbulence fluctuations were processed for each evaluation time.  Figure \ref{fig:temporal_rans_tke_reyStr} shows that all $\mathcal{L}_\mathrm{T}$-models closely approximate the \gls{dns} reference with respect to their distribution of resolved turbulence kinetic energy and Reynolds stresses along the cross-section, while the no-model simulation clearly deviates. Note that the mixing process causes a transfer of momentum from fast to slow moving sections through the effects of turbulent fluctuations. The shear layer growth is thus dominated by turbulent diffusion. Consequently, accurate estimates of the turbulent fluctuations are necessary for the correct evolution of the mixing layer. These fluctuations are most visible in the Reynolds stresses $u'v'$, and an accurate estimation is an indicator for well modelled turbulent momentum diffusion. The evaluations also reveal that unrolling more timesteps during training gains additional performance improvements. These effects are most visible when comparing the 10-step and 60-step model in a long temporal evolution, as seen in the Reynolds stresses in figure \ref{fig:temporal_rans_tke_reyStr}. The evaluation of resolved turbulence kinetic energies shows that the models correct for the numerical dissipation of turbulent fluctuations, while, in contrast, there is an underestimation of kinetic energy in the no-model simulation. While longer unrollments generally yield better accuracy, it is also clear that $30$ steps come close to saturating the model performance in this particular flow scenario. With the integral timescales mentioned earlier, it becomes clear that 30 simulation steps capture one integral timescale of the final simulation phase, i.e. the phase of the decaying simulation that exhibits the longest timescales.
One can conclude that an unrollment of one timescale is largely sufficient, and further improvements of unrolling 2 timescales with $60$ steps are only minor.

The resolved turbulence kinetic energy spectra are evaluated to assess the spatial scales at which the corrective models are most active. The spectral analysis at the centerline is visualised in figure \ref{fig:temporal_centerline_spectrum}, whilst the kinetic energy obtained from fluctuations across the cross-section with respect to streamwise averages is shown in figure \ref{fig:temporal_crosssection_spectrum}. These plots allow two main observations: Firstly, the deviation of kinetic energy mostly originates from medium-sized spatial scales, which are dissipated by the no-model simulation, but are accurately reconstructed by the neural network trained with $\mathcal{L}_T$. This effect is connected to the dampening of vorticity peaks in the snapshots in figure \ref{fig:temporal_vorticity}. Secondly, the fine-scale spectral energy of the no-model simulation has an amplitude similar to the \gls{dns} over long temporal horizons (figure \ref{fig:temporal_crosssection_spectrum}). This can be attributed to numerical oscillations rather than physical behaviour. These numerical oscillations, as also seen in the snapshots in figure \ref{fig:temporal_vorticity}, exist for the no-model simulation but are missing in the $\mathcal{L}_T$-modelled simulations. Training a model without the additional loss terms in $\mathcal{L}_T$ from equation \eqref{eq:total_loss}, i.e. only with the $\mathcal{L}_2$ from equation \eqref{eq:l2_loss}, yields a model that is inaccurate and results in unphysical oscillations. It does not reproduce the vorticity centers, and is also unstable over long temporal horizons. Herein, nonphysical oscillations are introduced, which also show up in the cross-sectional spectral energies and vorticity visualisations. We thus conclude that best performance can be achieved with a network trained with $\mathcal{L}_T$, which learns to dampen numerical oscillations and reproduces physical fluctuations across all spatial scales.

\begin{figure}
    \centering
    \includegraphics[width=\textwidth/2]{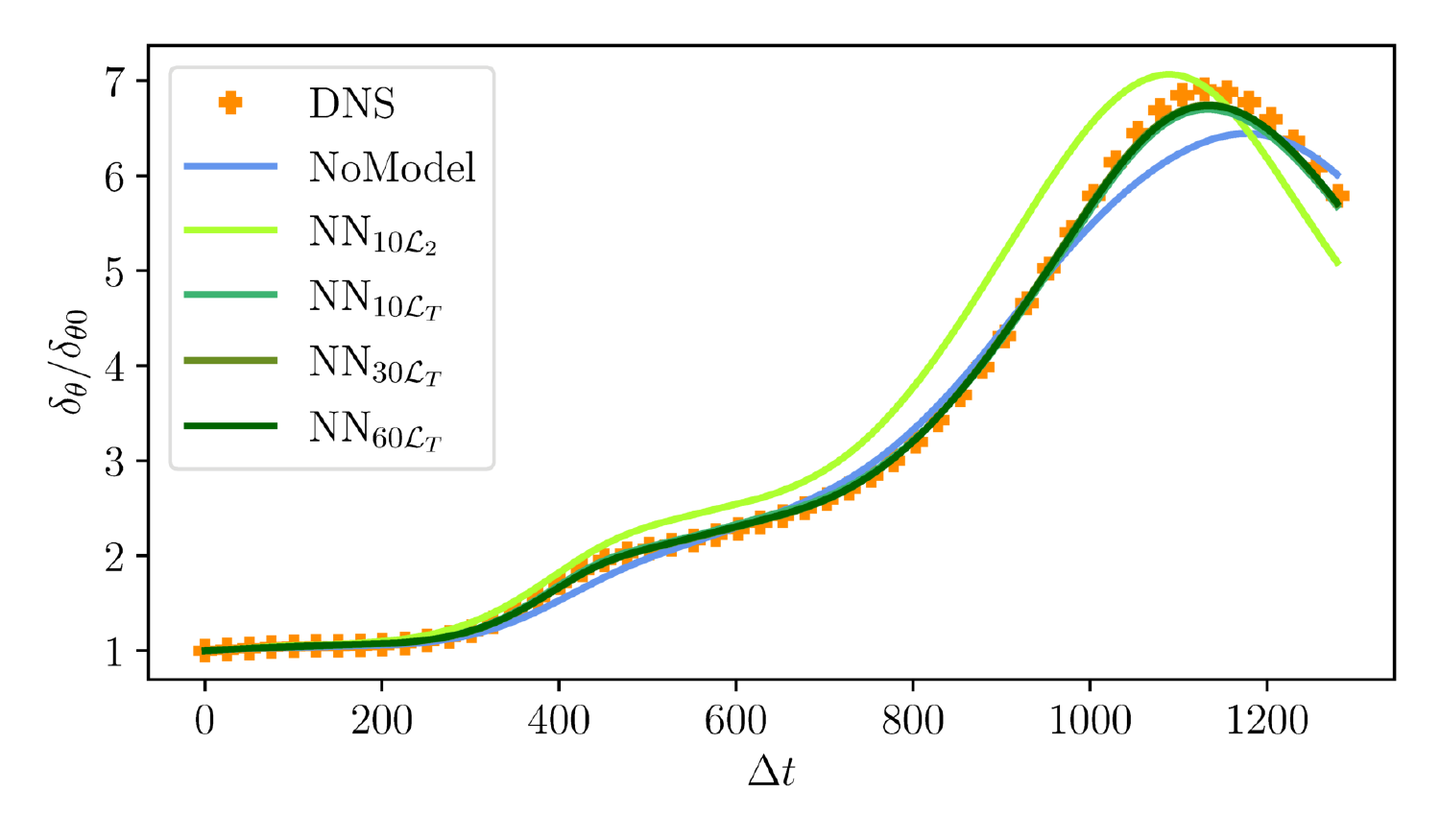}

    \caption{Momentum thickness of \gls{dns}, no-model, and learned model simulations, evaluated based on the streamwise averages}
    \label{fig:temporal_momentum_thickness}
\end{figure}

It is worth noting that our method is capable of enhancing an under-resolved simulation across a wide range of turbulent motions. The vortex-size in the validation simulation ranges from $7\delta_{\omega_0}$ at the starting frame to $60\delta_{\omega_0}$ after evolving for $1200 \Delta t$. This timespan encompasses two vortex merging events, both of which cannot be accurately reproduced with a no-model, or a $\mathcal{L}_2$-model simulation, but are captured by the $\mathcal{L}_T$-trained network models. This is shown in the comparison of the momentum thicknesses over time in figure \ref{fig:temporal_momentum_thickness}. The reproduction of turbulence statistics (figure \ref{fig:temporal_rans_tke_reyStr}) yields, in the long term, an accurate turbulent diffusion of momentum and mixing layer growth for the models trained with $\mathcal{L}_T$. On the contrary, the $\mathcal{L}_2$ model fails to reproduce the vortex cores and deviates with respect to the momentum thickness for long temporal horizons.

\section{Spatially Developing Planar Mixing Layers}
\label{subsection:spatial_mixing_layers}
In contrast to the temporally developing mixing layers investigated in section  \ref{subsection:temporal_mixing_layers}, the spatially developing counterpart features a fixed view on a statistically steady flow field,
which introduces a new set of challenges to the learning task.
While the main difficulty in previous transient simulations was the modelling of an evolving range of turbulent scales,
the statistically steady nature of the spatially developing mixing layer
requires a reproduction of the turbulent statistics in its own statistically steady state. This in turn necessitates long-term accuracy and stability.

Spatially mixing layers develop from an instability in the shear layer. This instability is driven by a disturbance at the inlet, whose characteristics have great effect on the mixing layer growth \citep{Ho1982}. In a simulation environment, these disturbances are realised by a Dirichlet inlet boundary condition, where temporally varying perturbations are added to a steady mean flow profile. As proposed by \cite{Ko2008SensitivityConditions}, a suitable inlet condition including perturbations can be written as
\begin{equation}
    u_{\text{in}}(y,t) = 1 + \frac{\Delta U}{2}\tanh(2y) + \sum_{d=1}^{N_d}\epsilon_d(1-\tanh^2(y/2))\cos(K_d y)\sin(\Omega_dt),
\end{equation}
where the number of perturbation modes $N_d=2$ holds for our simulations.
Furthermore, we used inviscid wall conditions for the two $y$-normal spanwise boundaries, and the outflow boundary was realised by a simple Neumann condition with a stabilising upstream sponge layer. For all simulations, we set the characteristic velocity ratio $\Delta U=1$ and the vorticity thickness to $\delta_\omega=1$. The vorticity-thickness Reynolds number is set to  $Re_{\delta_\omega}=\frac{\Delta U \delta_\omega}{\nu} = 500$. To generate the \gls{dns} dataset, this setup was discretised by a  uniform grid with $(N_x,N_y)=(2048,512)$ resolving the domain of size $(L_x,L_y)=(256,64)$. The timesteps were chosen such that \gls{cfl}$=0.3$ and the temporal evolution was run for $7$ periods of the slowest perturbation mode $i=2$ to reach a statistically steady state, before subsequent frames were entered into the dataset. A further $28$ periods of the slowest perturbation mode were simulated to generate $32000$ samples of the statistically steady state.
The training dataset consists of $5$ such simulations with different perturbations, as summarised in table \ref{table:spatial_perturbation_parameters}.
A downsampling ratio of $8\times$ in space and time was again chosen for the learning setup. The input to the network was set to include only the main simulation frame without the sponge layer region.
\begin{table}
\parbox{.49\textwidth}{
\centering
    \begin{tabular}{l r r r r r r}
    & \  $\epsilon_1 / \bar{U}$ & \  $K_1$ & \  $\Omega_1$ &
      \ \ $\epsilon_2 / \bar{U}$ &  \  $K_2$ &  \  $\Omega_2$ \\
    \hline
    \parbox[t]{8mm}{\multirow{7}{*}{train}}
    & $\mathbf{0.075}$ & $.4\pi$  & $.22$
    & $\mathbf{0.025}$ & $.3\pi$  & $.11$\\[.5ex]
    \cline{2-7}\rule{0pt}{2.6ex}
    & $\mathbf{0.060}$ & $.4\pi$  & $.22$
    & $\mathbf{0.040}$ & $.3\pi$  & $.11$\\[.5ex]
    \cline{2-7}\rule{0pt}{2.6ex}
    & $\mathbf{0.050}$ & $.4\pi$  & $.22$
    & $\mathbf{0.050}$ & $.3\pi$  & $.11$\\[.5ex]
    \cline{2-7}\rule{0pt}{2.6ex}
    & $\mathbf{0.040}$ & $.4\pi$  & $.22$
    & $\mathbf{0.060}$ & $.3\pi$  & $.11$\\[.5ex]
    \cline{2-7}\rule{0pt}{2.6ex}
    & $\mathbf{0.025}$ & $.4\pi$  & $.22$
    & $\mathbf{0.075}$ & $.3\pi$  & $.11$\\
    \hline
    \parbox[t]{8mm}{\multirow{1}{*}{test}}
    & $\mathbf{0.082}$ & $.4\pi$  & $.22$
    & $\mathbf{0.018}$ & $.3\pi$  & $.11$\\
    \end{tabular}
    \caption{Perturbation details for the inlet condition of training and test datasets}
    \label{table:spatial_perturbation_parameters}
}
\hfill
\parbox{.49\textwidth}{
\centering
    \begin{tabular}{ l c c r r }
    Name \ \  &  \ Loss \ &  Steps & \ $t_{f_\omega}$ \ &\  MSE at $t_e$ \\
    \hline
    $\text{NoModel}$& - & - & - & $2.03\mathrm{e}{-2}$\\
    $\text{NN}_{10,\mathcal{L}_\mathrm{T}}$& $\mathcal{L}_\mathrm{T}$ & 10 & $0.14$ &   $5.22\mathrm{e}{-3}$\\
    $\text{NN}_{30,\mathcal{L}_\mathrm{T}}$& $\mathcal{L}_\mathrm{T}$ & 30 & $0.42$ & $3.66\mathrm{e}{-3}$\\
    $\text{NN}_{60,\mathcal{L}_\mathrm{T}}$& $\mathcal{L}_\mathrm{T}$ & 60 & $0.85$ & $2.98\mathrm{e}{-3}$\\
    \end{tabular}
    \caption{Model details for unrollment study; MSE w.r.t. \gls{dns} from test-data at $t_e = 1000\Delta t$}
    \label{table:spatial_l2_comparison}
}
\end{table}%
Our best performing model applied the turbulence loss $\mathcal{L}_\text{T}$, with the loss factors set to $(\lambda_2 , \hspace{0.5em} \lambda_E, \hspace{0.5em} \lambda_\mathit{S}, \hspace{0.5em} \lambda_\text{MS})= (50, \hspace{0.5em} 0.5, \hspace{0.5em} 2, \hspace{0.5em} 0.5)$, and an unrollment of $60$ solver steps. The timespan covered by these $60$ solver steps is comparable to a full period of the slowest perturbation mode. Using the roll-up frequency of the spatial mixing layer as basis for the timescale $t_{f_\omega}=1/f_\omega$, $60$ solver steps unroll $0.85t_{f_\omega}$. As we detail in the following,
our test metrics show that this approach of unrolling roughly one integral timescale yields the best results.

\begin{figure}
\centering
\begin{subfigure}[b]{\textwidth}
    \centering
    \includegraphics[width=\textwidth]{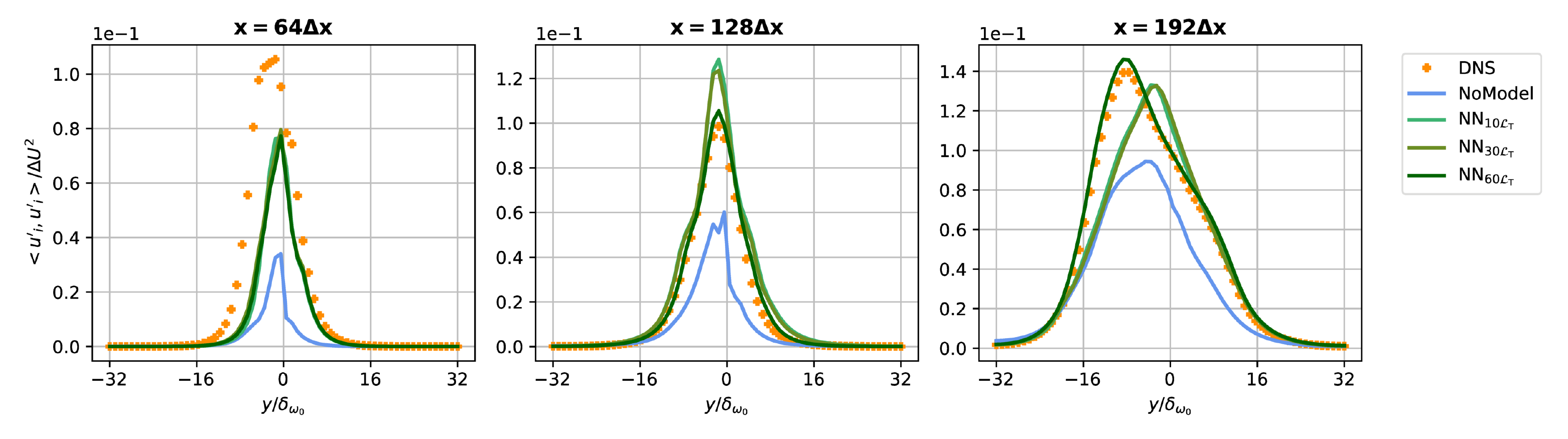}
    \caption{}
\end{subfigure}
\centering
\begin{subfigure}[b]{\textwidth}
    \centering
    \includegraphics[width=\textwidth]{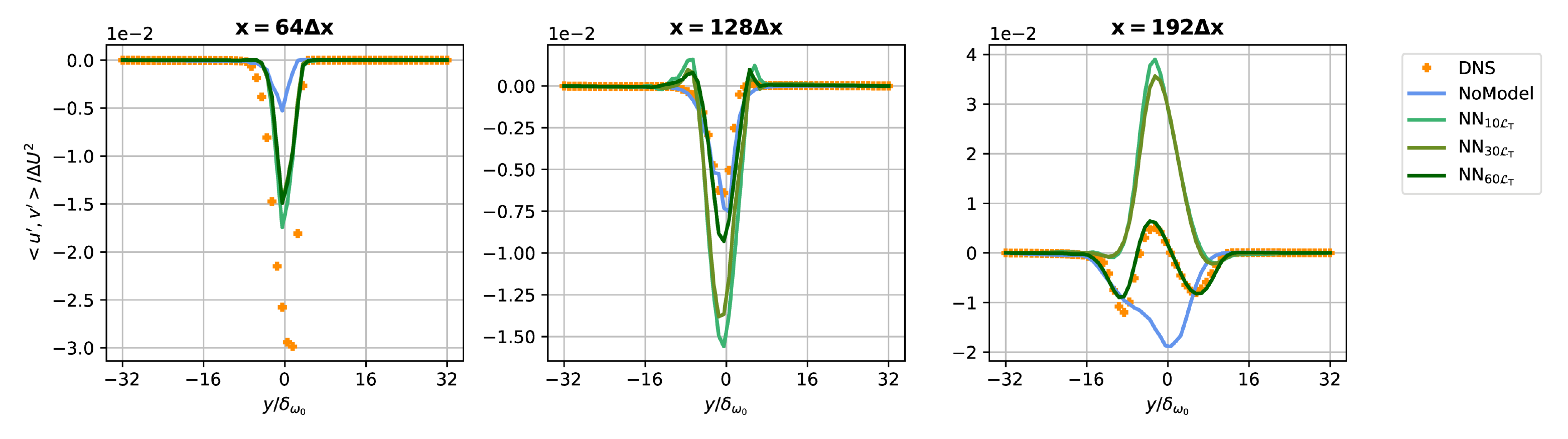}
    \caption{}
\end{subfigure}
\caption{Comparison of downsampled \gls{dns}, no-model, and learned model simulations with respect to Reynolds-averaged resolved turbulence kinetic energy (a); and Reynolds stresses (b)}
\label{fig:spatial_rans_tke_reyStr}
\end{figure}

First, we evaluate the influence of unrollment in this test case.
Once again, we show comparisons with additional setups, the parametric details of which can be found in table \ref{table:spatial_l2_comparison}. Similar to the temporal mixing layer, the $60$ step model was trained using a gradient stopping technique. A $30$-step back-propagation subrange performed best again by maintaining long-term information while avoiding instabilities in the gradient calculation. This model is described as $\text{NN}_{60,\mathcal{L}_\mathrm{T}}$ in this section. Details regarding the method are explained in section \ref{section:backpropagation}. The table shows that the simulation with the 60-step neural network outperforms the no-model baseline by 
an order of magnitude. For these evaluations, we assessed the model capabilities by running a \gls{cnn}-corrected forward simulation. This simulation was initialised with a downsampled frame from the \gls{dns} test dataset in its fully-developed state.
 This test dataset is generated with different inflow conditions, where the inlet forcing lies outside of the training range, making these evaluations an out-of-sample generalisation test. The variation in inlet forcing affects the location and intensity of the mixing layer roll-up and vortex merging.
The simulation was run for $5000 \Delta t$, or $36$ periods of the slowest perturbation mode in order to obtain data from a statistically stable state.
Despite this time frame being orders of magnitude longer than what is seen by the models at training time, the 60-step model retains a stable simulation that closely matches the behavior of the DNS reference. Interestingly, this longer unrollment on the order of one integral timescale is crucial to arrive at a stable model. The models trained with shorter unrollment exhibit various degrees of spurious oscillations, especially the $10$-step model. These oscillations most likely originate from slight deviations in turbulent structures (e.g. vortex roll-up) inferred by the network. Since short unrollment models have never seen any further development of these self-exited structures, applying said models eventually causes even stronger unphysical oscillations downstream.
As before, we omit purely data-driven models trained with pre-computed simulation states. These produce undesirable solutions within a few time steps of simulating the  test cases.
\begin{figure}[h]
\begin{subfigure}[b]{.48\textwidth}
    \centering
    \includegraphics[width=\textwidth]{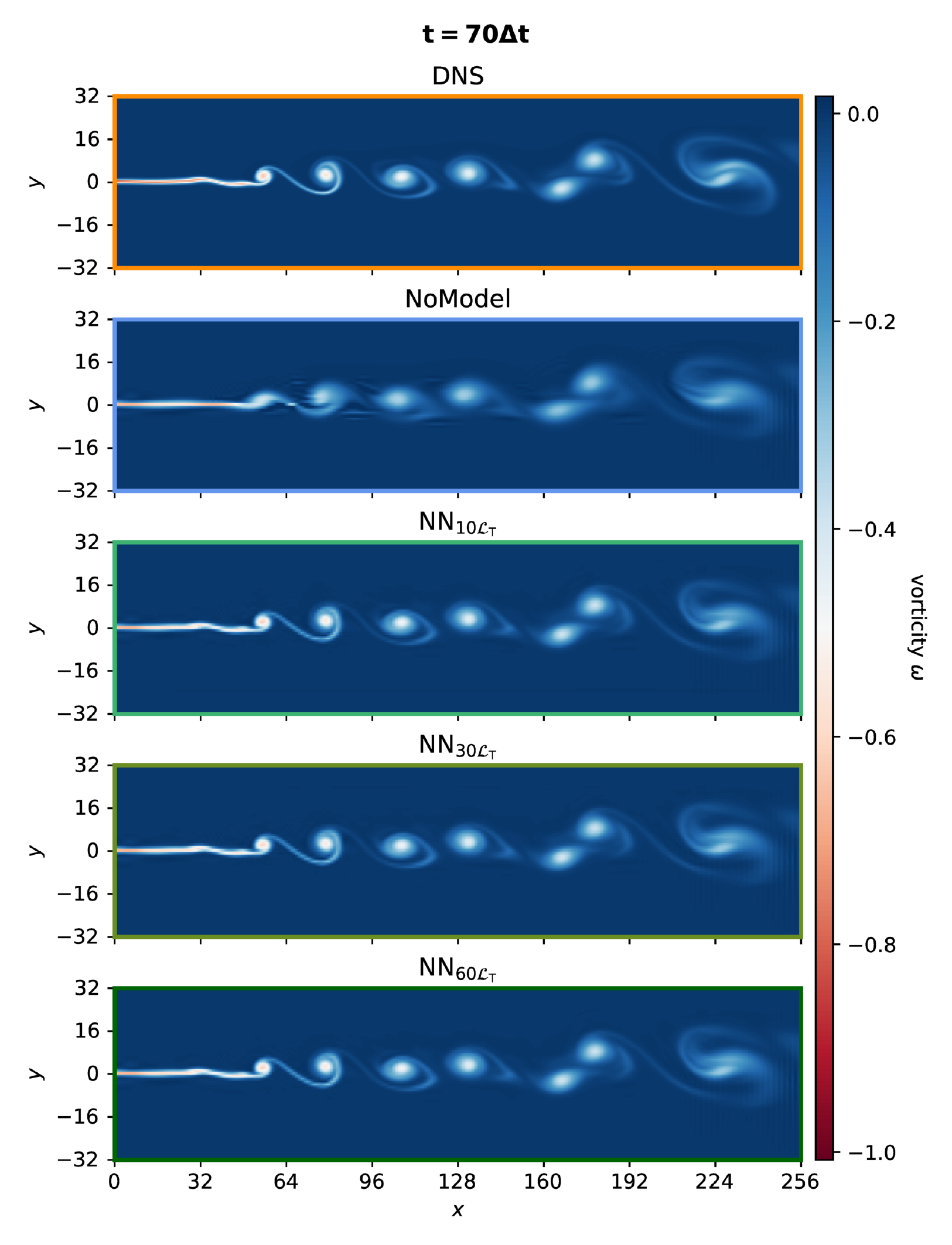}
    \caption{}
    \label{fig:spatial_vorticity_early}
\end{subfigure}
\begin{subfigure}[b]{.48\textwidth}
    \centering
    \includegraphics[width=\textwidth]{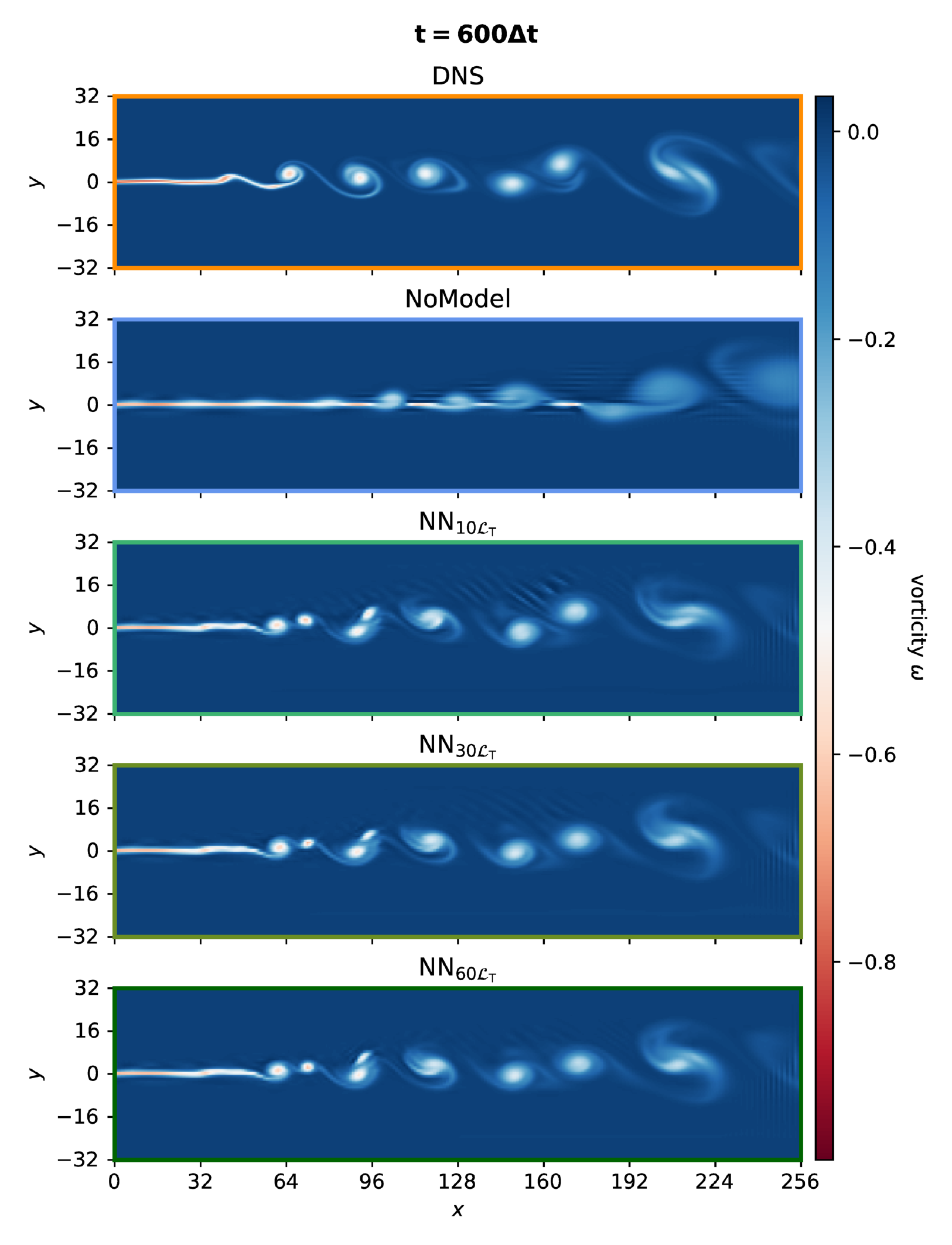}
    \caption{}
    \label{fig:spatial_vorticity_late}
\end{subfigure}
\caption{Vorticity heatmaps of the spatial mixing layer simulations at (a) $t = 70\Delta t$, and (b) $t = 600\Delta t$, on the test dataset}
\end{figure}
The vorticity visualisations after half a period of the slowest perturbation mode ($70 \Delta t$) and after $4$ periods or one flow through time ($600 \Delta t$) are shown in figure \ref{fig:spatial_vorticity_early} and figure \ref{fig:spatial_vorticity_late}, and compared to \gls{dns} and the no-model simulation. The early evaluation in figure \ref{fig:spatial_vorticity_early} reveals a severe loss of detail in the no-model simulation, even after a short time horizon. Over this time-span, the learned model achieves a close visual reproduction. Additionally, the later vorticity heatmap in figure \ref{fig:spatial_vorticity_late} shows a delayed roll-up in the no-model simulation, whereas the learned model maintains the roll-up location and shows improved accuracy. This behaviour is clarified by the Reynolds-averaged properties of the simulations, for which resolved Reynolds stresses and turbulence kinetic energies were calculated on the basis of the respective statistically steady simulations. As shown in figure \ref{fig:spatial_rans_tke_reyStr}, the no-model statistics severely deviate from the targeted \gls{dns}. 
In contrast, the corrective forcing inferred by the trained models approximates these statistics more accurately. The delayed roll-up of the no-model simulation and the improvement of the modelled ones is connected to the Reynolds stresses. The Reynolds stresses indicate turbulent diffusion of momentum, and figure \ref{fig:spatial_rans_tke_reyStr} shows that the \gls{cnn} learned to encourage turbulent fluctuations at the start of the mixing layer. The fluctuations trigger the shear layer instability and feed the roll-up, with decisive implications for the downstream development of the mixing layer. Especially the long unrollment of $60$ steps benefits the model performance. Evaluations at locations downstream of the initial roll-up see the accuracy of $10$ and $30$ step models deteriorate in direct comparison to the $60$-step model.

\begin{figure}
\centering
\includegraphics[width=.5\textwidth]{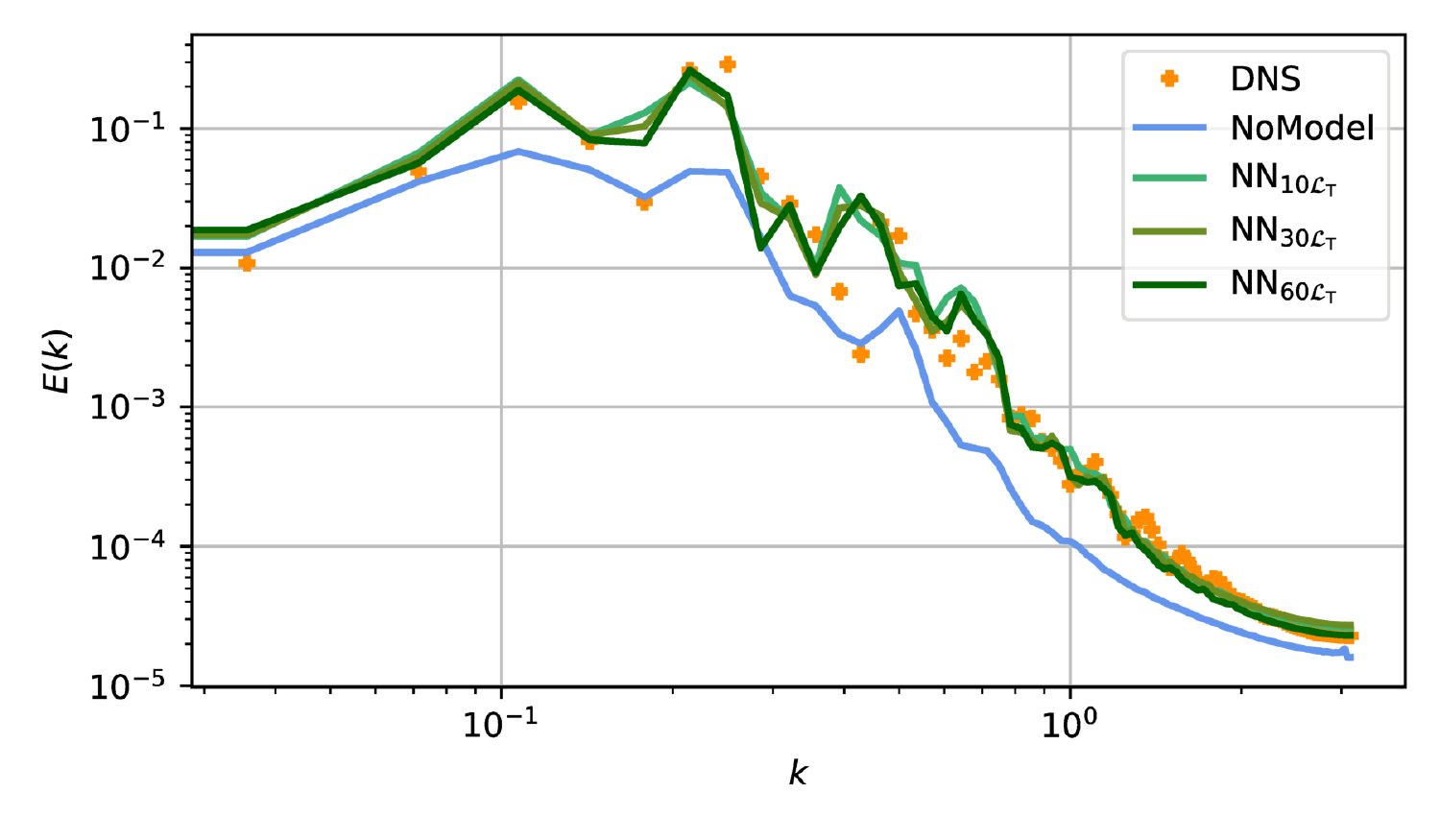}
\caption{Centerline kinetic energy spectra for downsampled \gls{dns}, no-model, and learned model simulations}
\label{fig:spatial_centerline_spectrum}
\end{figure}

These observations regarding the Reynolds stresses extend to the resolved turbulence kinetic energies (figure \ref{fig:spatial_rans_tke_reyStr}), where the same turbulent fluctuations yield an accurate reproduction of the \gls{dns}. The learned models are not limited to a specific spatial scale, but precisely match the \gls{dns} on all turbulent scales when comparing the center-line kinetic energy spectra in figure \ref{fig:spatial_centerline_spectrum}.

\begin{figure}
\centering
\begin{subfigure}[b]{.48\textwidth}
    \centering
    \includegraphics[width=\textwidth]{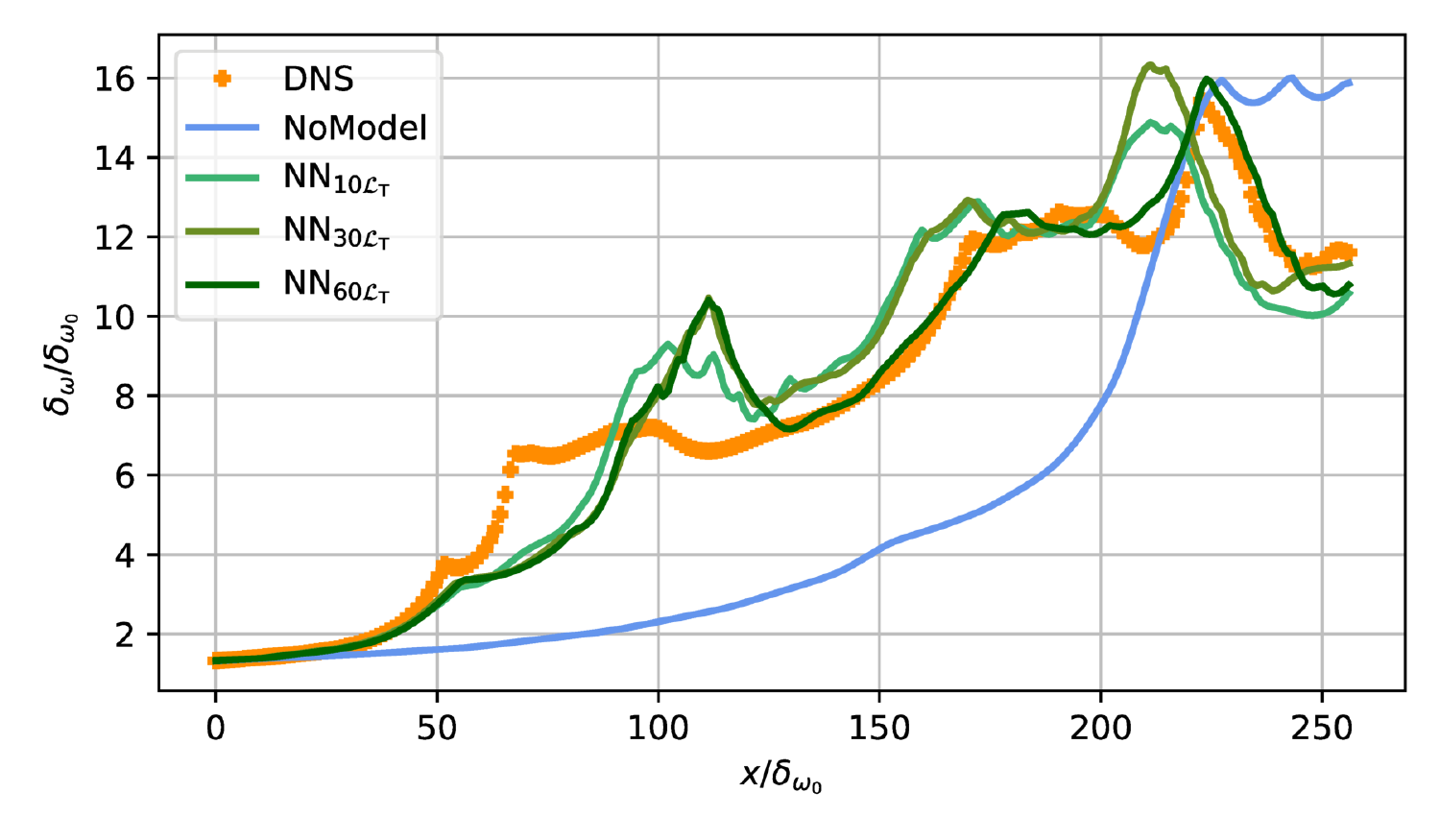}
    \caption{}
    \label{fig:spatial_vorticity_thickness}
\end{subfigure}
\hfill
\begin{subfigure}[b]{.48\textwidth}
    \centering
    \includegraphics[width=\textwidth]{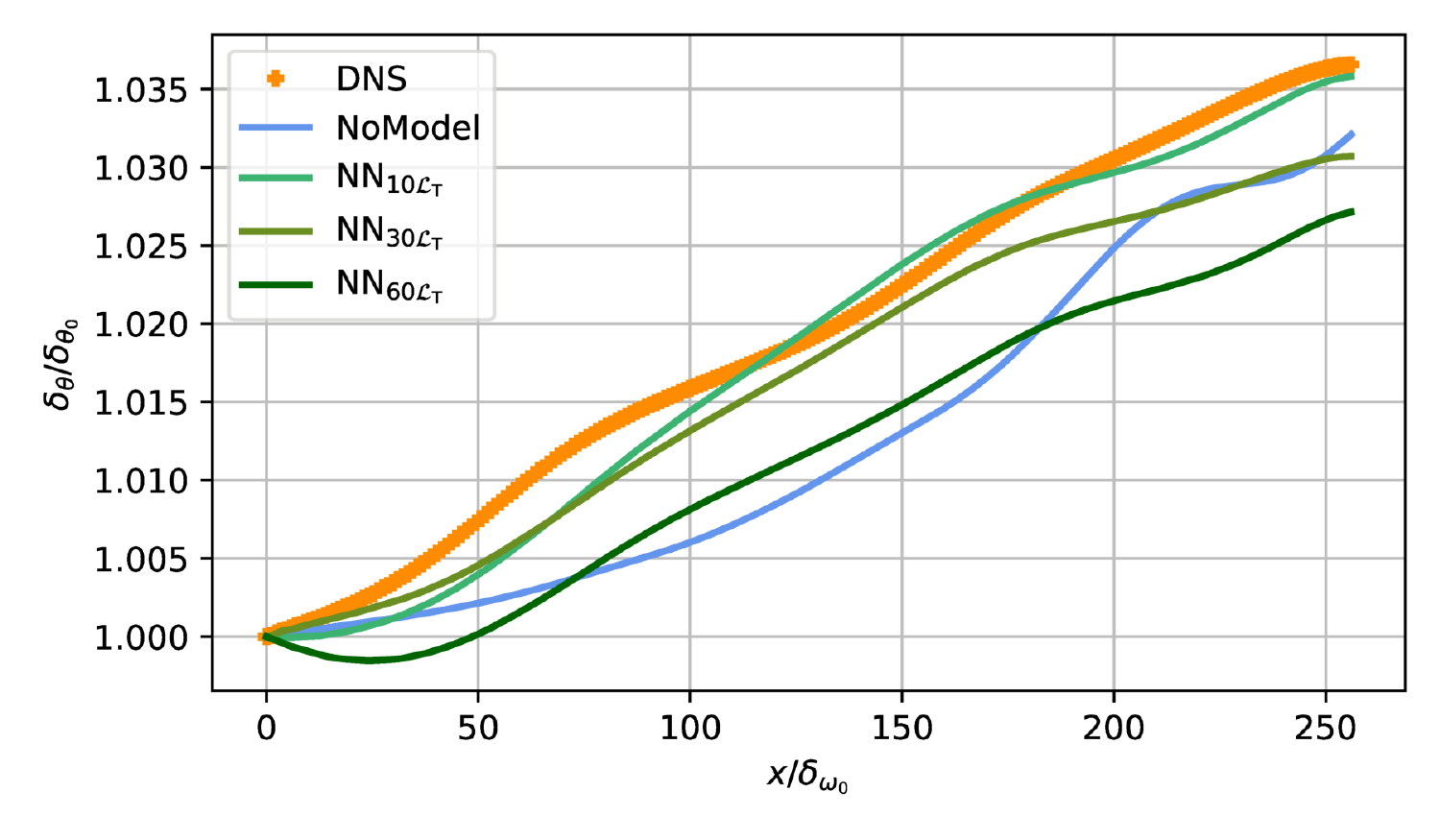}
    \caption{}
    \label{fig:spatial_momentum_thickness}
\end{subfigure}
\caption{Vorticity and momentum thickness of the downsampled \gls{dns}, no-model, and learned model simulations}
\end{figure}

The evaluations of vorticity and momentum thickness in figures \ref{fig:spatial_vorticity_thickness} and \ref{fig:spatial_momentum_thickness} capture a delayed mixing layer development. Especially early stages of the mixing layer immediately after the first roll-up are modelled inaccurately. While all models show this behaviour, the delay in terms of momentum thickness is more pronounced for the long unrollment $60$-step model.
Contrary, the roll-up inaccuracy results in a noticeable offset in the vorticity thickness around $x/\delta_{\omega_0}=100$ for all models, but the $60$-step model performs best further downstream by recovering the \gls{dns} behaviour. This recovery is lacking in $10$ and $30$ step models, causing the evaluation of Reynolds stresses at $x=192\Delta x$ (figure \ref{fig:spatial_rans_tke_reyStr}) to exhibit large discrepancies between \gls{dns} and learned model simulation for these models, with notable exception of the $60$-step model. Note however, that despite not being capable of exactly reproducing the entire mixing layer up to the finest detail, the learned models still greatly outperform a no-model simulation. Momentum thickness evaluations show beneficial results for the models trained with shorter unrollments. Due to the definition of momentum thickness as an integral quantity over the shear direction, an increase in this quantity is caused by strong deviations from the initial step-profile of the mixing layer. While the integral values for the momentum thickness of $10$ and $30$ step models are close to the \gls{dns}, the underlying turbulence fluctuations causing these values are not accurate to the \gls{dns}, which can be seen in turbulence kinetic energy and Reynolds stress evaluations in figure \ref{fig:spatial_rans_tke_reyStr}.
Considering these results jointly, we draw the conclusion that the $60$-step model yields the best performance.

Additionally, the evaluations show the benefits of training through multiple unrolled steps. The 10-step model develops instabilities after $500\Delta t$, which is equivalent to one flow-through time. From this time on, the learned model only sees self-exited instabilities in the mixing layer. This constitutes an extrapolation with respect to the temporal unrollment, as well as with respect to the inlet perturbation due to the use of a test dataset. This in turn can cause spurious oscillations and thus a deterioration of solution quality. The $30$-step model shows this behaviour to a lesser extent and generates a stable, statistically-steady progression of the mixing layer for this case of temporal extrapolation. Even better behaviour is only achieved by the $60$-step model. It practically eliminates the instabilities seen in other models.

\begin{figure}[h]
\centering
\includegraphics[width=\textwidth]{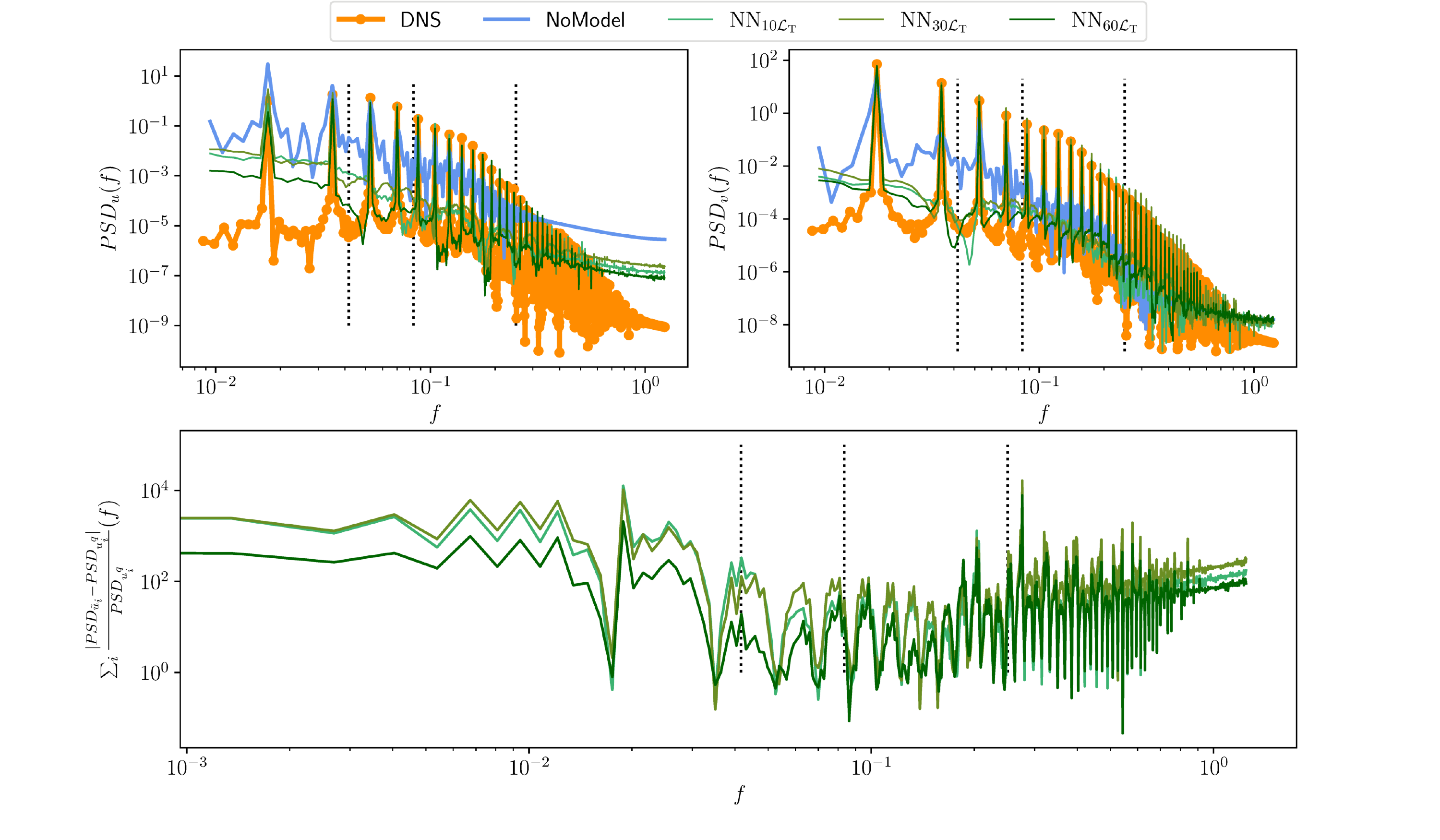}
\caption{Power Spectral Density of velocity fluctuations over time at sampling point $(x,y)=(192\Delta x,0)$ based on the \textit{training dataset} for \gls{dns}, no-model and learned model simulations at top; bottom figure displays the relative error of the power densities over frequencies, accumulated for both velocity components; frequencies to the right of a dotted vertical line are fully enclosed in a training iteration; vertical lines correspond to (60, 30, 10) unrolled steps from left to right
}
\label{fig:spatial_psd}
\end{figure}

While previous evaluations showcased the stability improvements gained by training through multiple solver steps, another benefit of this approach relates to the temporal fluctuations in \gls{dns} training data. As visualised in figure \ref{fig:spatial_psd}, only some of the interactions between \gls{cnn} and these temporal oscillations are covered in a training iteration. Consequently, the training loop imposes a high-pass cutoff on the observed frequencies that directly depends on the number of unrolled solver steps. To extract the temporal features that our models learned from the training dataset, we calculate the power-spectral density of the velocity fields at sampling point $(x,y)=(160,0)$ on training data. The sampling time-span for the learned models starts after one flow-through time and stops after the next $4$ flow-through times passed.
The resulting power-spectral densities are compared to a long-term evaluation of the \gls{dns} data, and a relative error between the spectra is computed. The results are shown in figure \ref{fig:spatial_psd}  and support the following observations. Firstly, all learned models can capture the discrete nature of the dominant frequencies quite well. Especially the $60$-step model shows good approximation of the \gls{dns} evaluation. In contrast, the no-model does not match the \gls{dns} characteristics. Secondly, the relative error of the power spectra generated by the $60$-step model is substantially lower for all but the highest frequencies. Since $30$- and $10$-step models only saw the interaction with fine scales during their training, these models perform worse on the lower frequencies, which results in higher relative errors for the relatively low vortex roll-up and vortex merging frequencies. These features operate on the order of one integral timescale and are better resolved by 60 unrolled steps.

%% file: sections/backpropagation.tex
\section{Gradient Back-propagation}\label{section:backpropagation}
\begin{table}
\parbox[t][][t]{.47\linewidth}{
    \centering
\begin{tabular}{l c r r r}
    Name  &  Loss  & Steps &  Grad & MSE at $ t_e$ \\
    \hline
    $\text{NoModel}$                 & -                      & - & - & $1.25\mathrm{e}{-3}$\\
    $\text{NN}_{60\text{-}10,\mathcal{L}_\text{T}}$   & $\mathcal{L}_\text{T}$ & 60 & 10 & $2.36\mathrm{e}{-5}$\\
    $\text{NN}_{60\text{-}20,\mathcal{L}_\text{T}}$   & $\mathcal{L}_\text{T}$ & 60 & 20 & $2.19\mathrm{e}{-5}$\\
    $\text{NN}_{60\text{-}30,\mathcal{L}_\text{T}}$   & $\mathcal{L}_\text{T}$ & 60 & 30 & $1.93\mathrm{e}{-5}$\\
    $\text{NN}_{60\text{-}60,\mathcal{L}_\text{T}}$   & $\mathcal{L}_\text{T}$ & 60 & 60 & $-$\\
    \end{tabular}
    \caption{Temporal mixing layer; model details for $60$-$X$ models; MSE w.r.t. \gls{dns} from test-data at $t_e = 512\Delta t$; training of $\text{NN}_{60\text{-}60,\mathcal{L}_\text{T}}$ is unstable }
    \label{table:temporal_60step_l2_comparison}
}
\hfill
\parbox[t][][t]{.47\linewidth}{
\centering
    \begin{tabular}{ l c r r r }
    Name    &   Loss  &  Steps &  Grad &  MSE at $t_e$ \\
    \hline
    $\text{NoModel}$& - & - & - & $2.03\mathrm{e}{-2}$\\
    $\text{NN}_{60\text{-}10,\mathcal{L}_\mathrm{T}}$& $\mathcal{L}_\mathrm{T}$ & 60 & 10 & $2.44\mathrm{e}{-3}$\\
    $\text{NN}_{60\text{-}20,\mathcal{L}_\mathrm{T}}$& $\mathcal{L}_\mathrm{T}$ & 60 & 20 & $2.73\mathrm{e}{-3}$\\
    $\text{NN}_{60\text{-}30,\mathcal{L}_\mathrm{T}}$& $\mathcal{L}_\mathrm{T}$ & 60 & 30 & $2.98\mathrm{e}{-3}$\\
    $\text{NN}_{60\text{-}60,\mathcal{L}_\mathrm{T}}$& $\mathcal{L}_\mathrm{T}$ & 60 & 60 & $1.19\mathrm{e}{-2}$\\
    \end{tabular}
    \caption{Spatial mixing layer; model details for $60$-$X$ models; MSE w.r.t. \gls{dns} from test-data at $t_e = 1000\Delta t$}
    \label{table:spatial_60step_l2_comparison}
}
\vspace{5mm}
\end{table}

Our evaluations on temporally and spatially developing mixing layers show significant performance gain by longer unrollment times, with the best accuracy given by a $60$-step model. However, long unrollments can cause stability problems. Repeated applications of neural networks are known to be problematic during training, where exploding or diminishing gradients can significantly deteriorate the quality of gradients \citep{pascanu2013OnThe}. To avoid this, we utilise a custom version of the gradient stopping technique: instead of discarding gradients generated by some (earlier) simulation steps, we split the gradient back-propagation into individually evaluated subranges. In other words, the training still exposes long temporal unrollments and preserves the gradient influence of all steps, but does not propagate gradients back to the first application of the network model. We use $60$-step models to study model accuracy with respect to the length of these back-propagation subranges on a range of $10$, $20$, $30$, and $60$ backward steps. We will use the notation $\text{NN}_{m\text{-}g}$ with two numbers $m$ and $g$, where $m$ describes the number of unrolled forward steps, and $g$ represents the length of the subranges for which gradients are calculated individually. In practice, this means that gradients of a $60$-$20$ model are only back-propagated through $3$ non-overlapping sections of $20$ steps each. The $60$-$60$ model recovers the standard differentiable training procedure used for previous models.

\begin{figure}
\centering
\begin{subfigure}[b]{\textwidth}
    \centering
    \includegraphics[width=\textwidth]{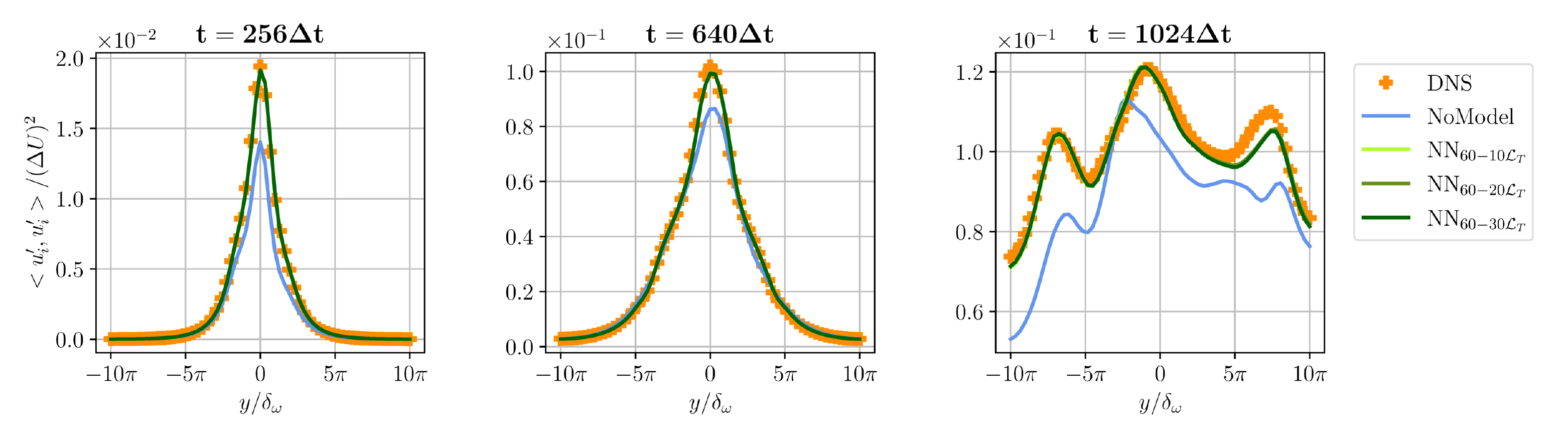}
    \caption{}
\end{subfigure}
\centering
\begin{subfigure}[b]{\textwidth}
    \centering
    \includegraphics[width=\textwidth]{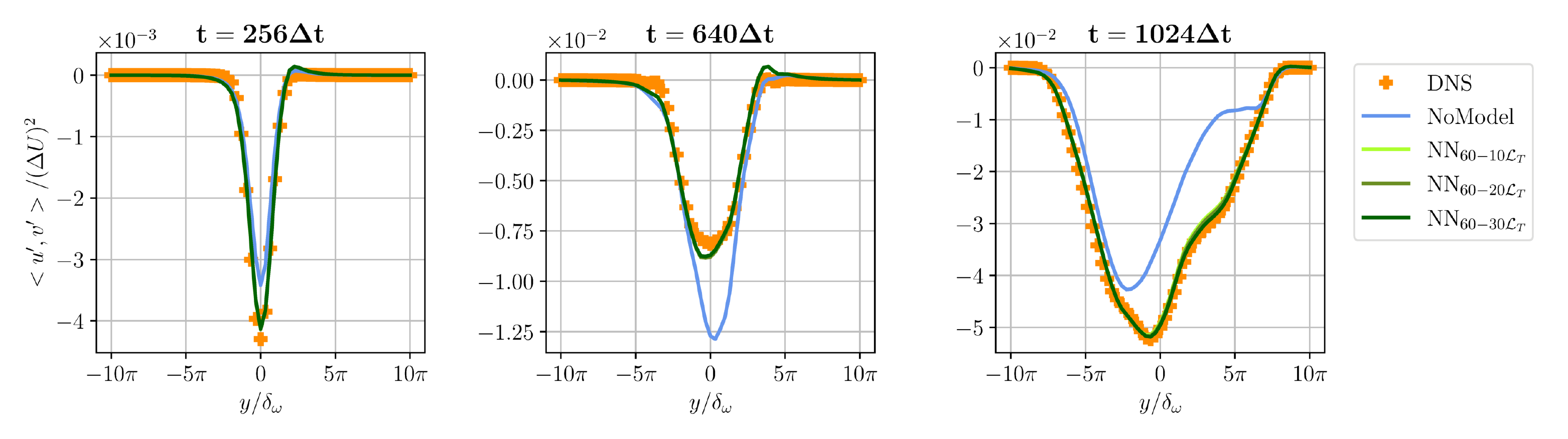}
    \caption{}
\end{subfigure}
\caption{Comparison of \gls{dns}, no-model, and $60$-step model simulations with respect to resolved turbulence kinetic energy (a), and Reynolds stresses (b)}
\label{fig:temporal_60step_rans_tke_reyStr}
\end{figure}

This procedure was applied to temporally and spatially developing mixing layers. Details of the trained models are found in tables \ref{table:temporal_60step_l2_comparison} and \ref{table:spatial_60step_l2_comparison}. Note that the training of the $\mathrm{NN}_{60-60,\mathcal{L}_\mathrm{T}}$ was not stable for the temporal mixing layer case, which we attribute to unstable gradients in the optimisation. In contrast, the subrange gradient models are stable during training. Additional evaluations of Reynolds stresses and turbulence kinetic energy for the temporal mixing layer indicate no performance differences between these models, as shown in figure \ref{fig:temporal_60step_rans_tke_reyStr}. We thus conclude that the method of subrange back-propagation makes the training of 60-step possible, but also that the model performance on the temporal mixing layer was already saturated by the $30$-step model, as previously mentioned in section \ref{subsection:temporal_mixing_layers}. The $\mathrm{NN}_{60-30,\mathcal{L}_\mathrm{T}}$ was used in the evaluation in section \ref{subsection:temporal_mixing_layers}.

\begin{figure}
    \centering
    \includegraphics[width=\textwidth]{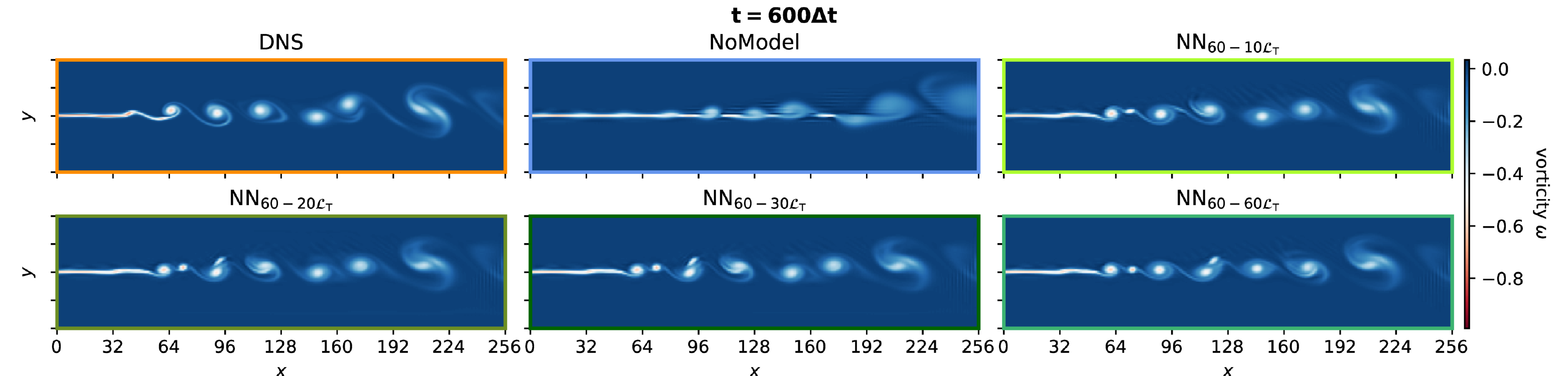}
    \caption{Vorticity comparison of $60$-step models on spatial mixing layer simulations at $t=700\Delta t$ on the test dataset}
    \label{fig:spatial_60step_vorticity_late}
\end{figure}

\begin{figure}
\centering
\begin{subfigure}[b]{\textwidth}
    \centering
    \includegraphics[width=\textwidth]{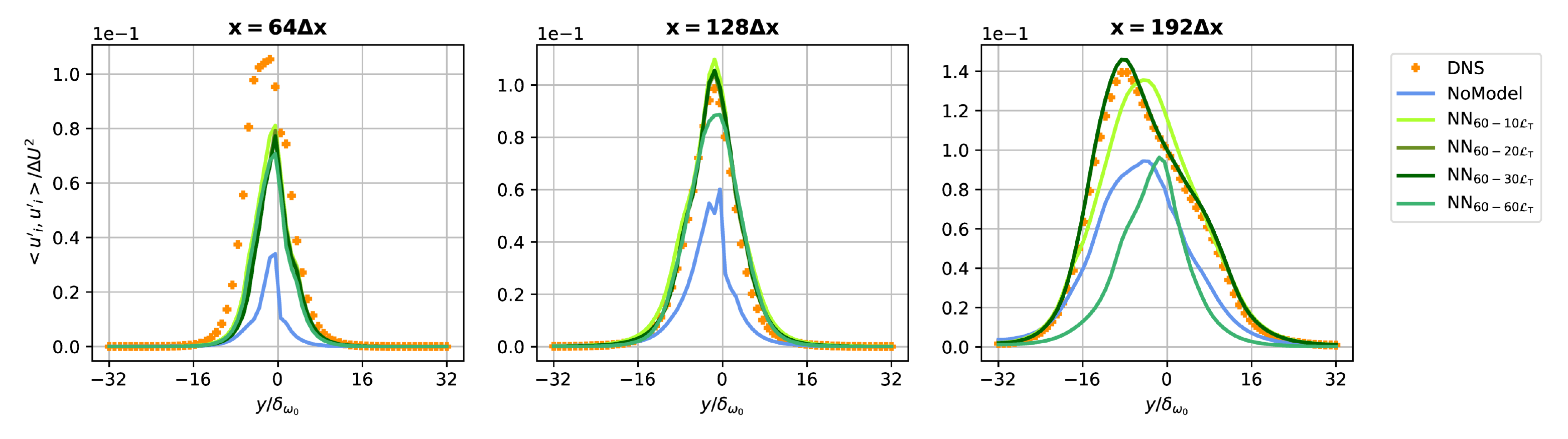}
    \caption{}
\end{subfigure}
\centering
\begin{subfigure}[b]{\textwidth}
    \centering
    \includegraphics[width=\textwidth]{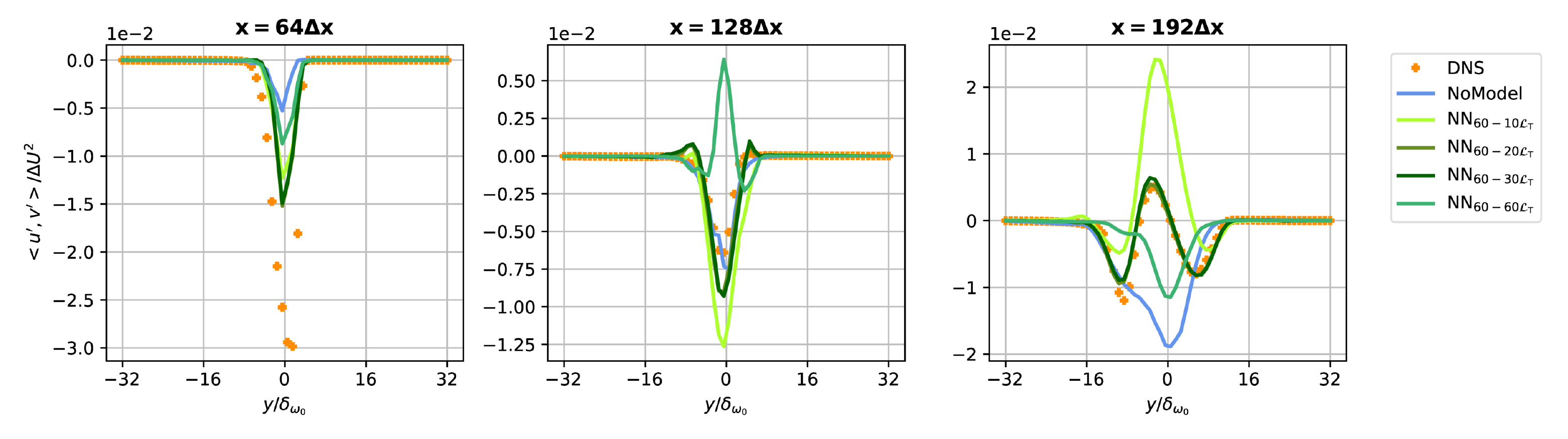}
    \caption{}
\end{subfigure}
\caption{Comparison of downsampled \gls{dns}, no-model, and learned model simulations with respect to Reynolds-averaged resolved turbulence kinetic energy (a); and Reynolds stresses (b)}
\label{fig:spatial_60step_rans_tke_reyStr}
\end{figure}

\begin{figure}
\centering
\begin{subfigure}[b]{.48\textwidth}
    \centering
    \includegraphics[width=\textwidth]{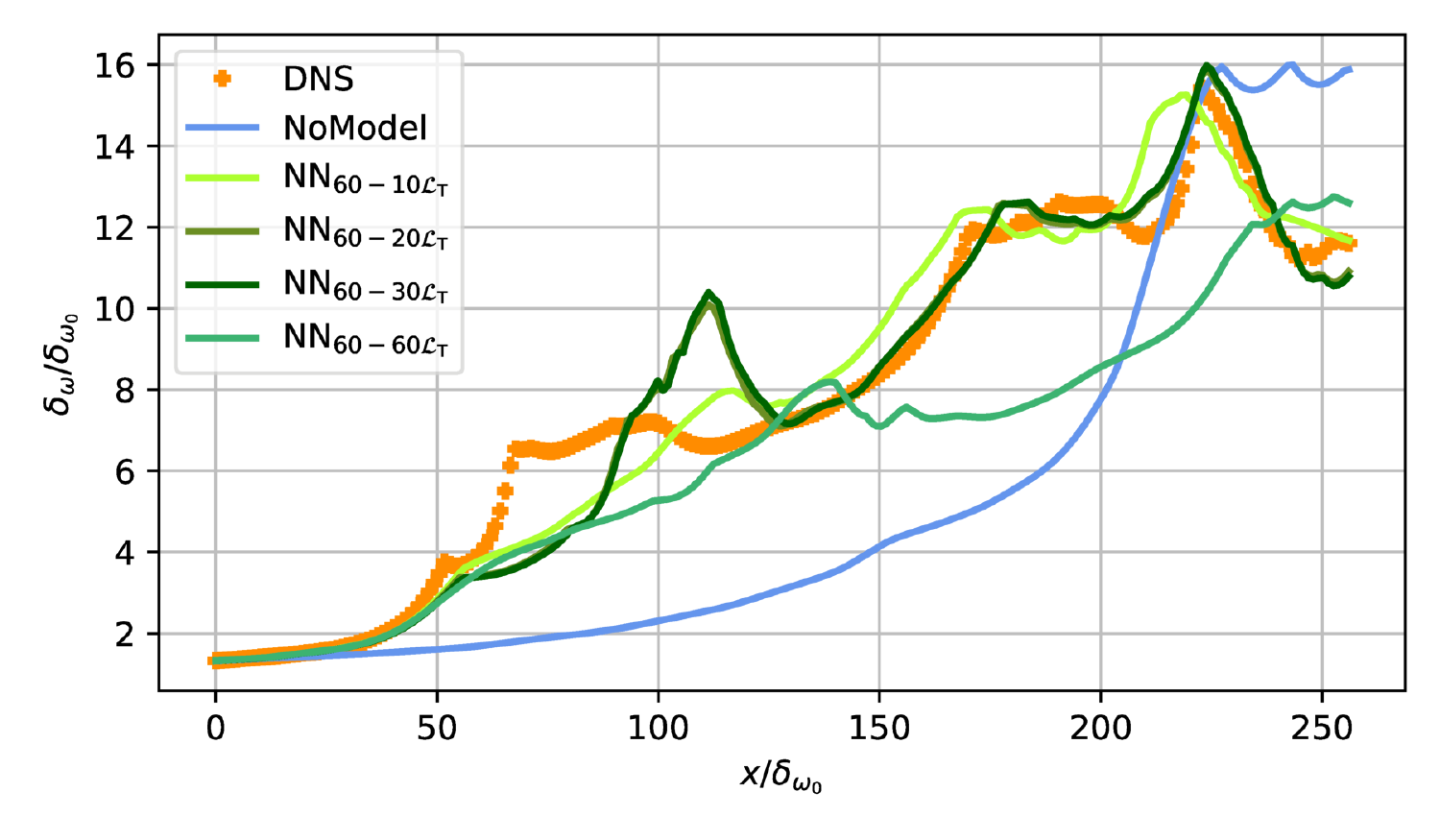}
    \caption{}
\end{subfigure}
\hfill
\begin{subfigure}[b]{.48\textwidth}
    \centering
    \includegraphics[width=\textwidth]{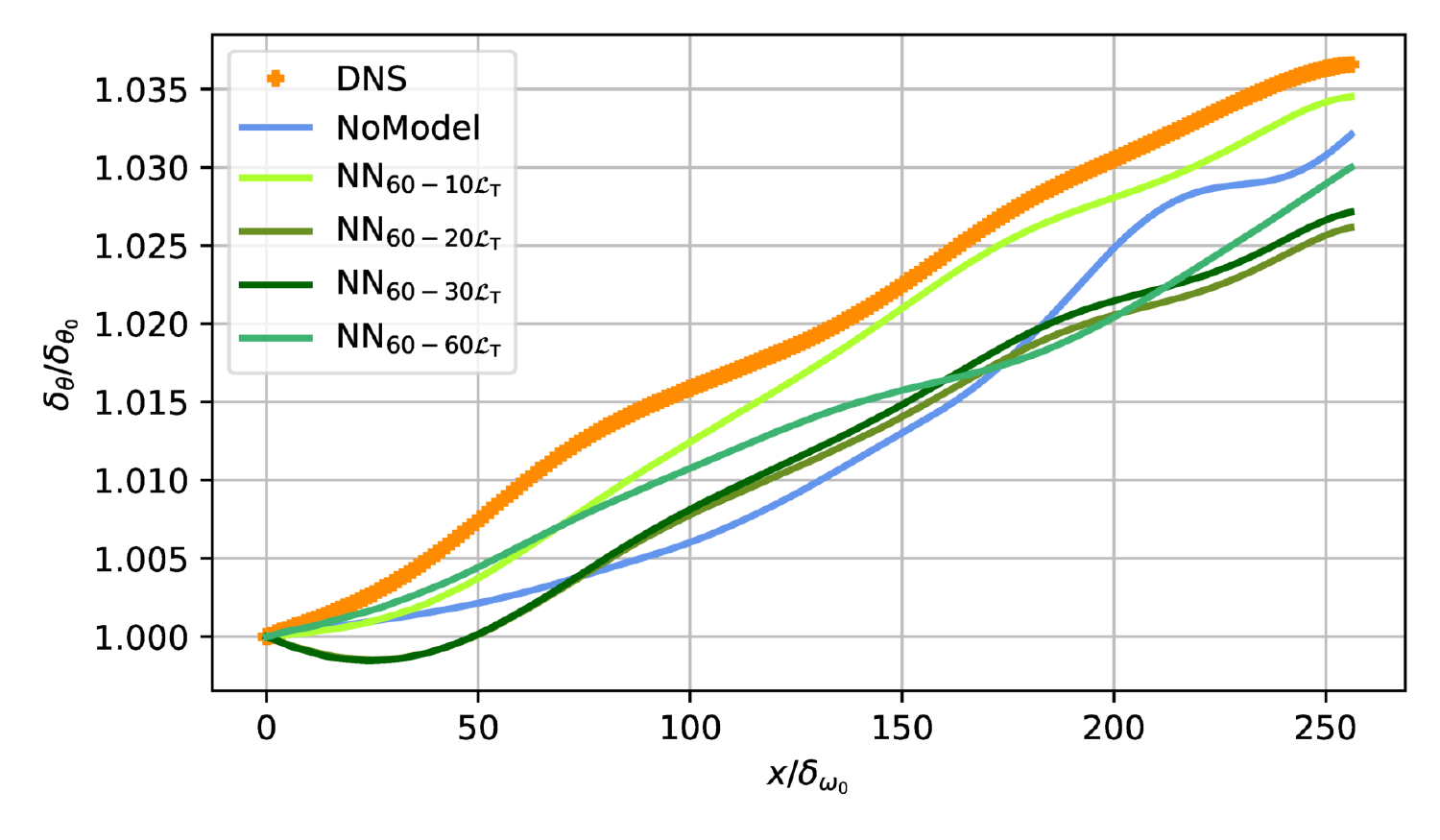}
    \caption{}
    \end{subfigure}
\caption{Vorticity and momentum thickness of the downsampled \gls{dns}, no-model, and learned model simulations}
\label{fig:spatial_60step_vorticity_momentum_thickness}
\end{figure}

The spatial mixing layer models are evaluated on vorticity snapshots in figure \ref{fig:spatial_60step_vorticity_late}, turbulence kinetic energy and Reynolds stresses in figure \ref{fig:spatial_60step_rans_tke_reyStr}, as well as vorticity and momentum thickness in figures \ref{fig:spatial_60step_vorticity_momentum_thickness}. These results indicate that there is a optimal number of consecutive back-propagation steps around $20$ to $30$, where the optimisation gradients contain long-term information while still maintaining good quality that is unaffected by risks of recurrent evaluation. The $\mathrm{NN}_{60-20,\mathcal{L}_\mathrm{T}}$ and $\mathrm{NN}_{60-30,\mathcal{L}_\mathrm{T}}$ model achieve best performance on all metrics except for the momentum thickness. We attribute the larger values of momentum thickness to some spurious oscillations exhibited by $\mathrm{NN}_{60-10,\mathcal{L}_\mathrm{T}}$ and $\mathrm{NN}_{60-60,\mathcal{L}_\mathrm{T}}$ models. The $\mathrm{NN}_{60-30,\mathcal{L}_\mathrm{T}}$ was used in earlier unrollment evaluations in section \ref{subsection:spatial_mixing_layers}.

\begin{figure}
\centering
  \begin{minipage}[b]{.43\textwidth}
  \subfloat
    []
    {\label{fig:figA}\includegraphics[width=\textwidth]{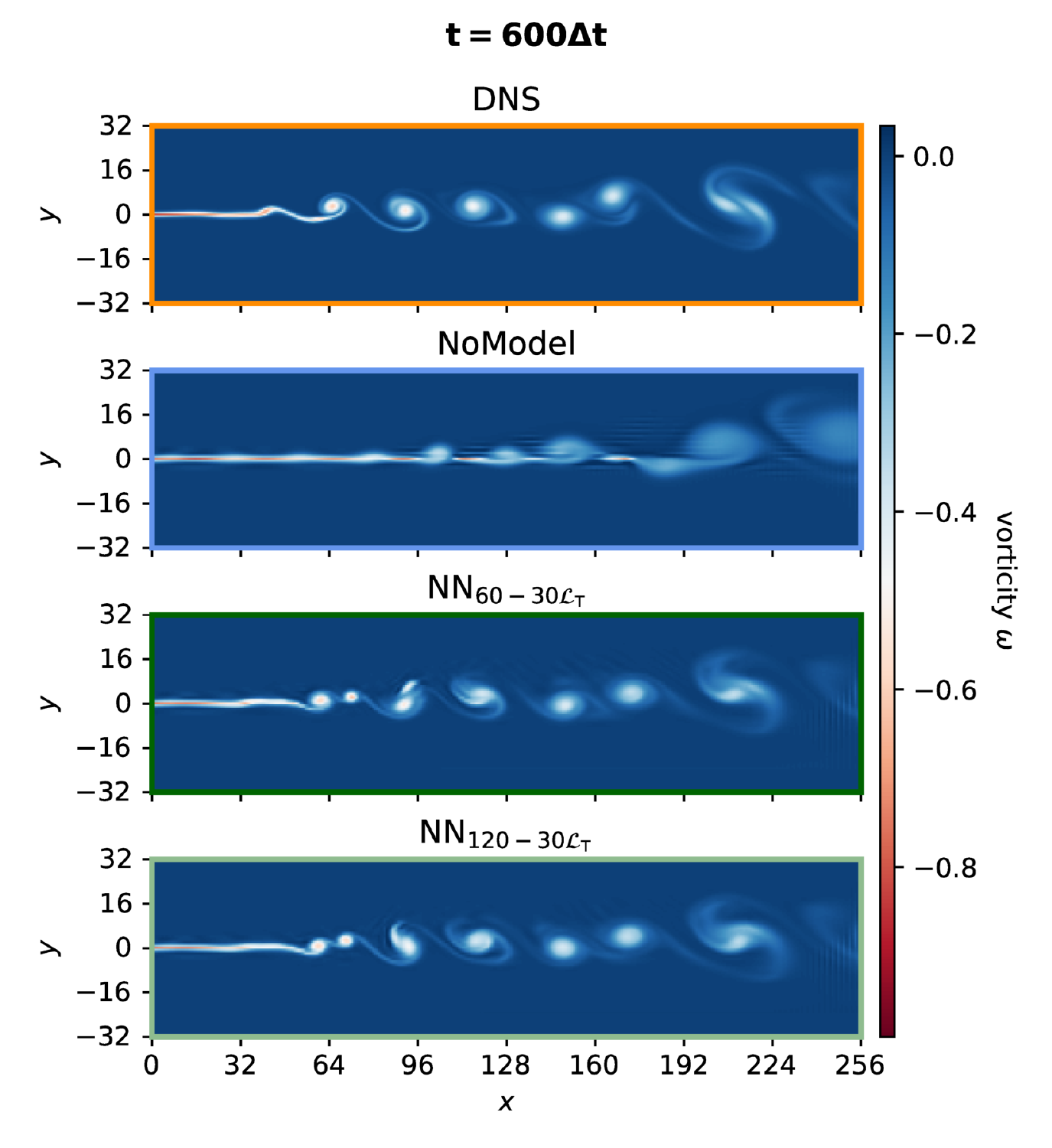}}
  \end{minipage}%
\begin{minipage}[b]{.57\textwidth}
\subfloat[]
  {\label{fig:figB}\includegraphics[width=\textwidth, ]{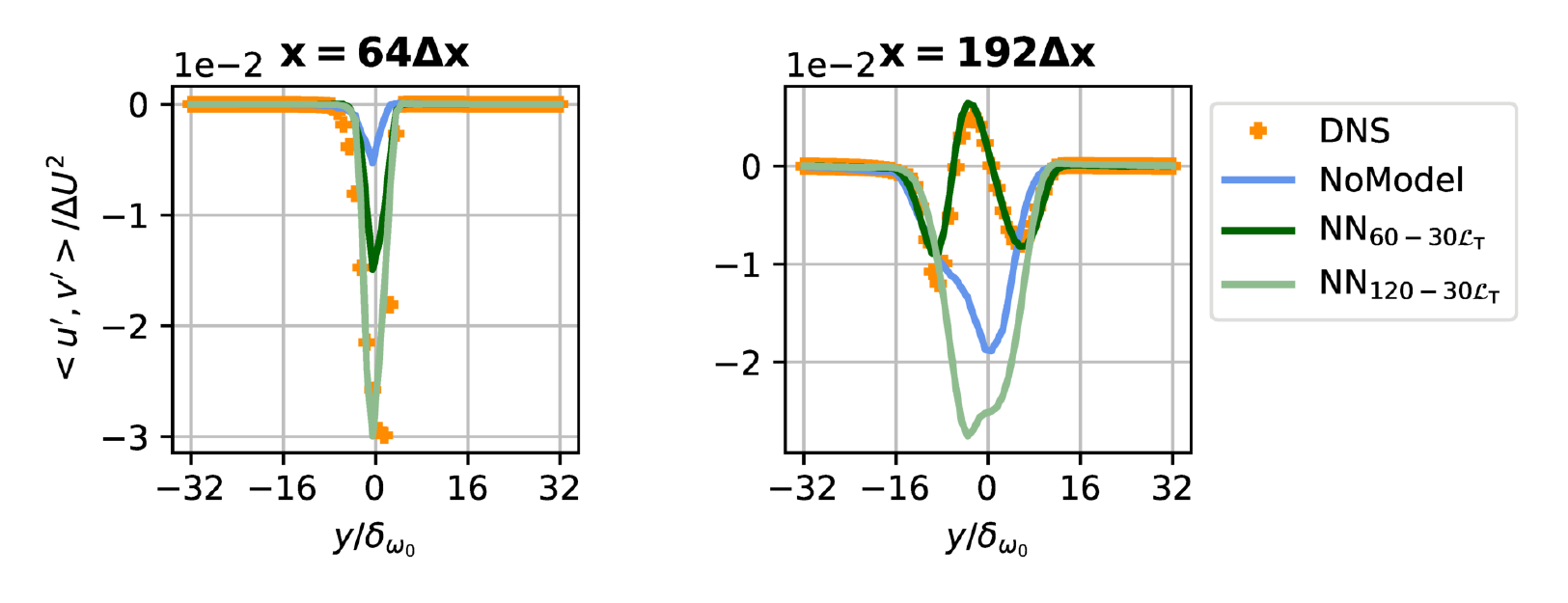}}

\vfill
\centering
\subfloat
  []
  {\label{fig:figC}\includegraphics[height=2.8cm]{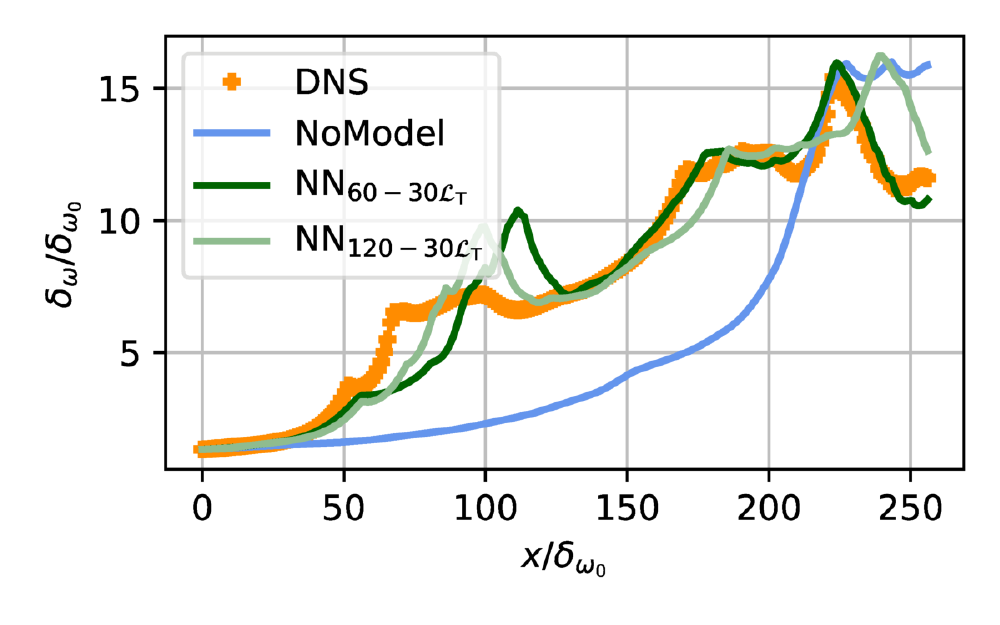}}
\end{minipage}%
\caption{Selected evaluations of the 120-step model with vorticity snapshots in (a), Reynolds stresses in (b), and vorticity thickness in (c)}
\label{fig:spatial_120step}
\end{figure}

Another potential problem could be caused by training towards matching frames separated by long time-spans. Turbulent flows could potentially loose correlation to the reference data over long temporal horizons, which would render this learning approach driven by simulated \gls{dns} data inapplicable. The unrollment times in this paper are, however, far from reaching an uncorrelated state. As shown in the previous evaluations, the $60$-step models perform better than their $30$-step counterparts, indicating that there is additional information provided by unrolling $60$ steps. This shows that the unrolled temporal horizons are far from exhibiting flow decorrelation. Further experiments with even longer unrollments on the spatial mixing layer revealed that no improvement is achieved beyond $60$ steps in this case. Figure \ref{fig:spatial_120step} depicts selected evaluations of a 120-step model, which lack improvements over the $60$ step counterpart. While the $120$-step model gains accuracy in early upstream cross-sections, the mixing layer shift downstream of the first roll-up is worse in direct comparison.
We also investigated yet longer horizons (180 and 240 steps), but these runs saw a reduced accuracy with respect to some of the evaluations.
One explanation is that the flow field is uncorrelated to the \gls{dns} data for these long horizons, leading to a diffused learning signal.
If the loss was computed on late, uncorrelated frames, we would expect generated gradients to resemble random noise. While earlier frames would still provide valuable information, the random noise from these later frames could prevent the learning of precise corrections.
In addition, the longer runs used the same set of hyperparameters as determined for the shorter unrollments, the long horizon runs could also profit from a broader hyperparameter search.

In this section, we have identified gradient instabilities as the main problem when unrolling long temporal horizons. We have introduced a gradient splitting technique that stabilised the training procedure. This is done by splitting the gradient calculation in to non-overlapping subranges. For the studied setups and $60$-step models, a split into $2$ subranges of $30$ steps each performed best. One can conclude that longer unrollments pay off in terms of modelling accuracy up to a certain saturation point. In our simulations this saturation point lies at circa $60$ steps, which coincides with the integral timescales of the respective scenarios. Unrolling beyond that saturation point is possible, but
leads to increased computational effort and may require special treatment such as a further tuning of the  hyperparameters.

%% file: sections/performance.tex
\section{Computational Performance}
\label{section:perfromance}

\begin{table}
 \centering
 \begin{tabular}{c l r r r r r r}
 & \multicolumn{1}{c}{Name} & \multicolumn{1}{c}{Resolution} & \multicolumn{1}{c}{Time until $t_e$} & \multicolumn{1}{c}{\ \ \# of $\Delta t$} & \multicolumn{1}{c}{\ \ \ \ Time per $\Delta t$} &\multicolumn{1}{c}{\ \ \ \ \ \ $\text{MSE}$} & \ \ Training\\
 \hline
 \parbox[t]{4mm}{\multirow{5}{*}{\rotatebox[origin=c]{90}{IDT}}} & DNS & $1024\times 1024$ & $14497.7s$ &$8000$&$1.812\pm 0.432s$&$0$ & -\\
 & NN &$128\times128$ & $71.2s$ &$1000$&$0.071\pm0.015s$&$1.96\mathrm{e}{-2}$ & 61h\\
 & NoModel &$128\times128$ & $65.5s$ &$1000$&$0.066\pm0.066s$&$5.45\mathrm{e}{-2}$ & -\\
 & - &$256\times256$ & $240.6s$ &$2000$&$0.120\pm0.027s$&$1.07\mathrm{e}{-2}$ & -\\
 & - &$512\times512$ & $1348.8s$ &$4000$&$0.337\pm0.049s$&$1.03\mathrm{e}{-3}$ & -\\
 \hline
 \parbox[t]{4mm}{\multirow{6}{*}{\rotatebox[origin=c]{90}{TML}}} & DNS & $1024\times 512$ & $12467.8s$ &$8000$& $1.559\pm0.764s$ & 0 & -\\
 & NN &$128\times 64$ & $155.6s$ &$1000$&$0.156\pm0.049s$&$1.66\mathrm{e}{-5}$ & 78h\\
 & NoModel &$128\times 64$ & $144.5s$ &$1000$&$0.145\pm0.052s$&$1.23\mathrm{e}{-3}$ & -\\
 & - &$256\times 128$ &  $593.5s$ &$2000$&$0.297\pm0.148s$&$2.30\mathrm{e}{-4}$ & -\\
 & - &$340\times 170$ & $1054.0s$ &$2656$&$0.395\pm0.216s$&$1.07\mathrm{e}{-4}$ & -\\
 & - &$512\times 256$ & $2154.9s$ &$4000$&$0.539\pm0.259s$&$2.51\mathrm{e}{-5}$ & -\\

  \hline
 \parbox[t]{4mm}{\multirow{6}{*}{\rotatebox[origin=c]{90}{SML}}} & DNS & $2048 \times 512$ & $81925.3s$ &$8000$&$10.242\pm1.144s$&$0$ & -\\
 & NN &$256\times 64$ & $1813.6s$ &$1000$&$1.815\pm0.254s$&$2.20\mathrm{e}{-3}$ & 240h\\
 & NoModel &$256\times 64$ & $1815.4s$ &$1000$&$1.817\pm0.450s$&$2.03\mathrm{e}{-2}$ & -\\
 & - &$512\times 128$ & $3971.3s$ &$2000$&$1.987\pm0.348s$&$6.22\mathrm{e}{-3}$ & -\\
 & - &$768\times 192$ & $6719.23s$ &$2667$&$2.240\pm0.273s$&$7.57\mathrm{e}{-4}$ & -\\
 & - &$1024\times 256$ & $12071.5s$ &$4000$&$3.019\pm0.245s$&$3.13\mathrm{e}{-4}$ & -\\
 \hline
 \end{tabular}
 \caption{Computational performance comparison over $t_e=1000\Delta t$ for the used flow scenarios, Isotropic Decaying Turbulence (IDT), Temporal Mixing Layer (TML) and Spatial Mixing Layer (SML); MSE values are evaluated on the velocity field at $500\Delta t$; Training time on one GPU}
 \label{table:performance}
\end{table}

The development of turbulence models is ultimately motivated by a reduced computational cost, which facilitates numerical simulations in flow scenarios where a \gls{dns} is prohibitively expensive. Preceding sections have outlined the corrective capabilities of our learned models. We now seek to put these improvements into perspective by studying the computational cost of our learned models at inference time. For all of our performance evaluations, an \textit{Intel Xeon E5-1650} CPU and a \textit{Nvidia GTX 1080Ti} \gls{gpu} are used. We use the computational setups from our model evaluation runs on test data in the Isotropic Turbulence, Temporal Mixing Layer and Spatial Mixing Layer cases in sections \ref{section:isotropic_turbulence}, \ref{subsection:temporal_mixing_layers} and \ref{subsection:spatial_mixing_layers} respectively.

Exactly as before, an $8\times$ scaling factor is deployed on both the spatial resolution and timestep size. We then run the simulations until the time $t_e=1000\Delta t$ is reached, while recording the required computational time for each timestep. The results are summarised in table \ref{table:performance}, where the total simulation time as well as per-timestep values are listed. We also assess the computational cost of a no-model simulation that matches the performance of our models.

The resulting data shows that the neural network incurs only a negligible cost of circa $10\%$ in comparison to no-model simulations at the same resolution.
The learned models clearly outperform the no-model variants in terms of MSEs, and incur only a fraction of the computational cost required for the \gls{dns} variants.

 \begin{figure}
\centering
\begin{subfigure}[b]{.32\textwidth}
    \centering
    \includegraphics[width=\textwidth]{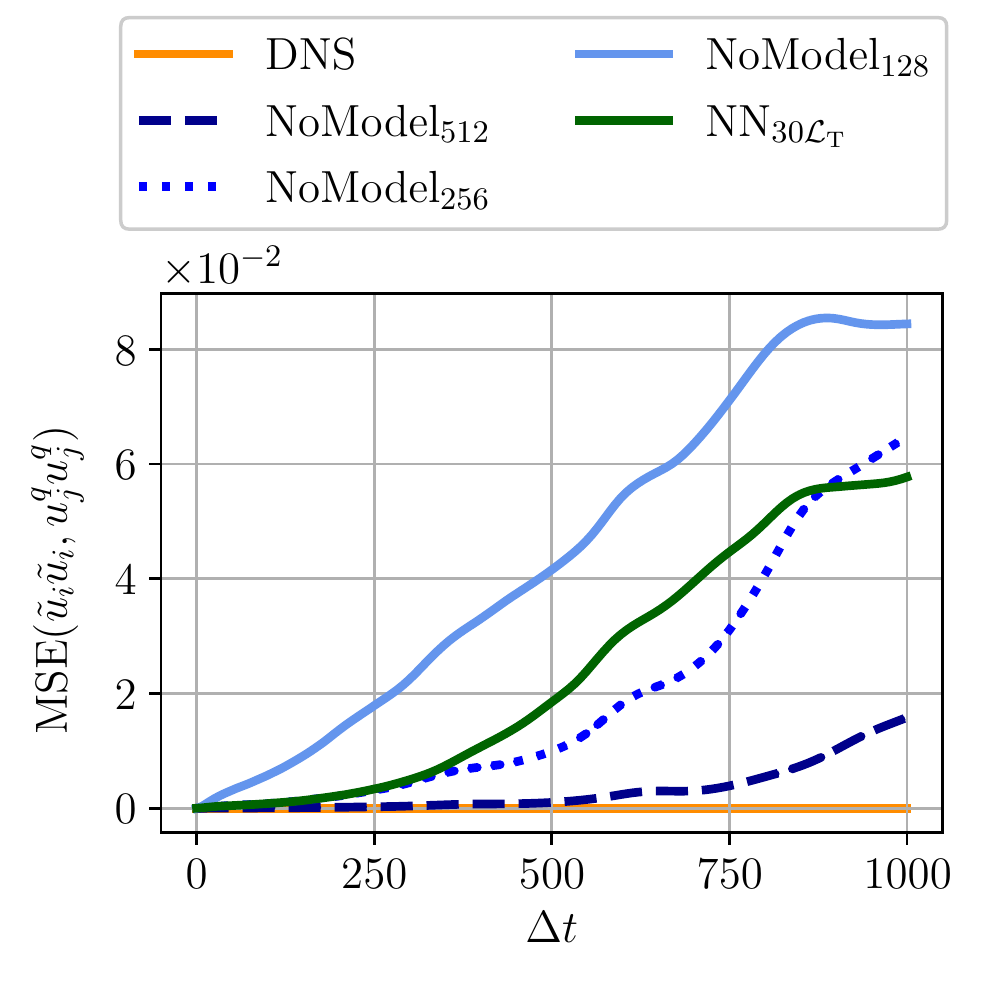}
    \caption{}
    \label{fig:performance_randomised_corr}
\end{subfigure}
\hfill
\begin{subfigure}[b]{.32\textwidth}
    \centering
    \includegraphics[width=\textwidth]{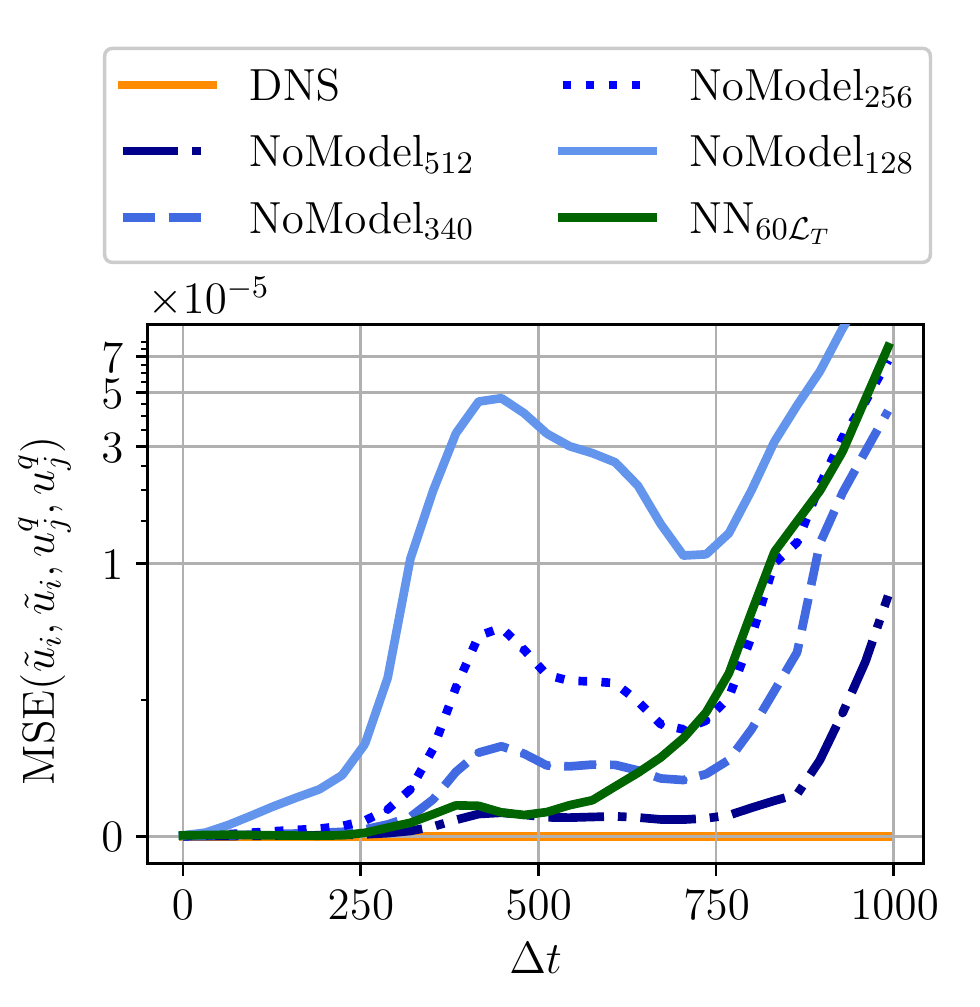}
    \caption{}
    \label{fig:performance_temporal_tke}
\end{subfigure}
\hfill
\begin{subfigure}[b]{.32\textwidth}
    \centering
    \includegraphics[width=\textwidth]{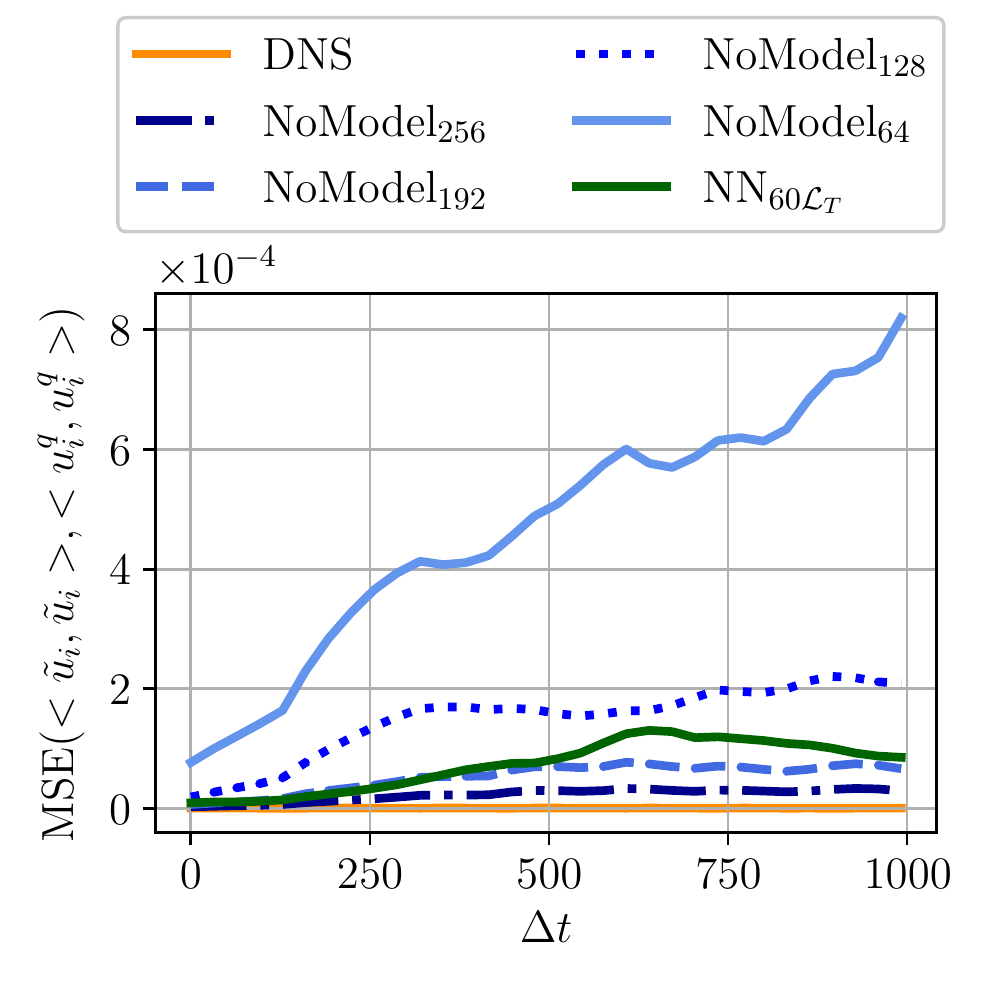}
    \caption{}
    \label{fig:performance_spatial_tke}
\end{subfigure}
\caption{Similarity evolutions over time measured by the MSE on resolved turbulence kinetic energy for randomised turbulence simulations (a), temporal mixing layer simulations (b), and spatial mixing layer simulations (c)}\label{fig:performance_all}
\end{figure}

In addition, we provide the temporal evolution of the MSE evaluated on resolved turbulence kinetic energies for all three scenarios in figure \ref{fig:performance_all}.
From this evaluation, we conclude that our method consistently outperforms simulations with a $2\times$ higher resolution in spatial and temporal dimensions. Additionally, we found our learned models to often be on-par with $4\times$ higher resolved simulations, e.g. in the first half of the temporal mixing layer case.
On the basis of the clock-times from table \ref{table:performance}, this corresponds to
a speedup of $3.3$ over $2\times$ isotropic turbulence simulations. For the mixing layer cases, the hybrid model is on average resembling the performance of $3\times$ reference simulations, which corresponds to a speed-up of $7.0$ for the temporal, and $3.7$ for the spatial mixing layer. For the former, our model even closely matches the performance of a $4\times$ simulation for several hundred time steps, which represents a speedup of $14.4$.

While other works have reported even larger performance improvements \citep{Kochkov}, we believe that our measurements are representative of real-world scenarios with higher-order solvers. Asymptotically, we also expect even larger payoffs for the high-resolution, three-dimensional simulations that are prevalent in real-world applications.

Naturally, the training of each neural network requires a substantial one-time cost. In our case, the network took 3 to 10 days of training, depending on the individual problem setup. The required \gls{gpu}-hours for the best-performing models are listed in table \ref{table:performance}. The longer unrolled temporal horizons and larger domain increase the required training time for the spatial mixing layer. For the three used setups, these training times are equivalent to [120, 118, 22] \gls{dns} solves of full length as used in dataset calculation.
However, under the assumption that the learned turbulence model can be employed by multiple users in a larger number of simulations to produce new outputs, this cost will quickly amortise. Especially the most complex spatial mixing layer case shows a favourable relation of training cost to simulation speedup. Additionally a successful application of this approach to 3D turbulence would make training cheaper in relation to \gls{dns} and speedups larger, due to the scaling through an additional spatial dimension. It is worth noting that our comparisons are based on \gls{gpu} solvers, and performance is likely to vary on CPU or mixed solvers, where parts of the computation are CPU-based and communication overheads could deteriorate gains.

%% file: sections/conclusion_arxiv_v2.tex
\section{Conclusion}
\label{section:conclusion}
In this paper, we studied adjoint training methods for hybrid solvers that integrate neural networks in a numerical solver. We targeted the modelling of the finest turbulent scales when these cannot be resolved by the simulation grid. The differentiable nature of our implementation of the PISO solver allows us to train the network through multiple unrolled steps. We deem this feature crucial, since we found strong dependence of the model performance and long-term stability on the number of unrolled steps. Our results indicate that covering one integral timescale yields the best performance. Shorter unrollments generally suffer from accuracy and stability issues, while for longer ones the model accuracy saturates and training becomes less efficient. 
We showcased the application of our method to three different flow scenarios, the two-dimensional isotropic decaying turbulence, the temporally developing mixing layer and the spatially developing mixing layer, whilst keeping the network architecture identical. The optimisation of network parameters yielded good results when optimising towards the $\mathcal{L}_2$-loss, but could be substantially improved through our formulation of the turbulence loss $\mathcal{L}_\text{T}$.

When run in inference mode, the simulation based on the learned models trained with our method remained stable for long periods and allowed us to run simulations vastly surpassing the initial training horizon. Our models proved to be in very good agreement with the \gls{dns} test datasets when compared on the basis of \textit{a-posteriori} statistics. These agreements were obtained despite the fact that the evaluation metrics were not a target of the training optimisation, and that the test datasets constitute an extrapolation from training data. Furthermore, our hybrid approach achieved good results on a wide range of scales, with the Reynolds number varying from $Re=126$ to $Re=296$ in the isotropic turbulence case, and the vortex sizes ranging from $7\delta_{\omega_0}$ to $60\delta_{\omega_0}$ in the temporal mixing layer. Similarly, our approach yielded a learned model simulation that remained accurate and stable in a statistically steady test-case of the spatial mixing layer. These spatial mixing layer models were trained with a range of perturbation parameters and demonstrated good extrapolation accuracy towards this quantity. In our test-cases, the learned model simulation accurately reproduced the turbulence kinetic energy in its spectral distribution as well as its temporal evolution. Furthermore, the learned models captured the turbulent fluctuations, which lead to a precise modelling of vortex roll-up and merging events. Our results also demonstrate the importance of unrolling simulator steps during training in achieving high accuracy and stability. Such models are effectively trained by our approach of optimising all subranges of a multi-step training loop divided by gradient stopping. This approach differs from the common practice in machine learning, where gradients of early evaluations of the neural network are usually discarded or re-scaled when gradient clipping is applied \citep{pascanu2013OnThe}.
Our learned models provide a significant increase in computational performance, where speedups in terms of computation time of a factor of up to $14$ are observed. The additional resources required for model inference are minor and can be justified with the gains in the solution accuracy.

Using the turbulence loss and large unrollment numbers is motivated by physical and numerical considerations. As introduced in section \ref{section:learning_turbulence_models}, the components of the turbulence loss are derived from fundamental equations in turbulence theory. As described above, our experiments deem the solver unrollment imperative for training a long-term stable model. On a theoretical level, these principles apply to both 2D and 3D flows, which is why we believe that our findings are also of interest to the development of learned turbulence models for 3D flows.

In its current form, our method has several limitations, such as the initial one time cost to train the neural network turbulence model. Also, our tests have focused on regular, Cartesian grids. However, more flexible convolutions \citep{sanchez2020learning, ummenhofer2019lagrangian} could be employed to use the presented method on more flexible mesh structures with irregular discretisations. Moreover, even regular \gls{cnn}s can be extended to take regular, non-uniform and stretched meshes into account \citep{chen2021highacc}.
For instance, this is highly important for wall-bounded flows and fluid-structure interactions.
Similarly, further interesting extensions could work towards a differentiable solver that directly trains towards \textit{a-posteriori} statistics, or study the modelling capabilities of different network architectures with respect to the modelled turbulent scales. 

To summarize, the improvements in accuracy and runtime of our approach render the proposed combination of neural network and numerical solver suitable for a variety of settings. 
As ground truth data is not restricted to originate from the same solver, it could stem from different numerical schemes such as higher order spectral methods or even experiments. Furthermore, the learned models offer significant savings when a large quantity of turbulent simulations is required. This is especially important for inverse problems such as flow optimisation tasks.
Due to the super-linear scaling of existing solvers, our method also could potentially provide even greater performance benefits when applied to three dimensional flow fields.

\bigskip
\bigskip
\noindent\textbf{Funding}    This work was supported by European Research Council (ERC) Consolidator Grant CoG-2019-863850 (SpaTe).

\noindent\textbf{Declaration of interests}   The authors report no conflict of interests.

\noindent\textbf{Further information}   \url{https://github.com/tum-pbs/differentiable-piso}

%% file: sections/appendix.tex
\appendix

\section{PISO Solver Details}\label{appendix:piso_solver}
The governing Navier-Stokes equations \eqref{equation:navier_stokes} were solved with a Finite-Volume approach, which naturally supports the staggered discretisation such that the velocity vector fields are stored at the cell faces, whereas the scalar pressure field is stored at the cell centers. All fluxes were computed to second order accuracy using a central difference scheme.

\subsection{Governing equations}\label{appendix:solver_equations}
The numerical solver follows the method introduced by \citep{Issa1986}. Our differentiable hybrid method includes a corrective network forcing $\mathbf{f}_\text{CNN}$ in the predictor step. In contrast, the supervised models cannot take advantage of any differentiable solver operations during training. The corrective forcing from a network trained with the supervised approach $\mathbf{f}_\text{CNN}^{\text{sup}}$ must thus be applied after a complete solver step. With the discrete velocity and pressure fields  $(\mathbf{u}_n, p_n)$ at time $t_n$, the equations of the PISO solver for both cases read as
\begin{align}
    M\mathbf{u}_n^* &=\mathbf{u}_n - \nabla p_n \Big[ +\mathbf{f}_\text{CNN}(\mathbf{u}_n,\nabla p_n|\theta) \Big] ,\\
    \nabla \cdot \big(A^{-1} \nabla p_n^* \big) &= \nabla \cdot \mathbf{u}_n^* ,\\
    \mathbf{u}_n^{**} &= \mathbf{u}_n^* - A^{-1}\nabla p^*_n ,\\
    \nabla \cdot \big(A^{-1} \nabla p_n^{**} \big) &= \nabla \cdot \big(H\mathbf{u}_n^{**}\big) ,\\
    \mathbf{u}_n^{***} &= \mathbf{u}_n^{**} + A^{-1} \big(H(\mathbf{u}_n^{**}-\mathbf{u}_n^{*}) - \nabla p^{**}_n\big) ,\\
    p_{n+1} &= p_n+p^*+p^{**} ,\\
    \mathbf{u}_{i+1} &= \mathbf{u}_n^{***}\Big[+\mathbf{f}_\text{CNN}^{\text{sup}}(\mathbf{u}_n^{***},\nabla p_{n+1}|\theta^\text{sup})\Big],
\end{align}%
where the corrective forcings $\mathbf{f}_\text{CNN}$ and $\mathbf{f}_\text{CNN}^{\text{sup}}$ are never applied at the same time, but share this set of equations for brevity. The matrix $M$ represents the discretised advection, diffusion, and temporal integration, and matrix $A$ contains the diagonal entries of $M$ such that $M=A+H$. The network weights are represented by $\theta$.

The optimisation loss is applied to the output of a solver step. Using the downsampling $(\Tilde{\mathbf{u}}_n,\Tilde{p}_n)=q(\mathbf{u}_n,p_n) = \Tilde{q}_{n}$
as introduced in section \ref{section:learning_turbulence_models}, we can abbreviate a solver step by
$\Tilde{q}_{n+1} = \mathcal{S}_\tau(\Tilde{q}_{n}, \Tilde{\mathbf{f}}_{\text{CNN},n})$
in case of the differentiable model, and by
$\Tilde{q}_{n+1} = \mathcal{S}_\tau(\Tilde{q}_{n})+\Tilde{\mathbf{f}}_{\text{CNN},n}^{\text{sup}}$
in case of the supervised model. The parameter $\tau$ describes the temporal increment of a solver step as $\Delta t=\tau\Delta t_\mathrm{\gls{dns}}$. At this stage, it becomes obvious that optimising
$\min_{\theta} [\mathcal{L}(\Tilde{q}_{n+\tau},\mathcal{S}_\tau(\Tilde{q}_{n},\mathbf{f}_{\text{CNN},n}))]$
with the differentiable model, as introduced in equation \eqref{eq:general_train}, requires the computation of
\begin{equation}
\frac{\partial\mathcal{L}}{\partial\theta}=
\frac{\partial\mathcal{L}}{\partial\Tilde{q}_{n+1}}
\frac{\partial\Tilde{q}_{n+1}}{\partial\mathbf{f}_\text{CNN}}
\frac{\partial\mathbf{f}_\text{CNN}}{\partial\theta},
\end{equation}
which in turn requires the differentiation of a solver step. In contrast, optimising a supervised model with $\min_{\theta} [\mathcal{L}(\mathbf{u}_{n+\tau},\mathcal{S}_\tau(\tilde{\mathbf{u}}_n,\Tilde{p}_n)+\mathbf{f}_\text{CNN}^\text{sup}(\mathcal{S}_\tau(\tilde{\mathbf{u}}_n,\Tilde{p}_n)))]$ has to compute
\begin{equation}\label{eq:app_supervised_gradient}
\frac{\partial\mathcal{L}}{\partial\theta^\text{sup}}=
\frac{\partial\mathcal{L}}{\partial\mathbf{f}_\text{CNN}^\text{sup}}
\frac{\partial\mathbf{f}_\text{CNN}^\text{sup}}{\partial\theta^\text{sup}},
\end{equation}
which can be achieved without a differentiable solver.

\begin{figure}
    \centering
    \includegraphics[width=\textwidth]{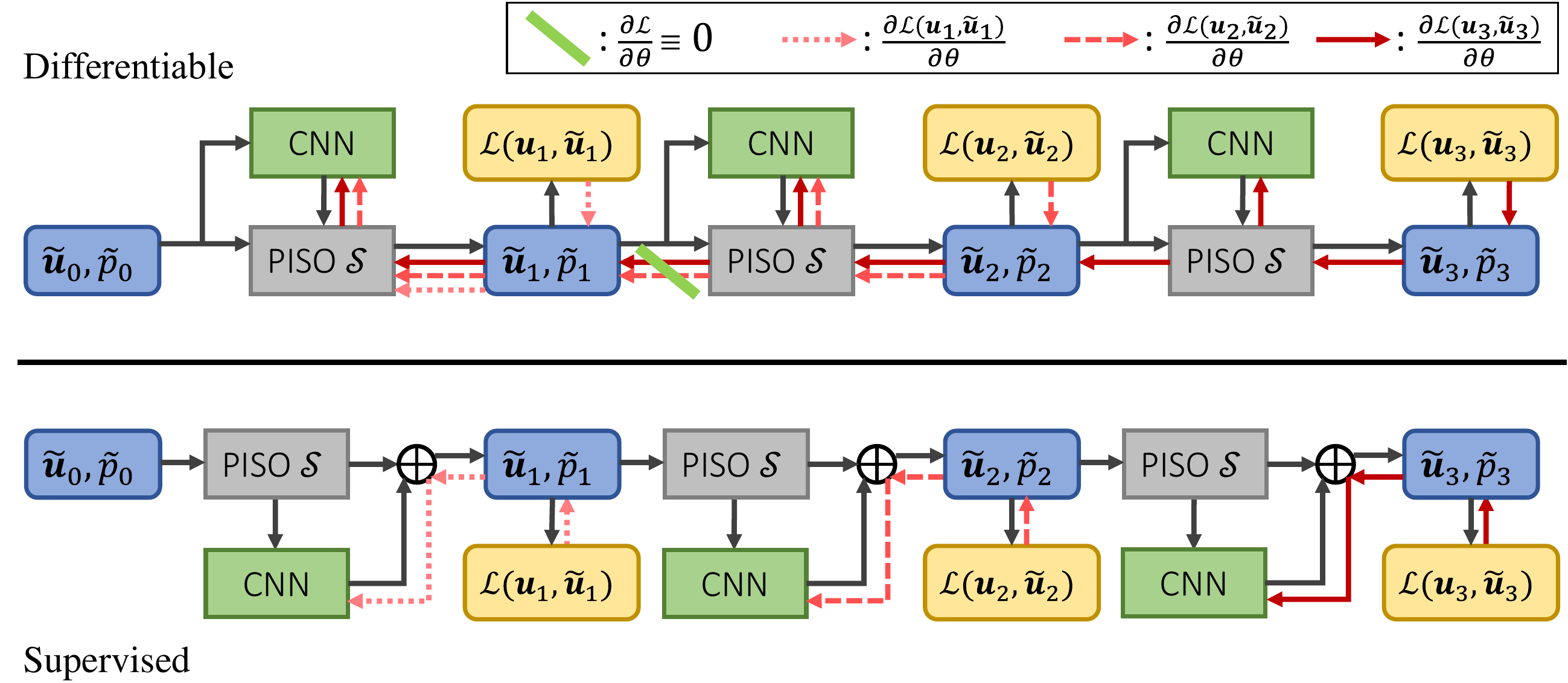}
    \caption{Visualisation of gradient back-propagation, comparing differentiable and supervised setups; displayed is a 3-step setup; the loss gradients from the last step are propagated through all previous steps and towards all previous network outputs; if the back-propagation is split into subranges, the gradients of the simulation state are set to zero, visualised by '\textcolor{green}{\textbf{\textbackslash}}'}
    \label{fig:unrollment_gradient}
\end{figure}

When $n$ solver steps are unrolled during training of differentiable models, this yields the optimisation procedure as introduced in equation \eqref{eq:multistep_train}
\begin{equation}
    \min_\theta\big(\sum_{s=0}^m\mathcal{L}(\tilde{q}_{n+s\tau},\mathcal{S}_\tau^s(\tilde{q}_n,\tilde{\mathbf{f}}_n))\big).
\end{equation}
During back-propagation, the gradients based on the losses at all (intermediate) steps are calculated and propagated through all previously unrolled forward steps, accumulating gradients with respect to all network forces on the way back. For a loss on an (intermediate) solver step $\mathcal{L}^s=\mathcal{L}(\tilde{q}_{n+s\tau},\mathcal{S}_\tau^s(\tilde{q}_n,\tilde{\mathbf{f}}_n))$, the following gradient calculation arises
\begin{equation}
\frac{\partial\mathcal{L}^s}{\partial\theta}= \sum_{B=1}^{s} \Big[
\frac{\partial\mathcal{L}^s}{\partial\Tilde{q}_{n+s}}
\Big(\prod_{b=s}^{B+1}\frac{\partial\Tilde{q}_{n+b}}{\partial\Tilde{q}_{n+b-1}}\Big)
\frac{\partial\Tilde{q}_{n+B}}{\partial\mathbf{f}^{B-1}_\text{CNN}}
\frac{\partial\mathbf{f}^{B-1}_\text{CNN}}{\partial\theta}\Big],
\end{equation}
where $\mathbf{f}^B_\text{CNN}$ denotes the network forcing in the $B^\text{th}$ step. As explained in section \ref{section:backpropagation}, we use a custom gradient splitting technique that splits the back-propagation into subranges. The gradients are only back-propagated within a subrange, and set to zero when they cross a subrange boundary. When using gradient subranges of length $r$, the gradient calculation gives
\begin{equation}
\frac{\partial\mathcal{L}^s}{\partial\theta}= \sum_{B=\lfloor s\backslash r\rfloor*r}^{s} \Big[
\frac{\partial\mathcal{L}^s}{\partial\Tilde{q}_{n+s}}
\Big(\prod_{b=s}^{B+1}\frac{\partial\Tilde{q}_{n+b}}{\partial\Tilde{q}_{n+b-1}}\Big)
\frac{\partial\Tilde{q}_{n+B}}{\partial\mathbf{f}^{B-1}_\text{CNN}}
\frac{\partial\mathbf{f}^{B-1}_\text{CNN}}{\partial\theta}\Big],
\end{equation}
where $\lfloor s\backslash r\rfloor$ denotes the integer division of $s$ by $r$. This formulation can be easily implemented by setting the gradient of the simulation state to zero at the subrange boundaries, as visiualised in figure \ref{fig:unrollment_gradient}.

Supervised models train on the optimisation
\begin{equation}
    \min_\theta\big(\sum_{s=0}^m\mathcal{L}(\tilde{q}_{n+s\tau},
    \big[\mathcal{S}_\tau(\tilde{q}_n) + \tilde{\mathbf{f}}^{\text{sup}}_{\text{CNN},n}\big]^s)\big),
\end{equation}
where the expression in $[...]^s$ denotes the recurrent application of a solver step with a supervised model. We abbreviate for simplicity $\mathcal{L}^{\text{sup},s}=\mathcal{L}(\tilde{q}_{i+s\tau},\big[\mathcal{S}_\tau(\tilde{q}_i) + \tilde{\mathbf{f}}^{\text{sup}}_{\text{CNN},i}\big]^s)$. The gradients of these losses are only calculated within the locality of an (intermediate) solution and are thus a trivial extension of equation \eqref{eq:app_supervised_gradient}
\begin{equation}
\frac{\partial\mathcal{L}^{\text{sup},s}}{\partial\theta^\text{sup}}=
\frac{\partial\mathcal{L}^{\text{sup},s}}{\partial\mathbf{f}_\text{CNN}^{\text{sup},s}}
\frac{\partial\mathbf{f}_\text{CNN}^{\text{sup},s}}{\partial\theta^\text{sup}}.
\end{equation}
The training unrollment and its gradient back-propagation for differentiable hybrid as well as supervised models is visualised in figure \ref{fig:unrollment_gradient}.

\subsection{Implementation}\label{appendix:implementation}
The presented method was implemented using the tensor operation library \textit{TensorFlow} \citep{199317}. This framework supports the \gls{gpu} based execution of various linear algebra operations, however does not support sparse matrix data at the time of this project. Consequently, a series of custom operations surrounding the linear solves for advection-diffusion and pressure in the PISO scheme were added to facilitate an efficient, \gls{gpu}-based execution of the solver. The back-propagation gradients of the custom linear solves $Ax=b$ were linearised around their respective matrices and thus read as $A^T\hat{b} = \hat{x}$, where $\hat{x}$ and $\hat{b}$ represent the incoming and outgoing back-propagation gradients of the linear solve operation. This yields a solver that can flexibly change the number of steps unrolled during training (only limited by \gls{gpu} memory and computation time), and account for any loss functions or network architectures. Access to our code is provided through the following GitHub page: \url{https://github.com/tum-pbs/differentiable-piso}

\subsection{Solver Verification} \label{appendix:verification}

\begin{figure}
    \centering
    \includegraphics[width=.5\textwidth]{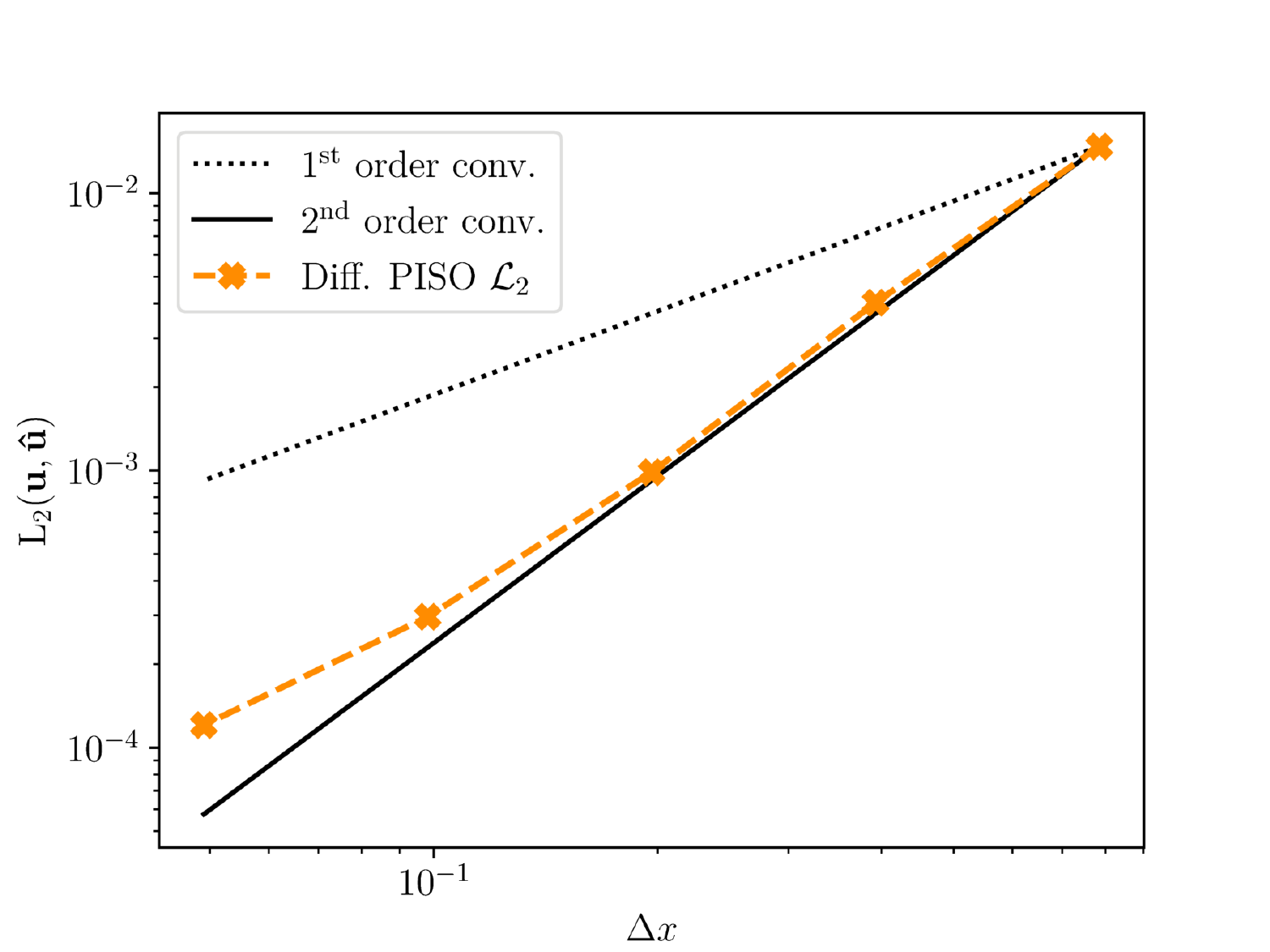}
    \caption{Grid convergence study, the numerical error on the Taylor-Green vortex with respect to analytical data converges with second order}
    \label{fig:taylor_green_convergence}
\end{figure}

\begin{figure}
\centering
\begin{subfigure}[b]{\textwidth}
    \centering
    \includegraphics[width=\textwidth]{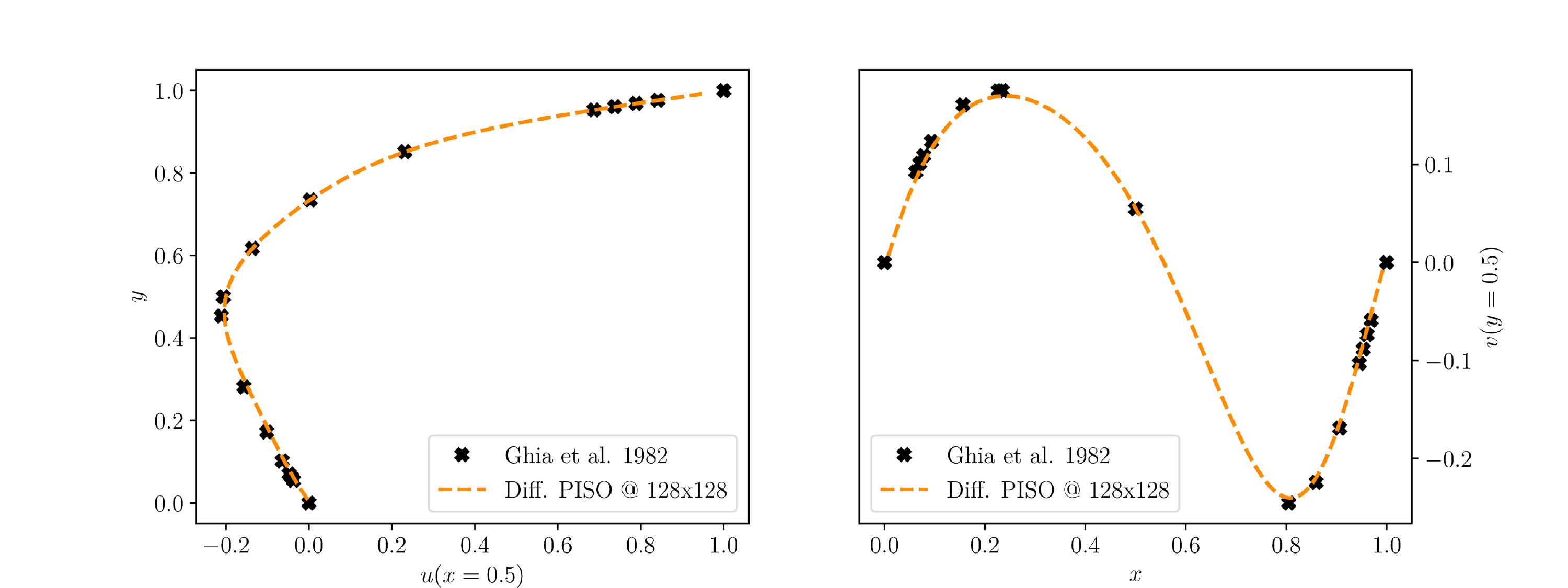}
    \caption{}
    \label{fig:lid_driven_cavity_Re100}
\end{subfigure}
\centering
\begin{subfigure}[b]{\textwidth}
    \centering
    \includegraphics[width=\textwidth]{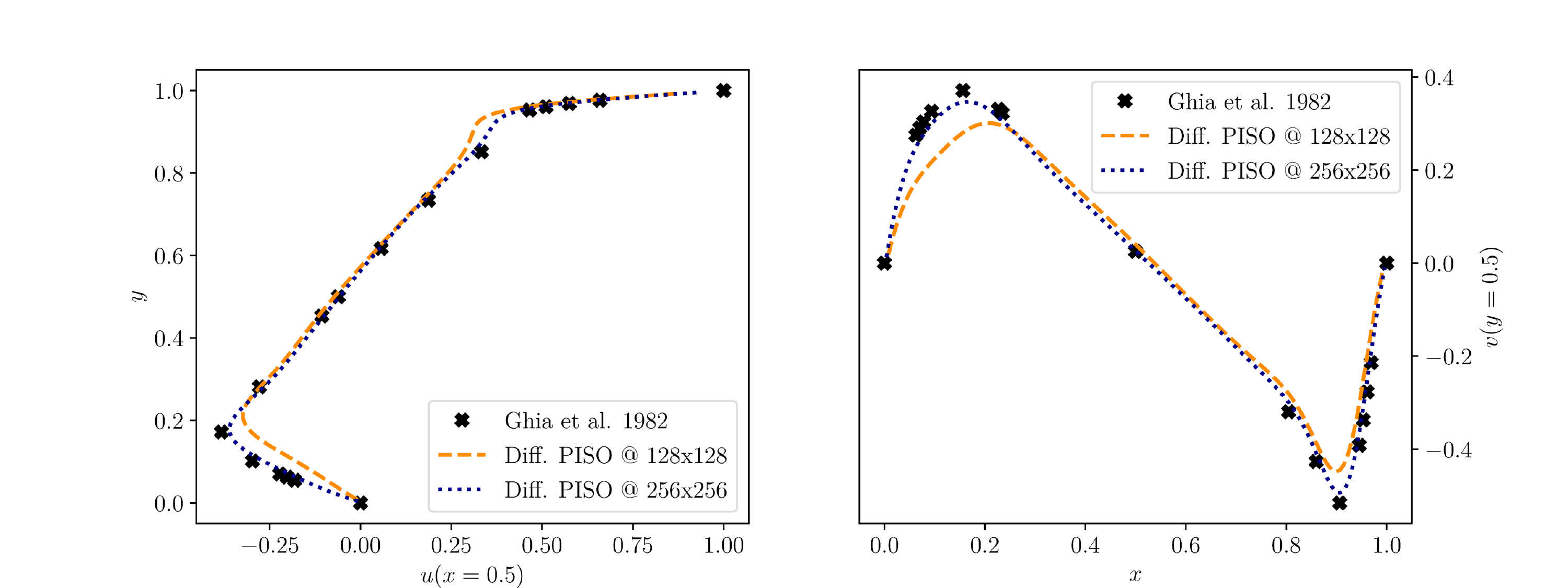}
    \caption{}
    \label{fig:lid_driven_cavity_Re1000}
\end{subfigure}
    \caption{Lid-driven cavity verification case, figures show the domain-center velocities for $Re=100$ in (a), and $Re=1000$ in (b), in comparison to numerical benchmark data by \cite{ghia1982high}}
\end{figure}

Our implementation is verified on two standardised simulations. Firstly, we study the grid convergence properties on the two-dimensional Taylor-Green vortex decay. This flow scenario is simulated on a periodic, square domain of size $(L_x,L_y)=(2\pi, 2\pi)$ and initialised with the analytical solution
\begin{equation} \label{equation:taylor_green}
\begin{aligned}
    \hat u(x,y,t) = -&\cos(x)\sin(y)\exp{\Big(\frac{-2t}{Re}\Big)},\\
    \hat v(x,y,t) = &\sin(x)\cos(y)\exp{\Big(\frac{-2t}{Re}\Big)},\\
    \hat p(x,y,t) = -&\frac{\cos(2x)+\cos(2y)}{4}\exp{\Big(\frac{-4t}{Re}\Big)},
\end{aligned}
\end{equation}
where the Reynolds number is set to $Re=10$. The grid resolution is varied between $N_x=N_y=[8,\hspace{0.2em} 16,\hspace{0.2em} 32,\hspace{0.2em} 64,\hspace{0.2em} 128]$. The governing equations \eqref{equation:navier_stokes} are integrated until $t=2$ is reached, while a small timestep of $\Delta t=10^{-3}$ is chosen for all resolutions. Figure \ref{fig:taylor_green_convergence} depicts the normalised error of the numerical solution $\mathbf{u}=(u,\hspace{0.2em} v)^T$ with respect to the analytical solution from equation \eqref{equation:taylor_green}, computed as $L_2=\frac{\sum_{i,j}{(u_{i,j}-\hat u_{i,j})^2+(v_{i,j}-\hat v_{i,j})^2}}{N_xN_y}$. This demonstrates second-order convergence of our implementation.
Secondly, we verify the solver on numerical benchmark data for a lid-driven cavity flow. This case consists of a fluid domain of size $(L_x,L_y)=(1,1)$ with no-slip wall boundaries enforcing $u(y=0)=0$, $v(x=0)=0$, $v(x=1)=0$, and $u(y=1)=1$ for the lid. Our simulations are performed at two different Reynolds numbers. For $Re=100$, the steady state is approximated by running the simulation until $t=10$ on a $(N_x,N_y)=128,128$ grid. We verify our solver by comparing the velocities at the domain-center cross-sections to the benchmark solutions reported by \cite{ghia1982high}. The results are shown in figure \ref{fig:lid_driven_cavity_Re100}. Similarly, the evaluations for simulations at $Re=1000$ on $128\times128$ and $256\times256$ grids are shown in figure \ref{fig:lid_driven_cavity_Re1000}. Both cases show good agreement with the benchmark data for sufficiently high resolutions.

\section{Convolutional Neural Network}\label{appendix:cnn}
Our turbulence models are parameterised by a \gls{cnn}, and thus formed by the kernel weights in each convolutional layer. Our setup utilises $7$ layers with kernel sizes $[7, \hspace{0.2em}5,\hspace{0.2em}5,\hspace{0.2em}3,\hspace{0.2em}3,\hspace{0.2em}1,\hspace{0.2em}1]$ and \textit{leaky ReLU} activation functions. The input to the network consists of the velocity and pressure gradient vector fields, yielding $4$ channels in total. The layers then operate on  $[8,\hspace{0.2em}8,\hspace{0.2em}16,\hspace{0.2em}32,\hspace{0.2em}32,\hspace{0.2em}32]$ channels respectively and output a forcing vector field with $2$ channels. Consequently, the network consist of $\sim 82\times10^3$ trainable weights contained in the kernels.

The structure of this network resembles an encoder network, where the larger kernel size in the first layers increases the receptive field of the convolution. The potential complexity of the function parameterised by the network is largely dependent on the channel widths and layer count. We have found the described architecture to work well for turbulence modelling, without overfitting to training data, as larger models are more likely to do.

By the nature of the discrete convolution operation, the output size shrinks with each layer. At periodic boundaries this can be counteracted by padding the input with real data. At other boundaries, where no periodicity is enforced, no padding procedure is used on the input to avoid feeding unphysical data. In these cases, the output of the \gls{cnn} does not coincide with the grid dimensions and is accordingly padded with zeros. Prior to training, the weights were initialised using the \textit{Glorot Normal} initialisation.

\section{Training Procedure}\label{appendix:training}
Our method trains neural networks to model the effect of turbulent motion. These effects are implicitly learnt from high-resolution \gls{dns} simulations by reproducing their behaviour. Our training procedure uses the commonly chosen Adam optimizer \cite{kingma2015adam}.
During one optimisation step $o$, Adam takes the loss gradient as specified in appendix \ref{appendix:implementation} and applies a weight update according to
\begin{algorithmic}
    \State $g_{o} \gets \frac{\partial\mathcal{L_o}}{\partial\theta_{o-1}}$
    \State $m_{o} \gets \beta_1 m_{o-1}+(1-\beta_1)g_{o}$
    \State $v_{o} \gets \beta_2v_{o-1}+(1-\beta_2)g_{o}^2$
    \State $\hat m_{o} \gets m_{o}/(1-\beta_1^o)$
    \State $\hat v_{o} \gets v_{o}/(1-\beta_2^o)$
    \State $\phi_{o} \gets \phi_{o} -\alpha \frac{\hat m_{o}}{\sqrt{\hat v_{o}}+\epsilon}$
\end{algorithmic}
where $m_{o}$ and $v_{o}$ are exponential moving averages approximating the mean and variance of the gradient. To account for the initialisation error in these approximates, the corrected variables $\hat m_{o}$ and $\hat v_{o}$ are introduced, see the original publication for further details. We set the bias corrections to the standard values $\beta_1=0.9$, $\beta_2=0.999$.
The networks were trained with a learning rate of $1\times10^{-5}$ and a learning-rate decay factor of $0.4$.
We found that the training procedure was stable for learning rates in the neighbourhood of that value, however no extensive hyper-parameter tuning was performed. Contrary, we found the unrollment number $s$ (see equation \eqref{eq:multistep_train}) to have great effect on the training procedure. Newly initialised models can cause the accumulation of erroneous structures and subsequently solver divergence in long unroll times. To mitigate this effect, the models trained on more than $10$ steps were initialised with a pre-trained network from a $10$-step model. The parameter optimisations were run until no further significant decrease in loss values is observed.

\section{Large Eddy Simulation with the Smagorinsky Model}\label{appendix:smagorinsky}
A series of tests were conducted to select an appropriate value for the Smagorinsky coefficient used in the isotropic decaying turbulence simulation in section \ref{section:isotropic_turbulence}. We ran simulations with our usual downscaling of $8\times$ in space and time and coefficients from $C_s = [0.17, \hspace{0.2em} 0.08, \hspace{0.2em} 0.02, \hspace{0.2em} 0.008, \hspace{0.2em} 0.002]$. The velocity-MSE of these simulations with respect to the \gls{dns} test-data after $100\Delta t$ were evaluated to $[12.21, \hspace{0.2em} 6.824, \hspace{0.2em} 4.320, \hspace{0.2em} 4.256, \hspace{0.2em} 4.364]\times10^{-3}$. Based on that analysis, $C_s=0.008$ was chosen for further consideration. This value is relatively low in comparison to other common choices, such as the default coefficient of $C_s=0.17$ for 3D turbulence \citep{pope2000turbulent}. Since 2D isotropic turbulence is largely dependent on the backscatter effect that transfers energy from small to large scales, lower $C_s$ are applicable \citep{smith1996crossover}. With the strictly dissipative behaviour of the Smagorinsky model, larger $C_s$ lead to an overly powerful dampening of fine scale motions that quickly decreases the turbulence kinetic energy. While backscatter is important to many flow scenarios \citep{biferale2021inverse}, especially 3D turbulence scenarios may rather have significant forward diffusion, which would be more favourable towards dissipative models like the Smagorinsky model \citep{kraichnan1967inertial}. Nevertheless, this showcases an inherent benefit of learned turbulence models, where no scenario dependent modelling assumptions are necessary.

\section{Supervised Models} \label{appendix:supervised_models}
\begin{figure}
\centering
\begin{subfigure}[c]{.5\textwidth}
    \centering
    \includegraphics[width=\textwidth]{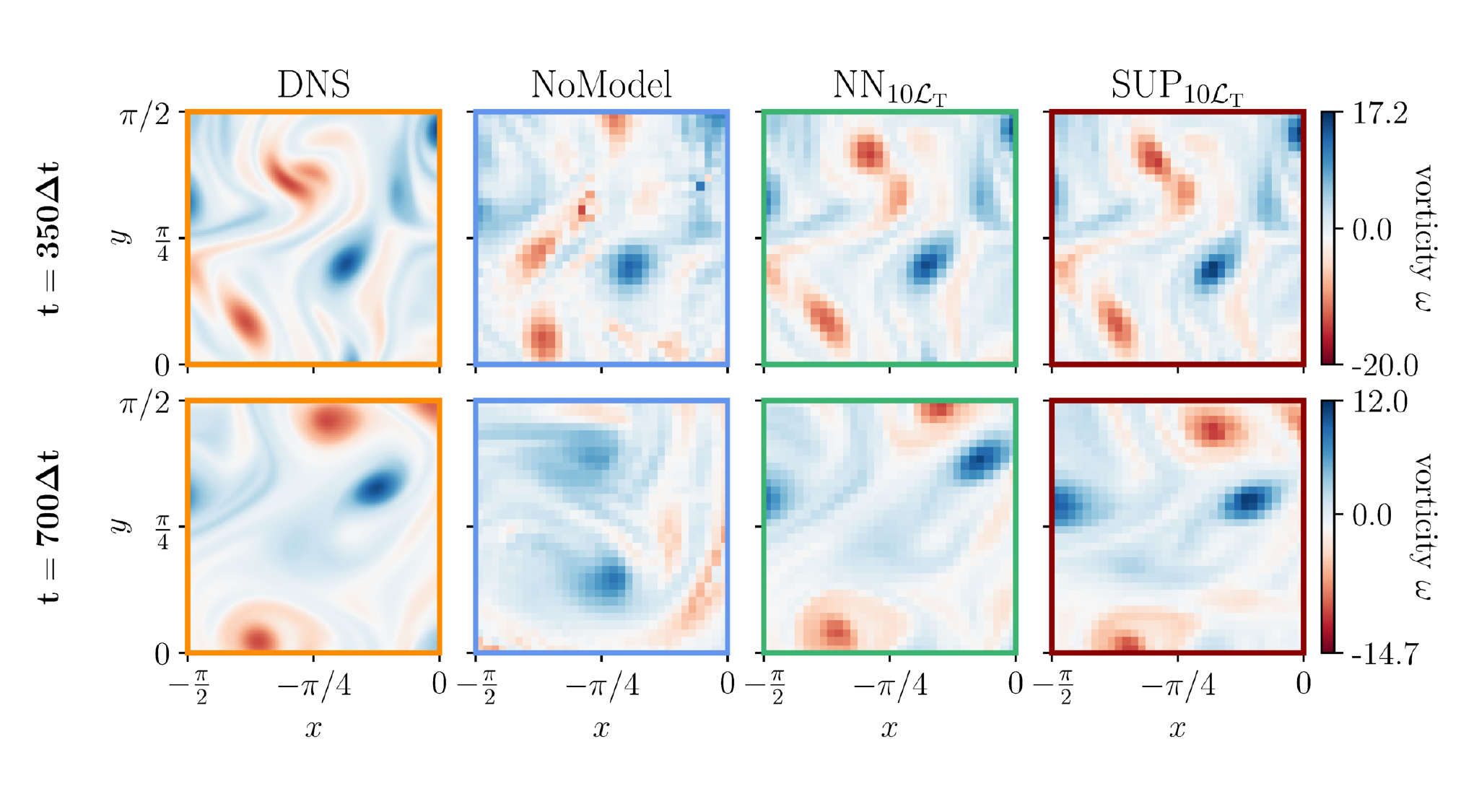}
    \caption{}
\end{subfigure}%
\begin{subfigure}[c]{.5\textwidth}
    \centering
    \includegraphics[width=\textwidth]{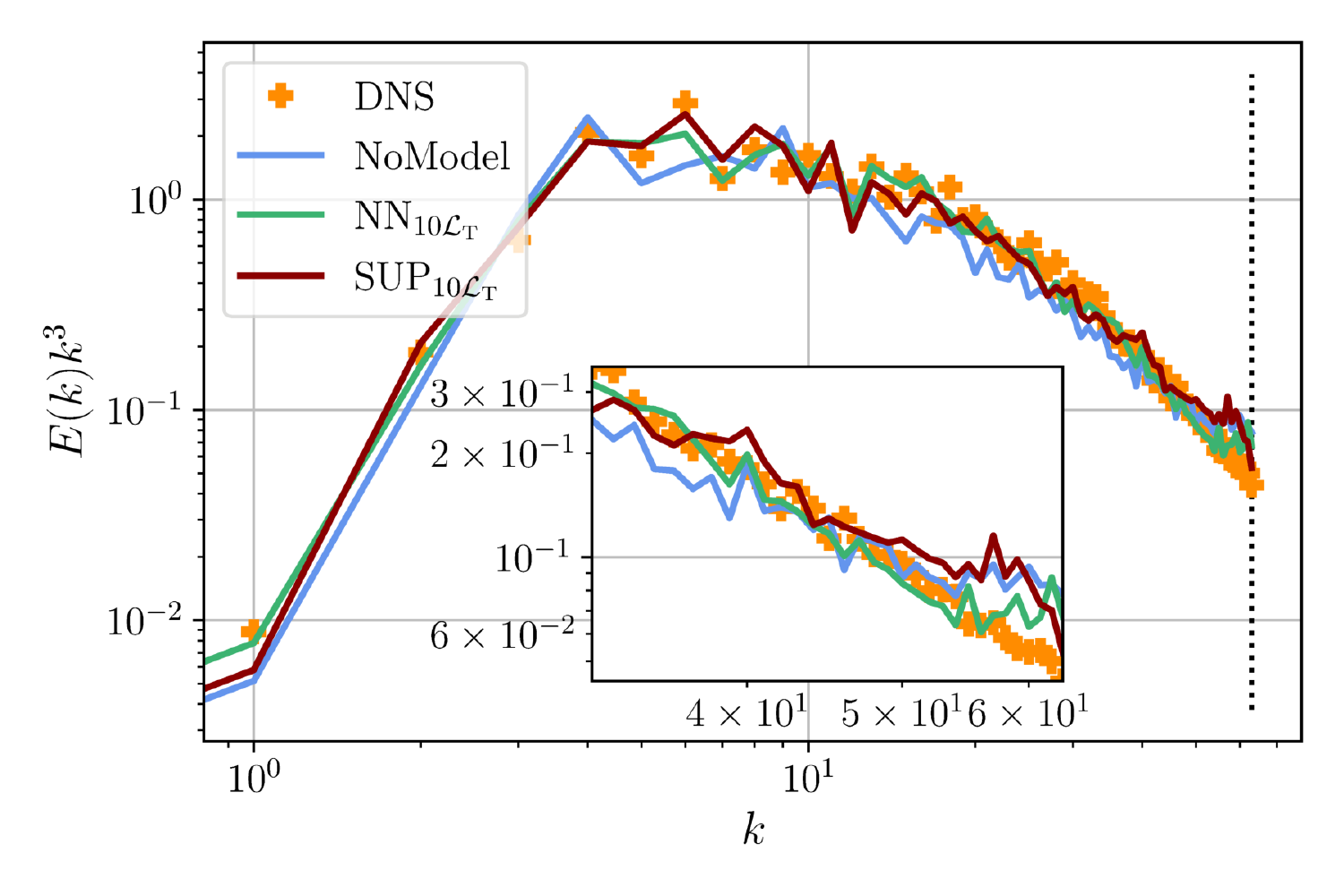}
    \caption{}
\end{subfigure}
    \caption{Comparison of \gls{dns}, no-model, and learned model simulations trained with the adjoint-based method and with a supervised method on isotropic decaying turbulence; evaluation with respect to vorticity (a); and resolved turbulence kinetic energy spectra (b)}
    \label{fig:appendix_supervised_isotropic}
\end{figure}

\begin{figure}
\centering
\begin{subfigure}[c]{.7\textwidth}
    \centering
    \includegraphics[width=\textwidth]{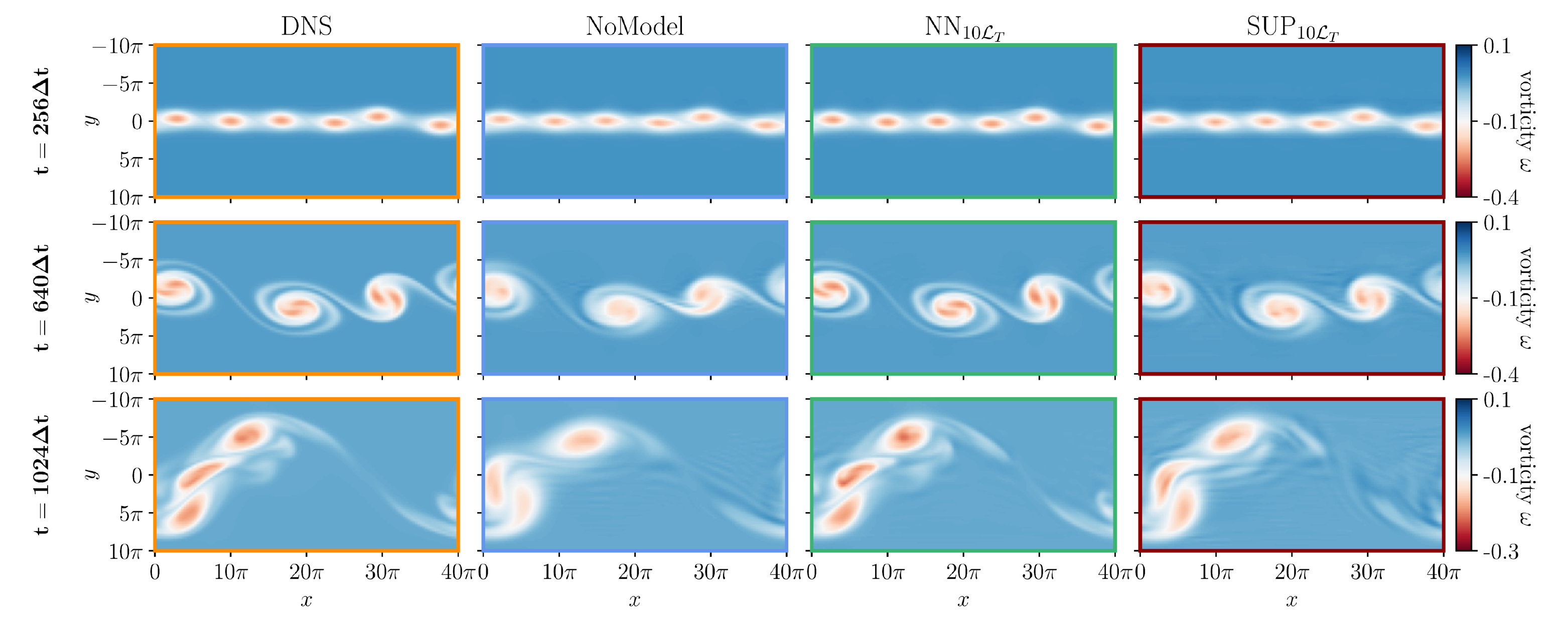}
    \caption{}
\end{subfigure}
\centering
\begin{subfigure}[c]{.8\textwidth}
    \centering
    \includegraphics[width=\textwidth]{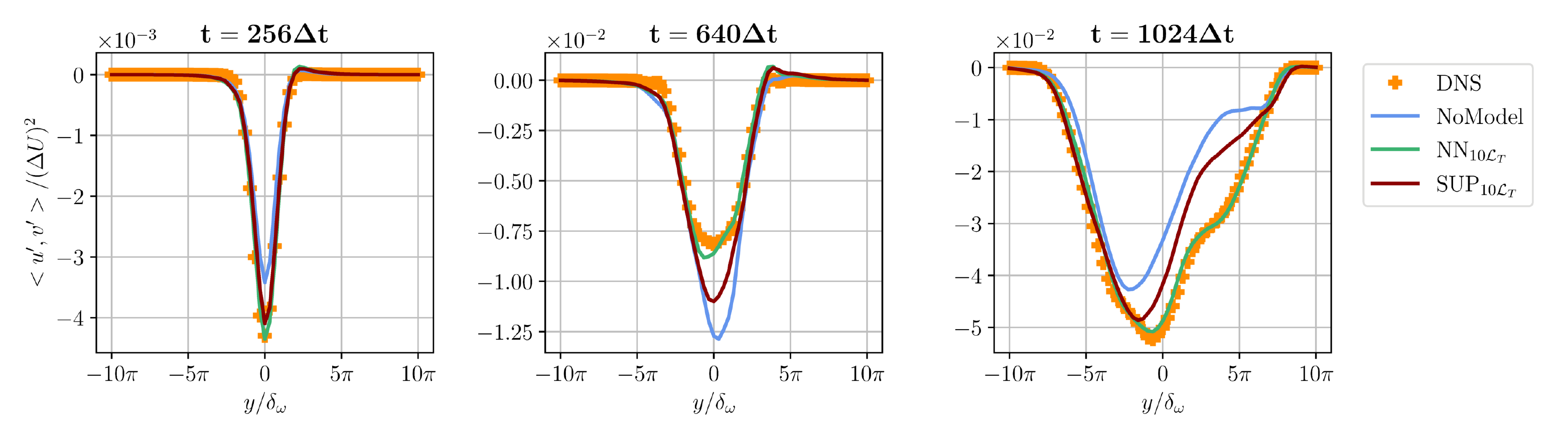}
    \caption{}
\end{subfigure}
\centering
\begin{subfigure}[c]{.8\textwidth}
    \centering
    \includegraphics[width=\textwidth]{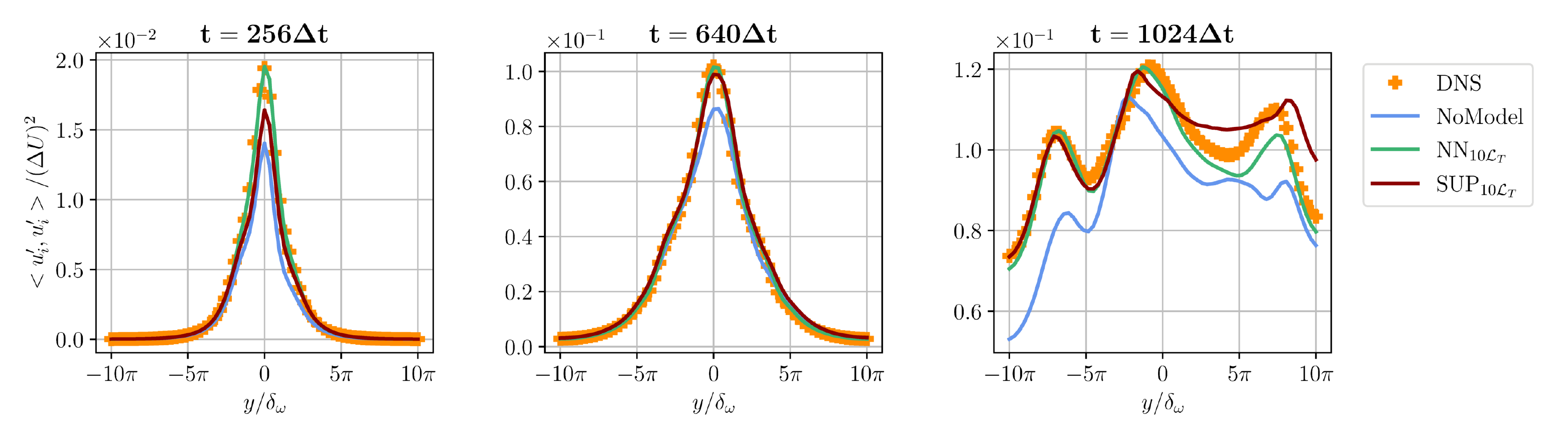}
    \caption{}
\end{subfigure}
    \caption{Comparison of \gls{dns}, no-model, and learned model simulations trained with the adjoint-based method and with a supervised method on temporal mixing layers; evaluation with respect to vorticity (a); resolved reynolds stresses (b); and resolved turbulence kinetic energy (c)}
    \label{fig:appendix_supervised_temporal}
\end{figure}

A core point of the experiments in the main section is the temporal unrollment during training, and substantial accuracy improvements of the differentiable models is achieved by this procedure. As illustrated in appendix \ref{appendix:solver_equations}, the temporal unrollment has less severe effects on the optimisation equations of supervised models. Despite this, considerable accuracy improvements are achieved by exposing the supervised training to multiple steps. Nevertheless, models trained with a differentiable approach outperform these improved supervised models, when all other parameters are kept constant, as revealed by our experiments on supervised models. For this, we trained 10-step supervised models for the isotropic decaying turbulence and temporal mixing layer cases. Figures \ref{fig:appendix_supervised_isotropic} and \ref{fig:appendix_supervised_temporal} depict evaluations on the spectral energy for isotropic turbulence, Reynolds stresses and turbulence kinetic energy for the temporal mixing layer, as well as vorticity visualisations for both. For the isotropic case, the supervised model comes remarkably close to the differentiable counterpart, and only shows slight over-estimation of fine-scale energies. For more complex flow like temporal mixing layers, it is clearer that differentiable models outperform supervised ones.

\section{Loss Ablation} \label{appendix:loss_ablation}
To test the effects of the loss terms introduced in section \ref{section:learning_turbulence_models}, we perform an ablation study on the loss term. A series of 10-step models are trained with identical initialisation, data shuffling, and learning rate, but variations in loss composure. These tests are conducted on all three flow scenarios. The loss factors $\lambda$ are identical to the ones used in the main sections, where the values are set to yield similar loss contributions for each loss term. An exception is $\lambda_2$, which was chosen to give a $10\times$ larger contribution in order to steer an initialised network into reproducing \gls{dns} structures. We then perform evaluations based on our out-of-sample test datasets. The results are summarised in table \ref{table:loss_ablation}. Our evaluations include three metrics. The first is an instantaneous MSE on the velocity field. Secondly, we assess the performance with respect to the turbulence kinetic energy by using an instantaneous MSE for isotropic turbulence, an MSE on spatially averaged energy for the temporal mixing layer, and the MSE on temporally averaged data for the spatial mixing layer. Lastly, we assess the energy distribution over spectral wavelengths, which is based on a 2D evaluation for isotropic turbulence, a cross-section analysis for the temporal mixing layer, and a centerline analysis for the spatial mixing layer. Additionally, two temporal snapshots were considered, a short $64\Delta t$ distance and a longer one, which was set to $1000 \Delta t$ for all setups except for the spatial mixing layer, where stability concerns limited the horizon to $500 \Delta t$.

The results indicate that the baseline $\mathcal{L}_2$ loss only performs well on short temporal horizons, while its performance deteriorates over longer evaluations. The tests on decaying turbulence and temporal mixing layers generally show best results with a combination of  $\mathcal{L}_2$, $\mathcal{L}_E$ and $\mathcal{L}_\mathcal{S}$ over longer temporal horizons. The only exception is the spectral energy analysis in the temporal mixing layer, where an addition of $\mathcal{L}_{\text{MS}}$ outperforms this combination by a small margin. Due to the fact that this margin is minor compared to the improvements of the $\mathcal{L}_2,\mathcal{L}_E, \mathcal{L}_\mathcal{S}$ combination on the long horizons, we conclude that including the temporal averaging loss is not beneficial in the flow scenarios that are not statistically steady. In contrast, the evaluations of the spatial mixing layer reveals that incremental additions of the turbulence loss terms $\mathcal{L}_E$, $\mathcal{L}_\mathcal{S}$ and $\mathcal{L}_{\text{MS}}$ yield better performance for each addition. Thus, we conclude that using all loss terms is beneficial in this case.

\begin{table}
 \centering
 \begin{tabular}{c l c c c c c c c}
 &  & \multicolumn{3}{c}{Time $t_1$} & & \multicolumn{3}{c}{Time $t_2$} \\
 \hline
 & Loss & $\text{MSE}(\mathbf{u})$ & $\text{MSE}(\mathbf{k})$ & $\sum\frac{E(k)_\mathbf{u}}{E(k)_\mathbf{\tilde{u}}}{-1}$& \ \ \ \ &
 $\text{MSE}(\mathbf{u})$ & $\text{MSE}(\mathbf{k})$ & $\sum{ \frac{E(k)_\mathbf{u}}{E(k)_\mathbf{\tilde{u}}}{-1}}$ \\
 \hline
 \parbox[t]{4mm}{\multirow{4}{*}{\rotatebox[origin=c]{90}{IDT}}} & $\mathcal{L}_2$
 & $\mathbf{5.45\mathrm{e}{-4}}$ & $\mathbf{1.96\mathrm{e}{-4}}$ & $7.256$ &
 & $0.182$ & $0.0317$ & $-19.27$\\

 & $\mathcal{L}_2$,$\mathcal{L}_E$
 & $5.80\mathrm{e}{-4}$ & $2.05\mathrm{e}{-4}$ & $\mathbf{3.598}$ &
 & $0.168$ & $0.0276$ & $-19.14$\\

 & $\mathcal{L}_2$,$\mathcal{L}_E$,$\mathcal{L}_\mathcal{S}$
 & $5.79\mathrm{e}{-4}$ & $2.04\mathrm{e}{-4}$ & $3.682$ &
 & $\mathbf{0.166}$ & $\mathbf{0.0271}$ & $\mathbf{-18.92}$\\

 & $\mathcal{L}_2$,$\mathcal{L}_E$,$\mathcal{L}_\mathcal{S}$,$\mathcal{L}_\text{MS}$  \ \
 & $5.70\mathrm{e}{-4}$ & $2.01\mathrm{e}{-4}$ & $4.373$ &
 & $0.182$ & $0.0332$ & $-21.20$\\
 \hline
 \parbox[t]{4mm}{\multirow{4}{*}{\rotatebox[origin=c]{90}{TML}}} & $\mathcal{L}_2$
 & $9.92\mathrm{e}{-7}$ & $\mathbf{1.22\mathrm{e}{-8}}$ & $\mathbf{-0.307}$ &
 & $1.39\mathrm{e}{-3}$ & $4.53\mathrm{e}{-5}$  & $5.109$ \\

 & $\mathcal{L}_2$,$\mathcal{L}_E$
 & $1.44\mathrm{e}{-6}$ & $2.28\mathrm{e}{-8}$ & $-0.613$ &
 & $1.10\mathrm{e}{-3}$ & $5.44\mathrm{e}{-5}$  & $4.951$ \\

 & $\mathcal{L}_2$,$\mathcal{L}_E$,$\mathcal{L}_\mathcal{S}$
 & $8.59\mathrm{e}{-7}$ & $1.97\mathrm{e}{-8}$ & $-1.261$ &
 & $\mathbf{2.15\mathrm{e}{-4}}$ & $\mathbf{1.78\mathrm{e}{-5}}$  & $4.155$ \\

 & $\mathcal{L}_2$,$\mathcal{L}_E$,$\mathcal{L}_\mathcal{S}$,$\mathcal{L}_\text{MS}$
 & $\mathbf{4.83\mathrm{e}{-7}}$ & $1.57\mathrm{e}{-8}$ & $-1.572$ &
 & $9.11\mathrm{e}{-4}$ & $2.85\mathrm{e}{-5}$  & $\mathbf{4.142}$ \\
 \hline
 \parbox[t]{4mm}{\multirow{4}{*}{\rotatebox[origin=c]{90}{SML}}} & $\mathcal{L}_2$
 & $2.29\mathrm{e}{-4}$ & $4.11\mathrm{e}{-6}$ & $62.01$ &
 & $0.0243$ & $2.67\mathrm{e}{-4}$ & $500.7$\\

 & $\mathcal{L}_2$,$\mathcal{L}_E$
 & $1.60\mathrm{e}{-4}$ & $4.14\mathrm{e}{-6}$ & $52.15$ &
 & $0.0260$ & $2.21\mathrm{e}{-4}$ & $454.5$\\

 & $\mathcal{L}_2$,$\mathcal{L}_E$,$\mathcal{L}_\mathcal{S}$
 & $1.07\mathrm{e}{-4}$ & $3.46\mathrm{e}{-6}$ & $50.85$ &
 & $0.0127$ & $6.72\mathrm{e}{-5}$ & $457.7$\\

 & $\mathcal{L}_2$,$\mathcal{L}_E$,$\mathcal{L}_\mathcal{S}$,$\mathcal{L}_\text{MS}$
 & $\mathbf{3.86\mathrm{e}{-5}}$ & $\mathbf{1.34\mathrm{e}{-6}}$ & $\mathbf{27.65}$ &
 & $\mathbf{0.0025}$ & $\mathbf{2.74\mathrm{e}{-5}}$ & $\mathbf{216.0}$\\
 \hline
 \end{tabular}
 \caption{Loss ablation study for the used flow scenarios, Isotropic Decaying Turbulence (IDT), Temporal Mixing Layer (TML) and Spatial Mixing Layer (SML);
 $t_1=64\Delta t=512\Delta t_\text{\gls{dns}}$ and $t_2=[1000,\hspace{0.2em} 1000 ,\hspace{0.2em} 500]\Delta t$ for IDT, TML, SML respectively;
 MSE$(k)$ is evaluated on \textit{instantaneous} turbulent kinetic energy fields for IDT, and on \textit{spatially/temporally} averaged fields for TML and SML;
 $\sum{ \frac{E(k)_\mathbf{u}}{E(k)_\mathbf{\tilde{u}}}{-1}}$ is evaluated  on 2-D spectral analysis for IDT, cross-sectional spectra for TML, and centerline spectra for SML }
 \label{table:loss_ablation}
\end{table}